\documentclass[prd,reprint, showpacs,nofootinbib,unsortedaddress]{revtex4}
\pdfoutput=1

\usepackage{amsfonts}
\usepackage{amsmath}
\usepackage{amssymb}
\usepackage{aas_macros}
\usepackage{upgreek}
\usepackage{bm}
\usepackage{dcolumn}
\usepackage{epsfig}
\usepackage{graphicx}
\usepackage{graphics}
\usepackage{placeins}
\usepackage{subfigure}

\usepackage{color} 

\def\be{\begin{equation}}
\def\ee{\end{equation}}
\def\bi{\begin{itemize}}
\def\ei{\end{itemize}}
\def\ben{\begin{enumerate}}
\def\een{\end{enumerate}}

\def\edth{\eth}

\newcommand{\mb}[1]{\mathbf{#1}}

\begin{document}

\title{Mapping gravitational-wave backgrounds using methods from CMB analysis:
Application to pulsar timing arrays}

\author{Jonathan Gair}
\affiliation{Institute of Astronomy, 
University of Cambridge, 
Madingley Road, Cambridge, CB3 0HA, UK}

\author{Joseph D.~Romano}
\affiliation{Department of Physics and Astronomy 
and Center for Gravitational-Wave Astronomy,
University of Texas at Brownsville, 
Brownsville, TX 78520, USA}

\author{Stephen Taylor}
\affiliation{Institute of Astronomy, 
University of Cambridge, 
Madingley Road, Cambridge, CB3 0HA, UK}

\author{Chiara M.~F.~Mingarelli}
\affiliation{School of Physics and Astronomy, 
University of Birmingham, 
Edgbaston, Birmingham B15 2TT, UK}
\affiliation{Max Planck Institute for Radio Astronomy, 
Auf dem H\"{u}gel 69, D-53121 Bonn, Germany }
\affiliation{Theoretical Astrophysics, California Institute of Technology, 
1200 E California Blvd., M/C 350-17, Pasadena, CA 91125, USA }

\date{\today}


\begin{abstract}
We describe an alternative approach to the analysis of
gravitational-wave backgrounds, based on the formalism used to
characterise the polarisation of the cosmic microwave background.  In
contrast to standard analyses, this approach makes no assumptions
about the nature of the background and so has the potential to reveal
much more about the physical processes that generated it. An arbitrary
background can be decomposed into modes whose angular dependence on
the sky is given by gradients and curls of spherical harmonics. We
derive the pulsar timing overlap reduction functions for the 
individual modes, which are given by simple combinations of spherical 
harmonics evaluated at the pulsar locations.  
We show how these can be used to
recover the components of an arbitrary background, giving explicit
results for both isotropic and anisotropic uncorrelated
backgrounds. We also find that the response of a pulsar timing array
to curl modes is identically zero, so half of the gravitational-wave
sky will never be observed using pulsar timing, no matter how many
pulsars are included in the array.  An isotropic, unpolarised and 
uncorrelated background can be accurately represented using only three modes,
and so a search of this type will be only slightly more complicated
than the standard cross-correlation search using the Hellings and
Downs overlap reduction function. However, by measuring the components of individual
modes of the background and checking for consistency with isotropy,
this approach has the potential to reveal much more information. Each
individual mode on its own describes a background that is correlated
between different points on the sky. A measurement of the components
that indicates the presence of correlations in the background on large
angular scales would suggest startling new physics.
\end{abstract}


\maketitle

\section{Introduction}

Near-future detections of gravitational waves (GWs) will open a new
window onto the cosmos by allowing astrophysical and cosmological
phenomena that generate only weak or difficult-to-detect electromagnetic signatures
to be probed for the first time and with an unprecedented precision. Within the next several years
a global network of advanced kilometre-scale laser interferometers
will come online, providing insights into stellar-mass compact binary
systems and stochastic gravitational-wave backgrounds in the kHz band
\cite{aLIGO,AdvVirgo,kagra2012,indigoupdate2013}. In $20$ years, the
launch of a $\sim\!10^9$ m arm-length space-based laser interferometer
will allow precision tests of fundamental physics, and perform
detailed demographic studies of massive black-holes throughout the
Universe \cite{eLISA:2012}.

Concurrently with these efforts are dedicated programs observing the
regular pulsed-emission from ensembles of Galactic millisecond pulsars
with the aim of detecting and characterising nanohertz gravitational
waves
\cite{vanHaasteren:2011,Demorest:2013,Shannon:2013,iptareview2013}. The
long-term stability of integrated pulse profiles allows incredibly
accurate models of the time-of-arrival (TOA) of pulses to be
constructed, and enables these pulsars to be used as standard clocks
in the sky. Potential gravitational-wave targets in the nHz band are
single resolvable sources (e.g., chirping supermassive black-hole
(SMBH) binaries
\cite{sesana-vecchio-volonteri-2009,lee2011,ellisoptimal2012} or
cosmic-string bursts
\cite{damour-vilenkin-2001,leblond2009,key-cornish-2009}) and
stochastic backgrounds from the superposition of many
inspiraling SMBH binary systems
\cite{rajaromani1995,jaffe-backer-2003,wyithe-loeb-2003}, decaying
cosmic-string networks
\cite{vilenkin-1981a,vilenkin-1981b,olmez-2010,sanidas-2012}, or even
backgrounds of primordial origin \cite{grishchuk-1976,grishchuk-2005}.

A pulsar timing array (PTA) can be thought of as a galactic-scale
gravitational-wave detector~\cite{foster-backer-1990}. When a
gravitational wave transits the Earth-pulsar line-of-sight it creates
a perturbation in the intervening metric, causing a change in the
proper separation, which manifests as a redshift in the
pulse frequency
\cite{sazhin-1978,detweiler-1979,estabrook-1975,burke-1975}. Standard
timing-models only factor in deterministic influences to the TOAs,
such that a subtraction of modelled TOAs from the raw observations
will result in a stream of {\it timing-residuals}, which encode the
influence of gravitational waves along with stochastic noise
processes. A PTA allows one to cross-correlate the residuals from many
pulsars, leveraging the common influence of a gravitational-wave
background against undesirable, uncorrelated noise processes.

In fact, for a Gaussian-stationary, 
isotropic, unpolarised stochastic background
composed of plus/cross gravitational-wave polarisation states, the
cross-correlation of timing-residuals is a unique smoking-gun
signature of the background's presence, and depends only on the
angular separation between pulsars on the sky: this is the famous {\it
  Hellings and Downs} curve \cite{HellingsDowns:1983}. Backgrounds
composed of non-Einsteinian polarisation states
\cite{lee2008,chamberlin2012}, or influenced by non-zero graviton mass
\cite{lee2014}, will induce different correlation signatures, as will
anisotropy in the background's energy density
\cite{Mingarelli:2013,TaylorGair:2013}, where the signature will
contain rich information on the distribution of gravitational-wave
power with respect to the position of pulsars on the sky.

Each of these standard analyses assumes a model describing the nature
of the background and then tries to measure a small number of model
parameters. For an isotropic, unpolarised and uncorrelated background 
there is just one
measurable parameter, which is the amplitude of the background. While
such analyses are optimal for the type of background being modelled,
they will not be as sensitive to alternative models and will not
indicate whether the model is correct. In this paper, we describe how
pulsar timing residuals can be used instead to construct a map of the
gravitational-wave background that makes no assumptions about its
nature. The properties of the observed background can be checked for
consistency with any particular model, e.g., to what extent it is
isotropic, unpolarised and uncorrelated, 
but this approach has the potential to
reveal much more, since we will extract all of the information that
can be determined about the background by a given pulsar timing
array. This information will not only tell us which out of our current
models provides the best description of the background, but will
clearly indicate if none of those models are accurate and therefore
that new physical models of the background are required.

The polarisation of the cosmic microwave background (CMB), which has
two independent components, can be represented as a transverse
traceless tensor field on the sky~\cite{KKS:1997}. In the analysis of
CMB data, the polarisation field is represented as a superposition of
gradients and curls of spherical harmonics, and CMB measurements
attempt to determine the individual components of those modes. A
gravitational-wave background is also a transverse traceless tensor
field on the sky and so the same formalism can be applied to the
analysis of a gravitational-wave background.  It is this that we
describe in this paper.  That the CMB approach can be readily applied
to gravitational waves is most easily seen from the fact that the
gradients and curls of spherical harmonics can also be written as the
real and imaginary parts of spin-$\pm2$ spin-weighted spherical
harmonics, which are widely used to decompose the gravitational-wave
emission from a source~\cite{Thorne:1980}.

Any gravitational-wave background can be decomposed as a sum of
gradient and curl modes. The components of this decomposition are the expansion coefficients of the
metric perturbation in terms of the gradient and curl spherical
harmonics, see Eq.~(\ref{e:habdecomp}).  The signature that arises in
the cross-correlation of the timing residuals of pairs of pulsars in a
PTA can therefore be computed as a sum of the cross-correlation curves
(overlap reduction functions) of each mode. For an unpolarised
statistically isotropic background the overlap reduction functions for
the individual models are just Legendre polynomials and the Hellings
and Downs curve can be recovered straightforwardly as a superposition
of these. Three modes are sufficient to represent the Hellings and Downs
correlation for reasonable assumptions about the PTA, so applying this
formalism to an isotropic, unpolarised and uncorrelated background 
will not be much more computationally challenging than the standard 
analysis.

The overlap reduction functions for individual modes can also be
computed for anisotropic backgrounds.  For pulsar timing arrays, the
resulting expression is relatively simple since the response of a
pulsar to curl modes is identically zero, while the response to a
gradient mode is proportional to the corresponding spherical harmonic
evaluated at the direction to the pulsar.  
For anisotropic, unpolarised and uncorrelated backgrounds, 
the integral expressions for the spherical
harmonic components of the overlap reduction function can be evaluated
analytically, allowing us to extend the results given for quadrupole
and lower backgrounds in~\cite{Mingarelli:2013}.

It is also relatively straightforward to reconstruct a map of the
gravitational-wave sky for that part of the background spanned by the
gradient modes visible to a PTA, and we describe how this can be
done. For a PTA consisting of $N$ pulsars, at any given frequency we
make two measurements---an amplitude and a phase---with each
pulsar. Since PTAs are static, the response function is frequency-independent 
and we would therefore not expect to be able to measure
more than $2N$ real components of the background. The fact that PTAs
are sensitive to only $2N$ components of the background is consistent
with recent unpublished results by Cornish and van~Haasteren (private
communication). We describe how we can recover these $N$ complex
combinations of gradient mode components and which components we expect to measure
most accurately (those for the low-$l$ modes). In practice, we can either
restrict our mapping search to fewer than $N$ low-$l$ modes or use
singular-value decomposition (SVD) of the mapping matrix to determine the
$N$ linear combinations to which the array is sensitive. Since we make
no assumptions about the properties of the underlying background in
this analysis, we can interpret the map that we obtain in terms of its
implications for fundamental physics, as described below. To
characterise an isotropic, unpolarised and uncorrelated background 
we need to reach
an angular resolution of $l_{\rm max} \sim 4$, which requires $21$
pulsars, well within reach of current PTA efforts. To reach the
angular resolution at which we expect to resolve individual sources
with a PTA we must probe $l_{\rm max} \sim 10$, which will require
$\sim 100$ pulsars. This should be achievable with the Square-Kilometre
Array (SKA)~\cite{SKApulsars}.

In our approach, each individual mode used in the decomposition describes a
background that is correlated between different points on the sky. 
By this we will mean a
correlation in the gravitational radiation coming from different
{\em angular} directions, which is different from the
correlation between the pulsar responses, present for all
types of background. It is also different from spatial correlations
that may exist between the metric perturbations evaluated at different
locations in space. A background that is spatially homogeneous and
isotropic can have spatial correlations provided the correlations
depend only on the distance $|\vec{x}-\vec{y}|$ between any two
points $\vec{x}$ and $\vec{y}$, and any background of this form will
be uncorrelated in Fourier (angle) space. We focus on angular
correlations because we will measure the gravitational-wave background
at a single point only and therefore cannot compute spatial
correlations from our data. Assumptions about the presence or absence
of spatial correlations are needed to compute the statistical
properties of a background in any particular physical model, but here
we will focus only on a measurement of the background and so the
angular correlation properties are the most important. 

The gravitational-wave background in the pulsar timing band is most likely
to be generated by a superposition of emission from many individual
astrophysical sources. Such a background will not show angular
correlations between different sky locations, but would show
anisotropy indicative of the spatial distribution of sources
contributing to the background. A background of cosmological origin
could in principle show angular correlations on some scale, and the
spectrum of modes present will be characteristic of the quantum
fluctuations that produced it. However, there are no mechanisms
currently known that would generate such correlations in the nanohertz
frequency band. Nonetheless, the power of the analysis described here
is that it can represent any background and it makes no assumptions about
the correlation properties or isotropy. It will allow us to derive a
map of the background, free from model assumptions, that will encode
all of the details about the underlying physical processes that
produced the background and that are possible to deduce from our
observations. If the map indicates the presence of correlated emission
or significant anisotropy, it will be a startling and profound result,
pointing either to unmodelled physics in the early Universe or an
unknown systematic affecting the timing data. In either case, the
result would be of great significance.

This paper is organised as follows: In Section~\ref{s:general} we
describe the general formalism, which is based on that used to
characterise CMB polarisation and can be used to describe arbitrary
gravitational-wave backgrounds.  We include a description of the basis
functions used to expand the backgrounds, and we give definitions of
the response functions and overlap reduction functions for arbitrary
gravitational-wave detectors.  In Section~\ref{s:PTAs} we specialise
to the case of PTAs, deriving the overlap reduction function for an
unpolarised statistically isotropic background, and show how the
Hellings and Downs curve for an isotropic, unpolarised and 
uncorrelated background is
well-approximated by a combination of the first three modes, $l=2, 3, 4$.  We show that
the response of a pulsar to the curl modes of a gravitational-wave
background is identically zero, while the response to an individual
gradient mode is simply proportional to the corresponding spherical
harmonic evaluated at the direction to the pulsar.  We also
demonstrate how the formalism can be used by recovering the
coefficients of the expansion from a simulated pulsar-timing data set.
In Section~\ref{s:arbback} we compute the overlap reduction functions
needed to represent arbitrary anisotropic backgrounds, giving explicit
expressions for a PTA.  In Section~\ref{s:mapping} we discuss how one
can reconstruct a map of the gravitational-wave sky in terms of the
gradient components visible to a PTA.  We show that an $N$-pulsar array
can measure $N$ (complex) combinations of the gradient components of
the background, but is blind to the curl component, irrespective of
the value of $N$.  Finally, in Section~\ref{s:discuss} we summarise
the results and discuss some of the implications if a measurement of
these parameters is made that is indicative of significant
correlations in the background.

We also include several appendices: Appendices~\ref{s:spinweightedY}
and \ref{s:legendre_polynomials} contain useful definitions and
identities for spin-weighted spherical harmonics and associated
Legendre functions and Legendre polynomials, respectively.  In
Appendix~\ref{s:pulsar_term}, we calculate the oscillatory behavior of
the pulsar term for an isotropic, unpolarised and uncorrelated 
stochastic background,
and show that it is negligible.  In Appendix~\ref{s:LIGO_response}, we
derive the grad and curl response for a static interferometer, and
find that the curl response is zero, similar to that for a PTA.  In
Appendix~\ref{app:UncorrAniso}, we derive analytic expressions for the
spherical harmonic components of the overlap reduction function for
anisotropic, unpolarised and uncorrelated backgrounds for all values of $l$ and $m$,
extending the analytical results of~\cite{Mingarelli:2013}.

\section{General formalism}
\label{s:general}

The gravitational-wave field is a symmetric transverse-traceless
tensor field, with two independent polarisation states, $h_+$ and
$h_\times$, which transform under rotations of the polarisation axes
defined at each point on the sky \cite{Maggiore}.  In the analysis of
the CMB, 
polarisation is characterised by a two dimensional, symmetric and
trace-free matrix, which is analogous to the symmetric
transverse-traceless metric perturbations 
describing a general gravitational-wave field. 
Therefore, our analysis will closely parallel the treatment of
polarisation in analyses of the CMB, see e.g. \cite{KKS:1997, Challinor:2000}.

\subsection{Gradient and curl spherical harmonics}

Any symmetric trace-free rank-two tensor field on the two-sphere $S^2$
can be written as the sum of the ``gradient'' of a scalar field
$A(\hat{k})$
\be
A_{;ab} - \frac{1}{2} g_{ab} A_{;c}{}^{c}\,,
\ee
plus the ``curl'' of another scalar field $B(\hat{k})$
\be
\frac{1}{2}\left( B_{;ac}\epsilon^c{}_b + B_{;bc} \epsilon^c{}_a \right)\,,
\ee
where a semi-colon denotes covariant differentiation, $g_{ab}$ is the
metric tensor on the sphere, and $\epsilon_{ab}$ is the Levi-Civita
anti-symmetric tensor
\be
\epsilon_{ab} 
= \sqrt{g} \left( \begin{array}{cc}0&1\\-1&0\end{array}\right)\,.
\ee
Following standard practice, we use the metric tensor
$g_{ab}$ and its inverse $g^{ab}$ to ``lower" and ``raise" tensor
indices---e.g., $\epsilon^{c}{}_b \equiv g^{ca}\epsilon_{ab}$. 
In standard spherical coordinates $(\theta,\phi)$,
\be
g_{ab}=\left(
\begin{array}{cc}
1&0\\
0&\sin^2\theta\\
\end{array}
\right)\,,
\qquad
\sqrt{g}=\sin\theta\,.
\ee
Since any scalar field on the two-sphere can be written as a sum of
spherical harmonics, $Y_{lm}(\hat{k})$, it follows that any symmetric
trace-free rank-two tensor field can be written as a sum of gradients
and curls of spherical harmonics \cite{KKS:1997, Zerilli:1970}.

Defining the gradient and curl spherical harmonics for $l\ge 2$ by:
\be
\begin{aligned}
Y^G_{(lm)ab} &= N_l 
\left(Y_{(lm);ab} - \frac{1}{2} g_{ab}  Y_{(lm);c}{}^{c} \right)\,,
\\
Y^C_{(lm)ab} &= \frac{N_l}{2} 
\left(Y_{(lm);ac}\epsilon^c{}_b +  Y_{(lm);bc} \epsilon^c{}_a \right)\,,
\end{aligned}
\ee
where
\be
N_l = \sqrt{\frac{2 (l-2)!}{(l+2)!}}\,,
\ee
it follows that
\begin{align}
\int_{S^2} {\rm d}^2\Omega_{\hat k}\> 
Y^{G}_{(lm)ab} (\hat{k}) Y^{G}_{(l'm')}{}^{ab\,*}(\hat{k}) 
&=\delta_{ll'}\delta_{mm'}\,, 
\label{e:YGGOrthog}
\\
\int_{S^2} {\rm d}^2\Omega_{\hat k}\> 
Y^{C}_{(lm)ab} (\hat{k}) Y^{C}_{(l'm')}{}^{ab\,*}(\hat{k}) 
&=\delta_{ll'}\delta_{mm'}\,,
\label{e:YCCOrthog}
\\
\int_{S^2} {\rm d}^2\Omega_{\hat k}\> 
Y^{G}_{(lm)ab} (\hat{k}) Y^{C}_{(l'm')}{}^{ab\,*}(\hat{k}) 
&=0\,.
\label{e:YGCOrthog}
\end{align}
Note that we have adopted the notational convention used in the
CMB literature, e.g., \cite{KKS:1997}, by putting parentheses around
multipole moment indices $l$ and $m$ to distinguish these indices
from spatial tensor indices $a$, $b$, etc.
 
\begin{widetext}
\subsection{Expanding the metric perturbations}

In transverse-traceless coordinates, the metric perturbations
$h_{ab}(t,\vec x)$ associated with a gravitational wave are transverse
to the direction of propagation $\hat k$ and hence define a
symmetric trace-free tensor field on the two-sphere.  The Fourier
components $h_{ab}(f,\hat{k})$ of the field can therefore be decomposed as
\be
h_{ab}(f,\hat{k}) = \sum_{l=2}^{\infty} \sum_{m=-l}^{l} 
\left[ a^G_{(lm)}(f) Y^G_{(lm)ab}(\hat{k}) 
+ a^C_{(lm)}(f) Y^C_{(lm)ab}(\hat{k}) \right]\,,
\label{e:habdecomp}
\ee
with
\be
\begin{aligned}
a^{G}_{(lm)}(f) &= \int_{S^2} {\rm d}^2\Omega_{\hat k}\>
h_{ab}(f,\hat{k}) Y^{G}_{(lm)}{}^{ab\,*}(\hat{k})\,,
\\
a^C_{(lm)}(f) &= \int_{S^2} {\rm d}^2\Omega_{\hat k}\>
h_{ab}(f,\hat{k}) Y^{C}_{(lm)}{}^{ab\,*}(\hat{k})\,. 
\label{e:aGCdef}
\end{aligned}
\ee
Note that the summation over $l$ starts at $l=2$ and not at $l=0$, as
would be the case if we were expanding a scalar function on the sphere in
terms of ordinary (i.e., undifferentiated) spherical harmonics
$Y_{lm}(\hat k)$.
In what follows we will use the shorthand notation
$\sum_{(lm)}$ for $\sum_{l=2}^\infty\sum_{m=-l}^l$.
From the above definitions it follows that
\be
Y^{G,C\,*}_{(lm)ab}(\hat k)=(-1)^m Y^{G,C}_{(l,-m)ab}(\hat k)\,,
\quad
Y^G_{(lm)}(-\hat k) = (-1)^l Y^G_{(lm)}(\hat k)\,,
\quad
Y^C_{(lm)}(-\hat k) = (-1)^{l+1} Y^C_{(lm)}(\hat k)\,,
\ee
and 
\be
a^{G,C\,*}_{(lm)}(f)=(-1)^m a^{G,C}_{(l,-m)}(-f)\,,
\quad
a^G_{(lm)}(f) \rightarrow (-1)^l a^G_{(lm)}(f)\,,
\quad
a^C_{(lm)}(f) \rightarrow (-1)^{l+1} a^C_{(lm)}(f)\,,
\label{e:acoeffsymmetry}
\ee
with respect to complex conjugation and parity 
(i.e., $\hat k\rightarrow -\hat k$) transformations.
Note that the gradient modes have ``electric-type" parity, 
while the curl modes have ``magnetic-type" parity. These are
sometimes referred to as ``$E$ modes'' and ``$B$ modes'',
respectively, in the CMB literature.

A general stochastic gravitational-wave background can be written as a
superposition of plane waves having frequency $f$ and propagation
direction $\hat k$. We assume that gravitational waves of different
frequencies are uncorrelated with one another, which follows if the
background is stationary with respect to time. 
Using the preceding
decomposition, we can therefore write the metric perturbation induced
by an arbitrary stochastic background in transverse-traceless
coordinates as
\begin{equation}
h_{ab}(t,\vec x)
=
\int_{-\infty}^\infty {\rm d}f\>
\int_{S^2} {\rm d}^2 \Omega_{\hat{k}}\>
\left\{ \sum_{(lm)} 
\left[a^G_{(lm)}(f) Y^G_{(lm)ab}(\hat{k}) + 
a^C_{(lm)}(f) Y^C_{(lm)ab}(\hat{k})\right]\right\}
e^{i 2\pi f(t-\hat k\cdot \vec x/c)} .
\label{e:habGC}
\end{equation}
Introducing the usual orthogonal coordinate axes on the sky
\be
\begin{aligned}
\hat k
&=\sin\theta\cos\phi\,\hat x+
\sin\theta\sin\phi\,\hat y+
\cos\theta\,\hat z
= \hat r\,,
\\
\hat l
&=\cos\theta\cos\phi\,\hat x+
\cos\theta\sin\phi\,\hat y-
\sin\theta\,\hat z
= \hat\theta\,, 
\\
\hat m
&=-\sin\phi\,\hat x+
\cos\phi\,\hat y
= \hat\phi\,, 
\label{e:klmdef}
\end{aligned}
\ee
and defining two polarization tensors by
\be
\begin{aligned}
e_{ab}^+(\hat k)
&=\hat l_a\hat l_b-\hat m_a\hat m_b\,,
\\
e_{ab}^\times(\hat k)
&=\hat l_a\hat m_b+\hat m_a\hat l_b\,,
\label{e:e+xdef}
\end{aligned}
\ee
the gradient and curl spherical harmonics can be written explicitly as
\cite{HuWhite:1997}:
\be
\begin{aligned}
Y^G_{(lm)ab}(\hat{k}) 
&= \frac{N_l}{2} \left[ W_{(lm)}(\hat{k}) e_{ab}^+(\hat k) 
+ X_{(lm)}(\hat{k}) e_{ab}^\times(\hat k) \right]\,, 
\\
Y^C_{(lm)ab}(\hat{k}) 
&= \frac{N_l}{2} \left[ W_{(lm)}(\hat{k})e_{ab}^\times(\hat k) 
- X_{(lm)}(\hat{k}) e_{ab}^+(\hat k) \right]\,,
\label{e:YGCdef}
\end{aligned}
\ee
where
\begin{align}
W_{(lm)}(\hat{k}) 
&= \left( \frac{\partial^2}{\partial \theta^2} 
- \cot\theta\frac{\partial}{\partial\theta} 
+\frac{m^2}{\sin^2\theta} \right)Y_{(lm)}(\hat{k}) 
= \left( 2 \frac{\partial^2}{\partial \theta^2} 
+ l(l+1) \right) Y_{(lm)}(\hat{k})\,,
\label{e:Wlm}
\\
X_{(lm)}(\hat{k}) 
&= \frac{2 i m}{\sin\theta} \left( \frac{\partial}{\partial\theta} 
- \cot\theta\right) Y_{(lm)}(\hat{k})\,. 
\label{e:Xlm}
\end{align}
These functions are related to spin-2 spherical harmonics 
\cite{NewmanPenrose:1966, Goldberg:1967} through the equation
\begin{equation}
{}_{\pm2}Y_{(lm)}(\hat{k})
=\frac{N_l}{\sqrt{2}} \left[
W_{(lm)}(\hat{k})\pm i X_{(lm)}(\hat{k})\right]\,,
\label{e:Ypm2}
\end{equation}
and can be written in terms of associated Legendre functions as
\begin{align}
W_{(lm)}(\hat{k}) 
&= +2 \sqrt{\frac{2l+1}{4\pi} \frac{(l-m)!}{(l+m)!}} 
G^+_{(lm)}(\cos\theta) e^{i m \phi}\,,
\label{e:Wdef}
\\
i X_{(lm)}(\hat{k}) 
&= -2 \sqrt{\frac{2l+1}{4\pi} \frac{(l-m)!}{(l+m)!}} 
G^-_{(lm)}(\cos\theta) e^{i m \phi}\,,
\label{e:Xdef}
\\
G^+_{(lm)}(\cos\theta) 
&= -\left(\frac{l-m^2}{\sin^2\theta} 
+ \frac{1}{2} l(l-1) \right) P^m_l(\cos\theta) 
+ (l+m) \frac{\cos\theta}{\sin^2\theta} P^m_{l-1}(\cos\theta)\,,
\label{e:G+def}
\\
G^-_{(lm)}(\cos\theta) 
&= \frac{m}{\sin^2\theta} 
\left[(l-1) \cos\theta P_l^m(\cos\theta) 
- (l+m) P^m_{l-1}(\cos\theta) \right]\,.
\label{e:G-def}
\end{align}

Using this explicit form for the gradient and curl spherical
harmonics, Eq.~(\ref{e:habGC}) becomes
\begin{multline}
h_{ab}(t,\vec x)
=\int_{-\infty}^\infty {\rm d}f\>
\int_{S^2} {\rm d}^2 \Omega_{\hat{k}}\>
\Bigg\{ \sum_{(lm)}\frac{N_l}{2} \left( a^G_{(lm)}(f) W_{(lm)}(\hat{k}) 
- a^C_{(lm)}(f) X_{(lm)} (\hat{k})\right) e^+_{ab}(\hat{k}) 
\\
+ \frac{N_l}{2} \left( a^G_{(lm)}(f) X_{(lm)}(\hat{k}) 
+ a^C_{(lm)}(f) W_{(lm)}(\hat{k}) \right) e^\times_{ab}(\hat{k})\Bigg\}
e^{i 2\pi f(t-\hat k\cdot \vec x/c)}\,.
\label{e:habWX}
\end{multline}
In terms of the more traditional ``plus" and ``cross"
decomposition of the Fourier components,
\be
h_{ab}(t,\vec x)
=\int_{-\infty}^\infty {\rm d}f\>
\int_{S^2} {\rm d}^2 \Omega_{\hat{k}}\>
\left[h_+(f,\hat k)e^+_{ab}(\hat{k}) +
h_\times(f,\hat k)e^\times_{ab}(\hat{k})\right]
e^{i 2\pi f(t-\hat k\cdot \vec x/c)}\,,
\label{e:hab+x}
\ee
we see that 
\be
\begin{aligned}
h_+(f,\hat k)&=
\sum_{(lm)}\frac{N_l}{2} \left[ a^G_{(lm)}(f) W_{(lm)}(\hat{k}) 
- a^C_{(lm)}(f) X_{(lm)} (\hat{k})\right]\,,
\\
h_\times(f,\hat k)&=
\sum_{(lm)}\frac{N_l}{2} \left[ a^G_{(lm)}(f) X_{(lm)}(\hat{k}) 
+ a^C_{(lm)}(f) W_{(lm)}(\hat{k}) \right]\,,
\label{e:h+x}
\end{aligned}
\ee
and, conversely, 
\be
\begin{aligned}
a^G_{(lm)}(f) = N_l\int_{S^2}{\rm d}^2\Omega_{\hat k}
\left[h_+(f,\hat k) W_{(lm)}^*(\hat k) 
+ h_\times(f,\hat k) X_{(lm)}^*(\hat k)\right]\,,
\\
a^C_{(lm)}(f) = N_l\int_{S^2}{\rm d}^2\Omega_{\hat k}
\left[h_\times(f,\hat k) W_{(lm)}^*(\hat k) 
- h_+(f,\hat k) X_{(lm)}^*(\hat k)\right]\,.
\label{e:aGC}
\end{aligned}
\ee
Finally, in terms of spin-weighted spherical harmonics:
%
%
%
%
%
%
\begin{align}
Y^G_{(lm)ab}(\hat k) \pm i Y^C_{(lm)ab}(\hat k)
&=\frac{1}{\sqrt{2}}
\left(e_{ab}^+(\hat k) \pm i e_{ab}^\times(\hat k)\right)
\,{}_{\mp 2}Y_{lm}(\hat k)\,,
\label{e:YG+iYC}
\end{align}
and
\begin{align}
h_+(f,\hat k)\pm i h_\times(f,\hat k)
&=\frac{1}{\sqrt{2}} \sum_{(lm)}
\left(a^G_{(lm)}(f)\pm i a^C_{(lm)}(f)\right)\,
{}_{\pm 2}Y_{(lm)}(\hat{k})\,,
\label{e:h++ihx}
\\
\frac{1}{\sqrt{2}}\left(
a^G_{(lm)}(f)\pm i a^C_{(lm)}(f)\right)
&= \int {\rm d}^2\Omega_{\hat{k}}\>
\left(h_+(f,\hat{k}) \pm i h_\times(f,\hat{k})\right)\,
{}_{\pm 2}Y_{(lm)}^*(\hat{k})\,.
\label{e:aG+iaC}
\end{align}
These latter expressions for $h_+(f,\hat k)$, $h_\times(f,\hat k)$,
$a^G_{(lm)}(f)$, and $a^C_{(lm)}(f)$ are convenient when one can make
use of relations derived for the spin-weighted spherical harmonics
${}_{\pm 2}Y_{(lm)}(\hat k)$ (see, e.g.,
Appendix~\ref{s:spinweightedY}).

\subsection{Statistical properties of the background}
\label{s:statprop}

The statistical properties of a Gaussian-stationary background are
encoded in the quadratic expectation values 
$\langle h_A(f,\hat k)h_{A'}^*(f',\hat k')\rangle$ or, equivalently, 
$\langle a^P_{lm}(f)a^{P'*}_{l'm'}(f')\rangle$, 
for $A, A'=\{+, \times\}$ and $P, P'=\{G,C\}$.  
For a statistically unpolarised and uncorrelated isotropic background
\be
\begin{aligned}
&\langle h_+(f,\hat{k}) h_+^*(f',\hat{k}') \rangle
= \langle h_\times(f,\hat{k}) h_\times^*(f',\hat{k}') \rangle
= \frac{1}{2}H(f) \delta^2(\hat{k},\hat{k}')\delta(f-f')\,,
\\
&\langle h_+(f,\hat{k}) h_\times^* (f',\hat{k}') \rangle
= \langle h_\times(f,\hat{k}) h_+^* (f',\hat{k}') \rangle
= 0\,,
\label{e:uncorrXV}
\end{aligned}
\ee
where $H(f)>0$.  
The factor of $1/2$ has been included so that $H(f)$ is the
two-sided gravitational-wave strain power, 
when summed over both polarizations.
Using Eq.~(\ref{e:aGC}) and assuming the above expectation values,
it follows that
%
%
\be
\begin{aligned}
\langle a^{G}_{(lm)}(f) a^{G*}_{(l'm')}(f') \rangle
&= N_l N_l'
\int {\rm d}^2\Omega_{\hat{k}}\int {\rm d}^2\Omega_{\hat{k}'}\>
\bigg\langle
\left[h_+(f,\hat{k})W^*_{(lm)}(\hat{k})  
+ h_\times(f,\hat{k})X^*_{(lm)}(\hat{k})\right] 
\\
&\hspace{2in}
\times\left[h^*_+(f',\hat{k}')W_{(l'm')}(\hat{k}')  
+ h^*_\times(f',\hat{k}')X_{(l'm')}(\hat{k}')\right] 
\bigg\rangle
\\
&= \frac{N_lN_{l'}}{2}
\int {\rm d}^2\Omega_{\hat{k}}\int {\rm d}^2\Omega_{\hat{k}'}\>
\delta^2(\hat k,\hat k')
\left[ W^*_{(lm)}(\hat k) W_{(l'm')}(\hat k')
+X^*_{(lm)}(\hat k) X_{(l'm')}(\hat k')\right]
H(f)\delta(f-f')
\\
&= \delta_{ll'}\delta_{mm'}H(f)\delta(f-f')\,,
\label{e:isocorr1}
\end{aligned}
\ee
where the last line follows from the orthogonality relation
\begin{align}
\int_{S^2} {\rm d}^2\Omega_{\hat k}\> 
\left[W^*_{(lm)}(\hat{k}) W_{(l'm')}(\hat k) +
X^*_{(lm)}(\hat{k}) X_{(l'm')}(\hat k) \right]
=\frac{2}{N_lN_l'}\,
\delta_{ll'}\delta_{mm'}\,, 
\label{e:WXOrthog1}
\end{align}
which is a consequence of Eqs.~(\ref{e:YGGOrthog}) and (\ref{e:YCCOrthog}).
In a similar way, one can show that
\be
\begin{aligned}
\langle a^{C}_{(lm)}(f) a^{C*}_{(l'm')}(f')\rangle
&=\delta_{ll'}\delta_{mm'}H(f)\delta(f-f')\,,
\\
\langle a^{G}_{(lm)}(f) a^{C*}_{(l'm')}(f') \rangle&=0
=\langle a^{C}_{(lm)}(f) a^{G*}_{(l'm')}(f') \rangle\,,
\label{e:isocorr23}
\end{aligned}
\ee
where the zero expectation values follow from
\begin{align}
\int_{S^2} {\rm d}^2\Omega_{\hat k}\> 
\left[X^*_{(lm)}(\hat{k}) W_{(l'm')}(\hat k) 
-W^*_{(lm)}(\hat{k}) X_{(l'm')}(\hat k) \right]=0\,,
\label{e:WXOrthog2}
\end{align}
which is a consequence of Eq.~(\ref{e:YGCOrthog}).
Thus, if we define 
\be
\langle a^{P}_{(lm)}(f) a^{P'*}_{(l'm')}(f') \rangle
=\delta_{ll'} \delta_{mm'} C_l^{PP'}(f)\delta(f-f')\,,
\label{e:acorr}
\ee
where the correlation functions have the form
\be
C_l^{PP'}(f) \equiv C_l^{PP'}H(f)\,,
\label{e:corr_factorization}
\ee
we deduce that an isotropic, unpolarised and uncorrelated 
background may be described by Eqs.~(\ref{e:acorr}) and
(\ref{e:corr_factorization})  with
\begin{equation}
C^{GG}_l=C^{CC}_l\equiv C_l=1\,,
\quad
C^{GC}_l=0=C_l^{CG}\,,
\label{e:corrforiso}
\end{equation}
for $l\ge 2$.

\subsection{Statistically isotropic backgrounds}
\label{s:statiso}

Stochastic backgrounds described by expectation values of the
form given in Eq.~(\ref{e:acorr}) are 
said to be {\em statistically isotropic}.
This means that there is no preferred direction on the sky, 
even though there can be non-trivial angular dependence in 
the distribution of gravitational-wave power via the $C_l^{PP'}(f)$.
The fact that the quadratic expectation values in Eq.~(\ref{e:acorr}) 
depend only on $l$ and not on $m$ is equivalent to the statement
that the angular distribution is independent of the orientation
of the reference frame in which it is evaluated.
In Sec.~\ref{sec:overlap_general}, we will extend our analysis 
to include more general (i.e., statistically anisotropic) backgrounds, 
allowing expectation values that can also depend on $m$, 
cf.~Eq.~(\ref{e:aniso_expectation_values}).
In principle, the correlation functions $C^{PP'}_l(f)$ for a 
statistically isotropic background are arbitrary, but if we impose 
additional physicality constraints the forms are restricted, 
as we shall discuss in Sec.~\ref{s:implications}. 
Requiring the background to be statistically unpolarised imposes 
the restrictions
\begin{align}
C_l^{GG}(f) = C_l^{CC}(f)\equiv C_l(f)\,,
\quad
C_l^{GC}(f) = -C_l^{CG}(f)\,,
\end{align}
which follow from invariance of the expectation values under 
rotations about a point on the sky. In addition, invariance of 
the expectation values under a parity transformation 
($\hat k\rightarrow -\hat k$) further requires
\be
C_l^{GC}(f) = 0 = C_l^{CG}(f)\,.
\ee
To see that this is indeed the case, recall that under a 
parity transformation, cf.~Eq.~(\ref{e:acoeffsymmetry}),
\begin{equation}
a^G_{(lm)}(f) \rightarrow (-1)^l a^G_{(lm)}(f)\,,
\quad
a^C_{(lm)}(f) \rightarrow (-1)^{l+1} a^C_{(lm)}(f)\,,
\end{equation}
for which
\be
\begin{aligned}
\langle a^G_{(lm)}(f) a^{C\,*}_{(l'm')}(f')\rangle
&\rightarrow
(-1)^{l+l'+1}
\langle a^G_{(lm)}(f) a^{C\,*}_{(l'm')}(f')\rangle
\\
&=(-1)^{l+l'+1}\delta_{ll'}\delta_{mm'}C_l^{GC}(f)\delta(f-f')
\\
&=-\delta_{ll'}\delta_{mm'}C_l^{GC}(f)\delta(f-f')
\\
&= -\langle a^G_{(lm)}(f) a^{C\,*}_{(l'm')}(f')\rangle\,.
\end{aligned}
\ee
Thus, invariance under a parity transformation requires
\begin{equation}
\langle a^G_{(lm)}(f) a^{C\,*}_{(l'm')}(f')\rangle
= -\langle a^G_{(lm)}(f) a^{C\,*}_{(l'm')}(f')\rangle
\quad
\Rightarrow
\quad
\langle a^G_{(lm)}(f) a^{C\,*}_{(l'm')}(f')\rangle=0\,,
\end{equation}
so $C_l^{GC}(f)=0$.
Similarly, one can show $C_l^{CG}(f)=0$.
Hence, a statistically isotropic, unpolarised and parity-invariant 
background is completely characterised by the single correlation 
function $C_l(f)\equiv C_l^{GG}(f)= C_l^{CC}(f)$.

\subsection{Detector response functions}

The response of a detector to a passing gravitational wave is given by
the convolution of the metric perturbations $h_{ab}(t,\vec x)$ with
the impulse response $R^{ab}(t,\vec x)$ of the detector:
\be
r(t) 
= \int_{-\infty}^\infty {\rm d}\tau\int {\rm d}^3y\,
R^{ab}(\tau,\vec y)
h_{ab}(t-\tau,\vec x-\vec y)\,.
\ee
If we expand the metric perturbations in terms of the plus and cross 
Fourier modes $h_A(f,k)$, where $A=\{+,\times\}$, we can write the
response as
\begin{align}
r(t)
=\int_{-\infty}^\infty {\rm d}f\int_{S^2} {\rm d}^2\Omega_{\hat k}
\sum_A
R^A(f,\hat k)
h_A(f,\hat k)
e^{i2\pi f t}\,, 
\label{e:habRA}
\end{align}
where
\begin{align}
R^A(f,\hat k) 
=
e^{-i2\pi f\hat k\cdot\vec x/c}\,
e^A_{ab}(\hat k)
\int_{-\infty}^\infty {\rm d}\tau\int {\rm d}^3 y\> R^{ab}(\tau,\vec y)\,
e^{-i2\pi f(\tau-\hat k\cdot\vec y/c)}\,.
\end{align}
Alternatively, if we expand the metric perturbations in terms of the gradient 
and curl spherical harmonic modes $a^P_{(lm)}(f)$, where $P=\{G,C\}$, we have
\begin{align}
r(t)
=\int_{-\infty}^\infty {\rm d}f
\sum_{(lm)}\sum_P
R^P_{(lm)}(f)a^P_{(lm)}(f)
e^{i2\pi f t}\,, 
\label{e:habRP}
\end{align}
where
\begin{align}
R^P_{(lm)}(f) 
= \int_{S^2}{\rm d}^2\Omega_{\hat k}\>
e^{-i2\pi f\hat k\cdot\vec x/c}\,
Y^P_{(lm)ab}(\hat k)
\int_{-\infty}^\infty {\rm d}\tau\int {\rm d}^3 y\> R^{ab}(\tau,\vec y)\,
e^{-i2\pi f(\tau-\hat k\cdot\vec y/c)}\,.
\label{e:RPlm}
\end{align}
The detector response functions implicitly depend on the assumptions made about
the choice of polarisation axes, but we will assume these are
consistent with the definitions used in 
Eqs.~(\ref{e:klmdef}) and (\ref{e:e+xdef}) above.
Note that the response functions for the two different mode decompositions are
related by:
\be
\begin{aligned}
R^+(f,\hat k)
&=\sum_{(lm)} N_l
\left[R^G_{(lm)}(f)W^*_{(lm)}(\hat k) - R^C_{(lm)}(f)X^*_{(lm)}(\hat k)\right]\,,
\\
R^\times(f,\hat k)
&=\sum_{(lm)} N_l
\left[R^G_{(lm)}(f)X^*_{(lm)}(\hat k) + R^C_{(lm)}(f)W^*_{(lm)}(\hat k)\right]\,,
\label{e:R+xintermsofRGC}
\end{aligned}
\ee
and, conversely,
\be
\begin{aligned}
R^G_{(lm)}(f)
&=\frac{N_l}{2}
\int_{S^2}{\rm d}^2\Omega_{\hat k}\>
\left[R^+(f,\hat k)W_{(lm)}(\hat k) + R^\times(f,\hat k)X_{(lm)}(\hat k)\right]\,,
\\
R^C_{(lm)}(f)
&=\frac{N_l}{2} 
\int_{S^2}{\rm d}^2\Omega_{\hat k}\>
\left[R^\times(f,\hat k)W_{(lm)}(\hat k) - R^+(f,\hat k)X_{(lm)}(\hat k)\right]\,,
\label{e:RGC}
\end{aligned}
\ee
which follow from Eq.~(\ref{e:YGCdef}).

\subsection{Overlap reduction function}

Using Eqs.~(\ref{e:acorr}) and (\ref{e:corr_factorization})
for a statistically isotropic background,
and assuming $C_l^{GG}=C_l^{CC}\equiv C_l$ and $C_l^{GC}=0=C_l^{CG}$, 
the expectation value of the correlation between two detectors, 
labeled by $1$ and $2$, can be written as
\begin{equation}\label{e:r1r2corr}
\langle r_1(t) r_2(t')\rangle
=\int_{-\infty}^\infty {\rm d}f\>
e^{i2\pi f(t-t')}H(f)
\Gamma_{12}(f)\,,
\end{equation}
where 
$\Gamma_{12}(f)$ is the overlap reduction function
(see, e.g., \cite{Christensen:1992, Flanagan:1993, Finn-et-al:2009}),
and is given by
\begin{align}
\Gamma_{12}(f) &= \sum_{l=2}^\infty C_l \Gamma_{12,l}(f)\,,
\label{e:gamma(f)}
\end{align}
with
\begin{align}
\Gamma_{12,l}(f)
&\equiv
\sum_{m=-l}^l 
\sum_P
R^P_{1(lm)}(f)R^{P*}_{2(lm)}(f)\,,
\label{eq:gamma_12,l}
\end{align}
where $R^P_{I(lm)}(f)$ are the
gradient and curl response functions for the two detectors, $I=1,2$.
In Sec.~\ref{sec:overlap_general}, we will extend our analysis to 
compute overlap reduction functions for general anisotropic backgrounds.

\section{Application to pulsar timing arrays}
\label{s:PTAs}

In this section, we apply the above formalism to PTAs, deriving the
overlap reduction function for statistically isotropic backgrounds,
and showing how one can recover the Hellings and Downs curve.  The
same approach can also be used to characterise gravitational-wave
backgrounds in other frequency bands, relevant to ground-based or
space-based detectors.  Although the overlap reduction functions in
those cases will be different due to the different detector response
functions, they can be calculated in a similar way to the pulsar
timing response derived here.

\subsection{Detector response functions}
\label{sec:detresponse}

As a plane gravitational wave transits the Earth-pulsar line-of-sight, it
creates a perturbation in the intervening metric, causing a change in
the proper separation, which is manifested as a redshift in the pulse
frequency \cite{sazhin-1978,detweiler-1979,estabrook-1975,burke-1975}:
\be
z(t,\hat k) \equiv 
\frac{\Delta v(t)}{\nu_0} = \frac{1}{2} 
\frac{ u^a u^b}{1+\hat k \cdot \hat u} \Delta h_{ab}(t,\hat{k})\,,
\ee
where $\hat k$ is the direction of propagation of the gravitational wave,
$\hat u$ is the direction to the pulsar, and $\Delta h_{ab}(t,\hat k)$
is the difference between the metric perturbation at Earth, $(t,\vec
x)$, and at the pulsar, $(t_p,\vec x_p)= (t-L/c, \vec x+L\hat u)$:
\be
\begin{aligned}
\Delta h_{ab}(t,\hat k) 
&\equiv \int_{-\infty}^\infty {\rm d}f\>
h_{ab}(f,\hat k)
\left[e^{i2\pi f(t-\hat k\cdot\vec x/c)}-e^{i2\pi f(t_p-\hat k\cdot\vec x_p/c)}\right]
\\
&=\int_{-\infty}^\infty {\rm d}f\>
h_{ab}(f,\hat k)
e^{i2\pi f(t-\hat k\cdot\vec x/c)}
\left[1-e^{-i2\pi fL(1+\hat k\cdot\hat u)/c}\right]\,.
\label{eq:ETPT}
\end{aligned}
\ee
For a gravitational-wave background, which is a superposition of plane
waves from all directions on the sky, the pulsar redshift integrated
over $\hat k$ is given by
\be
\label{eq:res_wpulsar}
z(t) = \int_{-\infty}^\infty {\rm d}f\int_{S^2} {\rm d}^2\Omega_{\hat k}\>
\frac{1}{2}\frac{u^a u^b}{1+\hat k\cdot \hat u}
h_{ab}(f,\hat k)e^{i2\pi f (t-\hat k\cdot\vec x/c)}
\left[ 1-e^{-i2\pi fL(1+\hat k\cdot \hat u)/c}\right]\,.
\ee
The quantity that is actually observed by a pulsar timing 
measurement is the timing residual $r(t)$, which is 
related to the redshift $z(t)$ via
\be
\begin{aligned}
r(t)
&\equiv \int_0^t {\rm d}t'\> z(t')
\\
&= 
\int_{-\infty}^\infty {\rm d}f\>
\frac{1}{i2\pi f}
\int_{S^2} {\rm d}^2\Omega_{\hat k}\>
\frac{1}{2}\frac{u^a u^b}{1+\hat k\cdot \hat u}
h_{ab}(f,\hat k)e^{i2\pi f (t-\hat k\cdot\vec x/c)}
\left[ 1-e^{-i2\pi fL(1+\hat k\cdot \hat u)/c}\right]\,.
\label{e:r(t)}
\end{aligned}
\ee
If we expand $h_{ab}(f,\hat k)$ in terms of either 
$h^A(f,\hat k)$ or $a^P_{(lm)}(f)$ 
(see Eqs.~(\ref{e:hab+x}) and (\ref{e:habGC})),
and then compare the above expressions with 
Eqs.~(\ref{e:habRA}) and (\ref{e:habRP}), we see that
the detector response functions for the timing residuals
$r(t)$ are given by
\begin{align}
R^A(f,\hat k) 
&= 
\frac{1}{i2\pi f}
\frac{1}{2}\frac{u^a u^b}{1+\hat k\cdot \hat u}e_{ab}^A(\hat k)
e^{-i2\pi f\hat k\cdot\vec x/c}
\left[1-e^{-i2\pi fL(1+\hat k\cdot\hat u)/c}\right]\,,
\label{e:RA_pulsars_exact}
\\
R^P_{(lm)}(f)
&= 
\frac{1}{i2\pi f}
\int_{S^2}{\rm d}^2\Omega_{\hat k}\>
\frac{1}{2}
\frac{u^a u^b}{1+\hat k\cdot \hat u}Y^P_{(lm)ab}(\hat k)
e^{-i2\pi f\hat k\cdot\vec x/c}
\left[1-e^{-i2\pi fL(1+\hat k\cdot\hat u)/c}\right]\,.
\label{e:RP_pulsars_exact}
\end{align}
The detector response functions for the redshift $z(t)$ are
the above expressions {\em without} the factors of $1/(i2\pi f)$.

In what follows, we will make the approximations
\begin{align}
&R^A(f,\hat k)=(i2\pi f)^{-1} F^A(\hat k)\,,
\quad{\rm where}\quad
F^A(\hat k)
\equiv\frac{1}{2}\frac{u^a u^b}{1+\hat k\cdot \hat u}e_{ab}^A(\hat k)\,,
\label{e:RA_pulsars}
\\
&R^P_{(lm)}(f)=(i2\pi f)^{-1} F^P_{(lm)}\,,
\quad{\rm where}\quad
F^P_{(lm)}
\equiv\int_{S^2}{\rm d}^2\Omega_{\hat k}\>
\frac{1}{2}
\frac{u^a u^b}{1+\hat k\cdot \hat u}Y^P_{(lm)ab}(\hat k)\,.
\label{e:RP_pulsars}
\end{align}
This amounts to: (i) choosing a reference frame with the origin 
at the solar-system barycentre (SSB), for which a detector 
(i.e., a radio telescope on Earth) has $\vec x\approx \vec 0$, and
(ii) omitting the pulsar term, which is proportional to 
$\exp[-i2\pi f L(1+\hat k\cdot\hat u)/c]$.
In the case of an uncorrelated background, the contribution from this 
term averages to zero in the limit $fL/c \rightarrow \infty$, except for the
auto-correlation of each individual pulsar, which this term increases
by a factor of $2$.  The integrand that enters the cross-correlation 
$\langle r_1(t) r_2(t')\rangle$ for an uncorrelated background contains the factor
(see e.g.,~\cite{Mingarelli:2013}):
\begin{equation}
\kappa_{12}(f,\hat k)=
\left[1 - e^{-i2\pi f L_1(1 + \hat{k}\cdot\hat{u}_1)/c}\right]
\left[1 - e^{i2\pi f L_2(1 + \hat{k}\cdot\hat{u}_2)/c}\right] \,.
\end{equation}
For the autocorrelation of data from pulsar 1 with itself, this factor becomes
\begin{equation}
\label{eq:oscillation}
|1-e^{-i2\pi fL_1(1 + \hat{k}\cdot\hat{u}_1)/c}|^2
=2-2\cos[2\pi fL_1(1 + \hat{k}\cdot\hat{u}_1)/c]\,.
\end{equation} 
It is clear that for $fL_1/c\gg1$ the contribution of the 
oscillatory term to the integral for the overlap reduction function
will be suppressed by a factor of a least $1/(fL_1/c)$ and can hence
be ignored. In fact, for an isotropic uncorrelated background, the
contribution from the oscillatory term is always small as
it is suppressed by a factor of $1/(fL_1/c)^2$. (Details of this
calculation are given in Appendix~\ref{s:pulsar_term}.)  As the angular
separation between pulsar pairs increases from zero, the value of the
overlap reduction function decreases rapidly to the Earth-term only
value. This is a continuous transition; however for $fL_1/c\gg1$, it is
well modeled by a step function at zero angular separation of the
pulsar pair.  In the following analysis we will consider only
inter-pulsar correlations and can therefore ignore the pulsar term. For a 
more rigorous investigation of when the pulsar term can be ignored, 
see e.g. \cite{MingarelliSidery:2014}.

\subsection{Antenna beam patterns and pulsar response sky maps}

Figure~\ref{f:FpFc} shows Mollweide projections of the 
frequency-independent response functions
$F^+(\hat k)$, $F^\times(\hat k)$ 
for a pulsar located in direction $(\theta,\phi)=(50^\circ, 60^\circ)$,
which corresponds to ($40^\circ$N, $60^\circ$E) in the two plots.
\begin{figure*}[htbp]
\begin{center}
\includegraphics[trim=3cm 6.5cm 3cm 3.5cm, clip=true, angle=0, width=0.4\textwidth]{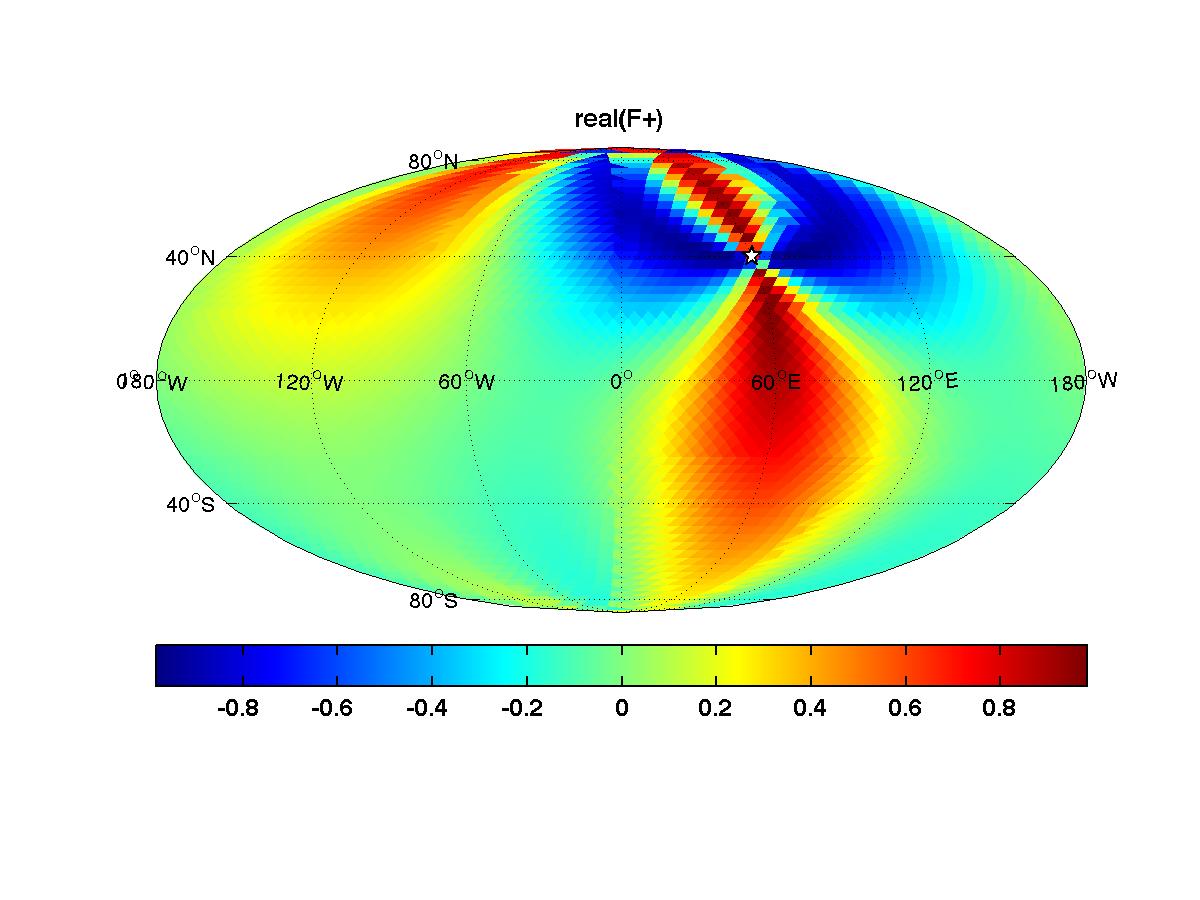}
\hspace{0.05\textwidth}
\includegraphics[trim=3cm 6.5cm 3cm 3.5cm, clip=true, angle=0, width=0.4\textwidth]{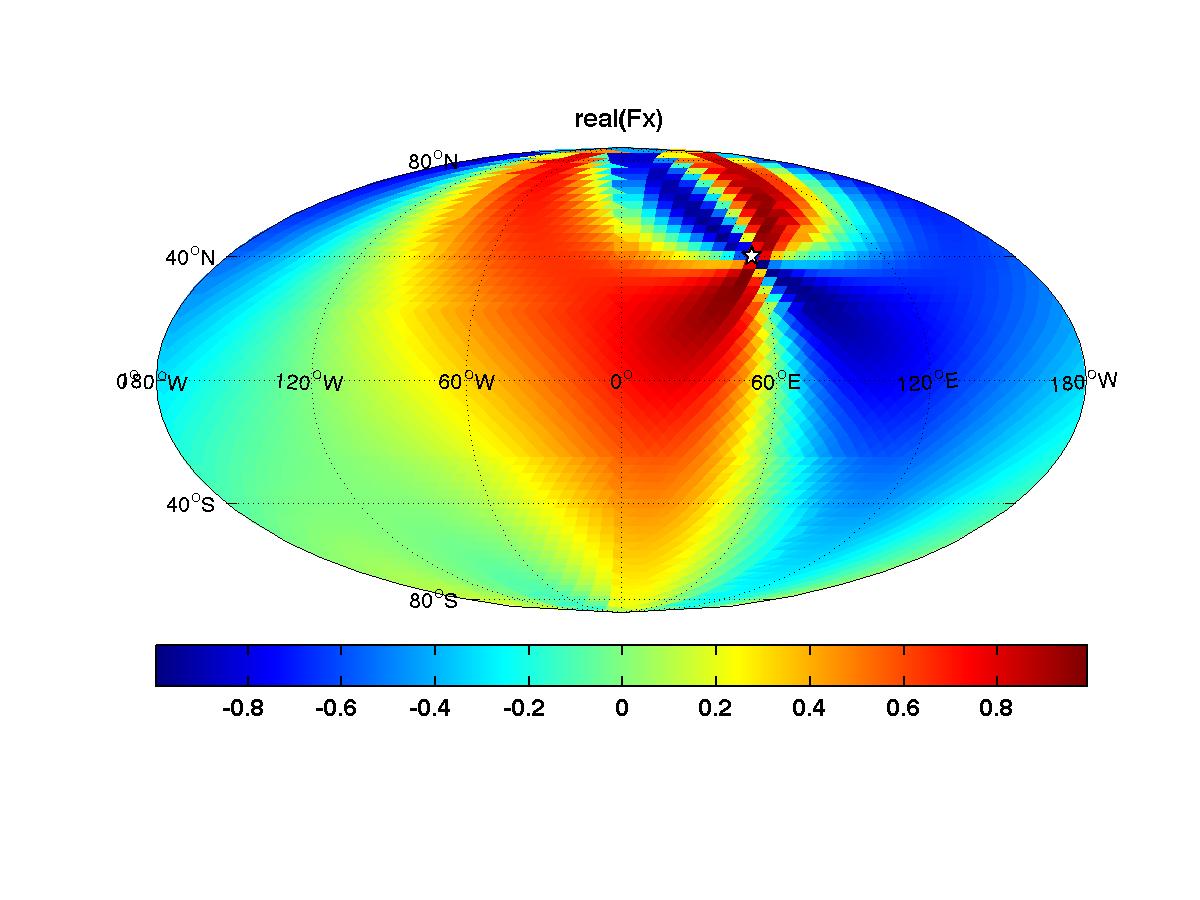}
\caption{Mollweide projections of the frequency-independent
response functions $F^+(\hat k)$, $F^\times(\hat k)$, 
for a pulsar located in direction $(\theta,\phi)=(50^\circ, 60^\circ)$,
indicated on the plots by a white star.
The imaginary parts of both response functions are identically 
zero, and are not shown above.}
\label{f:FpFc}
\end{center}
\end{figure*}
Note that the two sky maps are related by a rotation of $45^\circ$ 
around an axis passing through the direction to the pulsar.
Plots of $F^+(\hat k)$, $F^\times(\hat k)$ are sometimes referred
to as ``antenna beam patterns" in the literature, e.g., \cite{Mingarelli:2013}.
In Fig.~\ref{f:responseMap_point}, we plot the real and imaginary 
parts of the Fourier transform of the redshift response
\be
\tilde z(f)= \int_{S^2}{\rm d}^2\Omega_{\hat k}\>
\left[
F^+(\hat k) h_+(f,\hat k) + F^\times(\hat k) h_\times(f,\hat k)
\right]
\ee
to a $+$-polarized point source 
$h_+(f,k)\equiv\delta(\hat k,-\hat k_0)$, $h_\times(f,\hat k)\equiv 0$
located in direction $\hat k_0$ having angular
coordinates $(\theta_0,\phi_0)=(50^\circ, 60^\circ)$.
\begin{figure*}[htbp]
\begin{center}
\includegraphics[trim=3cm 6.5cm 3cm 3.5cm, clip=true, angle=0, width=0.4\textwidth]{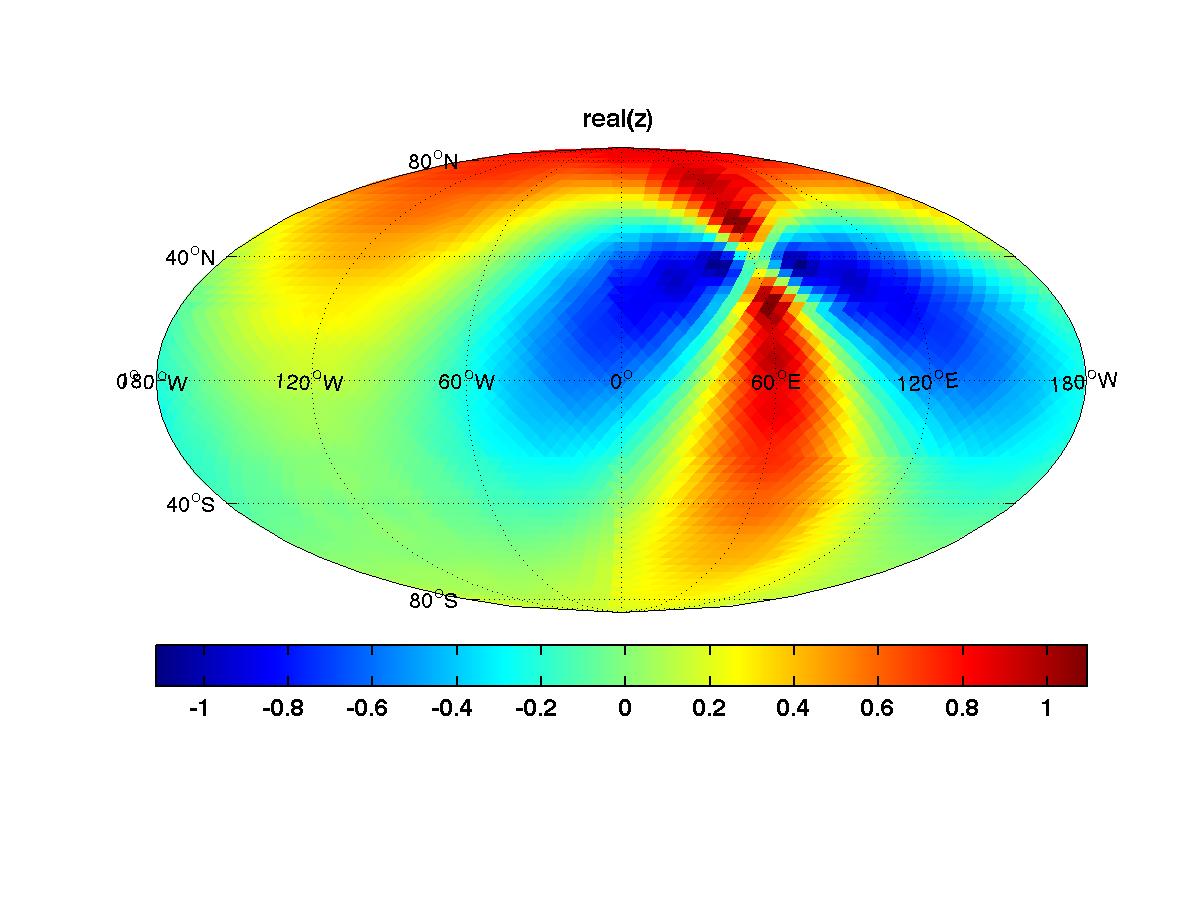}
\hspace{.05\textwidth}
\includegraphics[trim=3cm 6.5cm 3cm 3.5cm, clip=true, angle=0, width=0.4\textwidth]{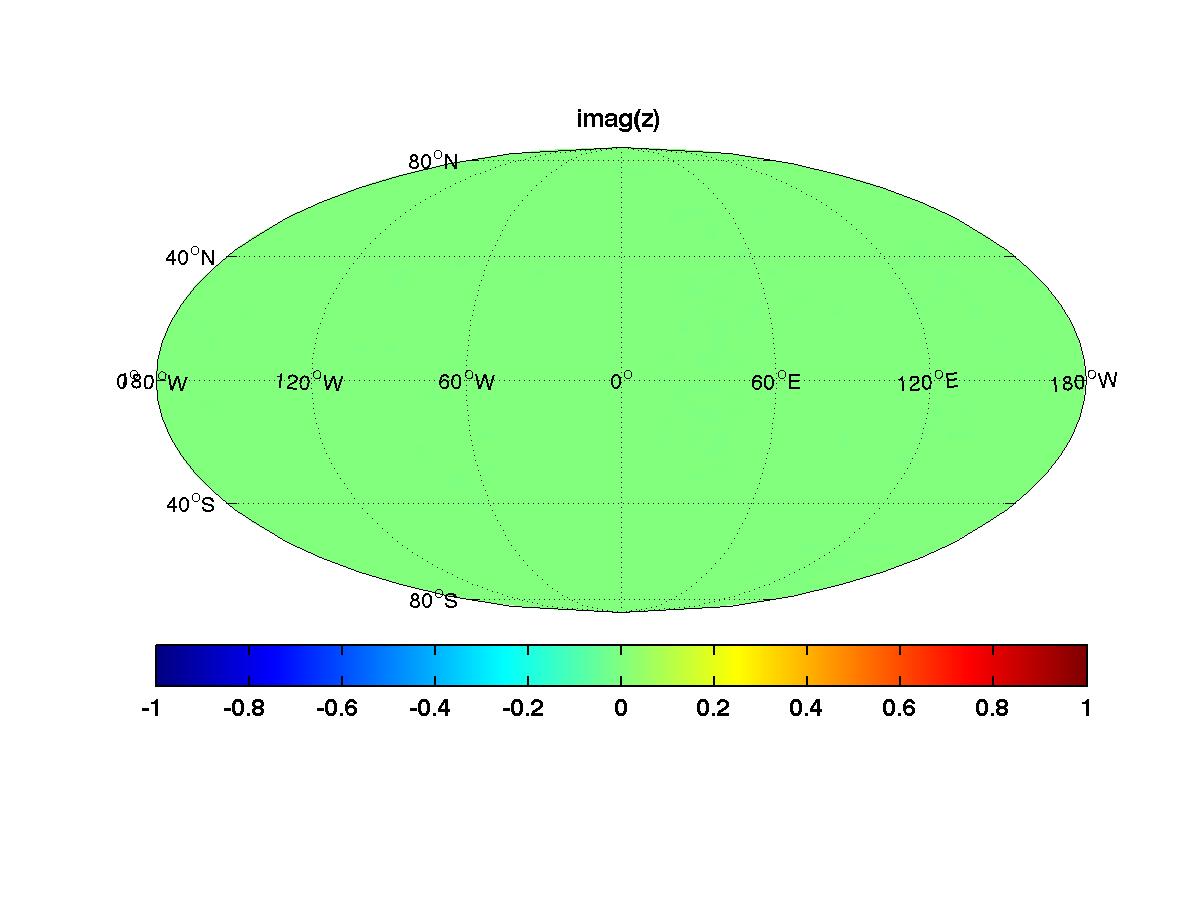}
\caption{Mollweide projections of the real and imaginary parts of
the Fourier transform of the redshift response $\tilde z(f)$ 
to a $+$-polarized point source located in direction 
$(\theta_0,\phi_0)=(50^\circ, 60^\circ)$.
The sky location of the pulsar is variable.
The imaginary part of the response is identically zero, indicated by
the solid green color of the second plot.}
\label{f:responseMap_point}
\end{center}
\end{figure*}
In this pulsar response sky map, the point source is fixed, but
the direction to the pulsar is variable, specified by the angles $(\theta$, $\phi)$. 
If we had the ability to move pulsars on the sky, this map shows us 
where the pulsars should ideally be placed to maximise the magnitude 
of the response (the dark red and dark blue areas). 
As a simple example, one can show that 
$\tilde z(f) = \frac{1}{2}(1+\cos \theta)\cos(2\phi)$
for a $+$-polarized point source located at the North Pole.
For a more complicated gravitational-wave background, 
the pulsar response map is shown in Fig.~\ref{f:responseMap_background}.
\begin{figure*}[htbp]
\begin{center}
\includegraphics[trim=3cm 6.5cm 3cm 3.5cm, clip=true, angle=0, width=0.4\textwidth]{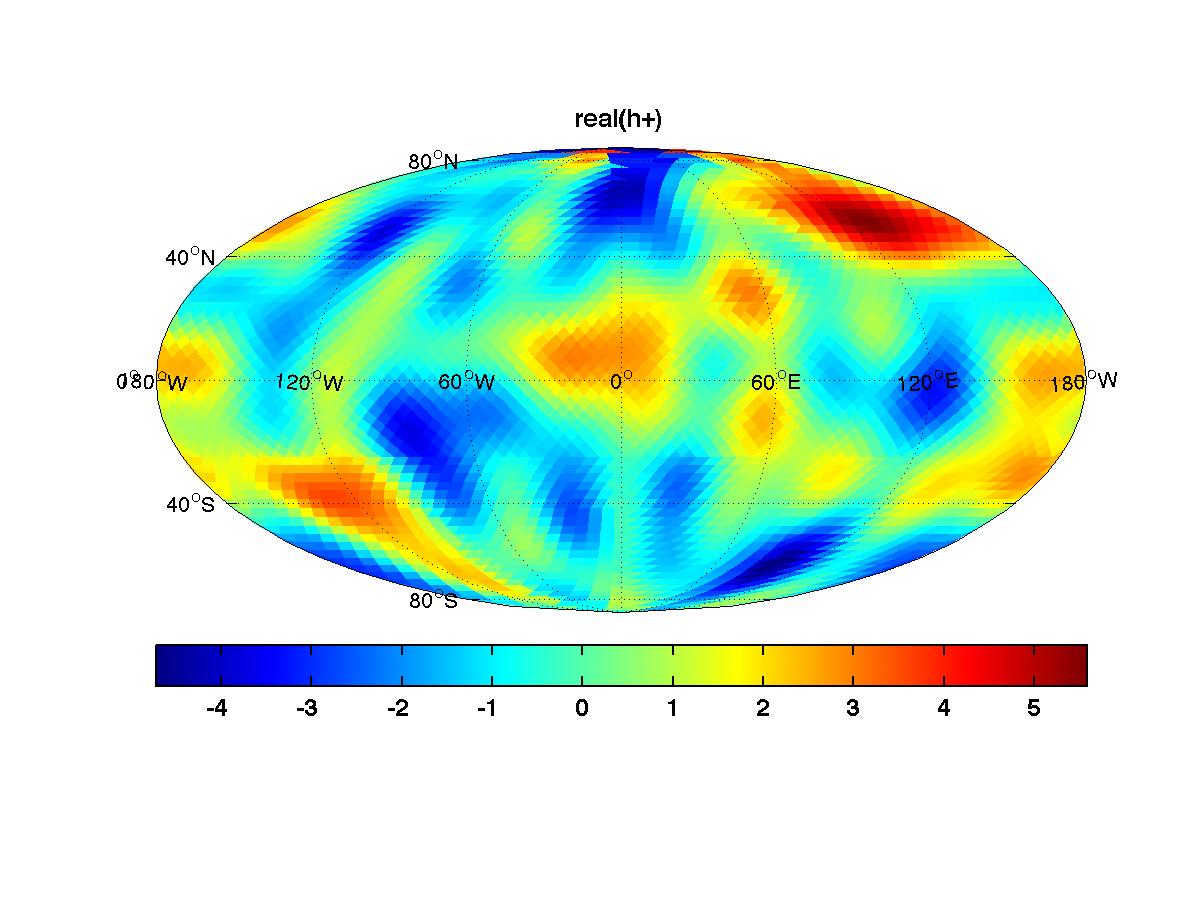}
\hspace{.05\textwidth}
\includegraphics[trim=3cm 6.5cm 3cm 3.5cm, clip=true, angle=0, width=0.4\textwidth]{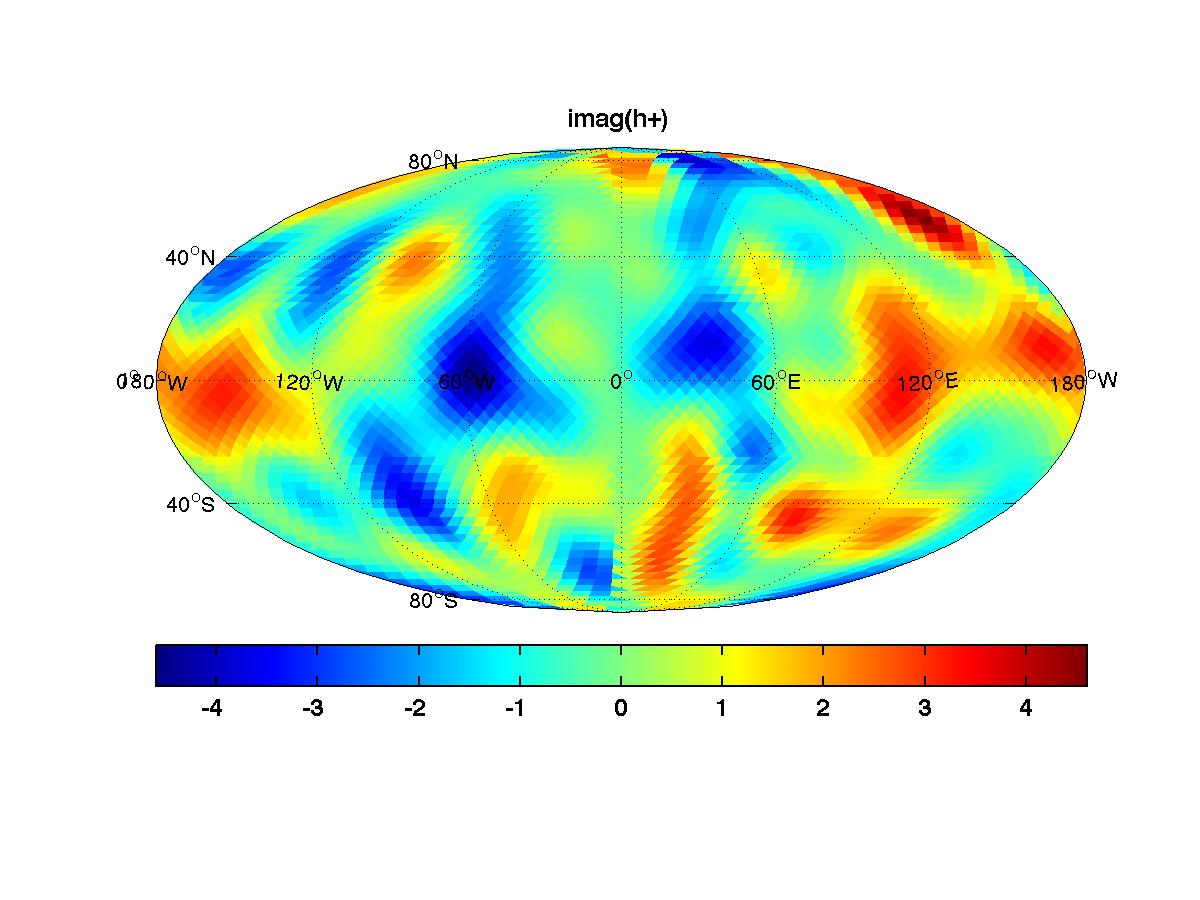}
\includegraphics[trim=3cm 6.5cm 3cm 3.5cm, clip=true, angle=0, width=0.4\textwidth]{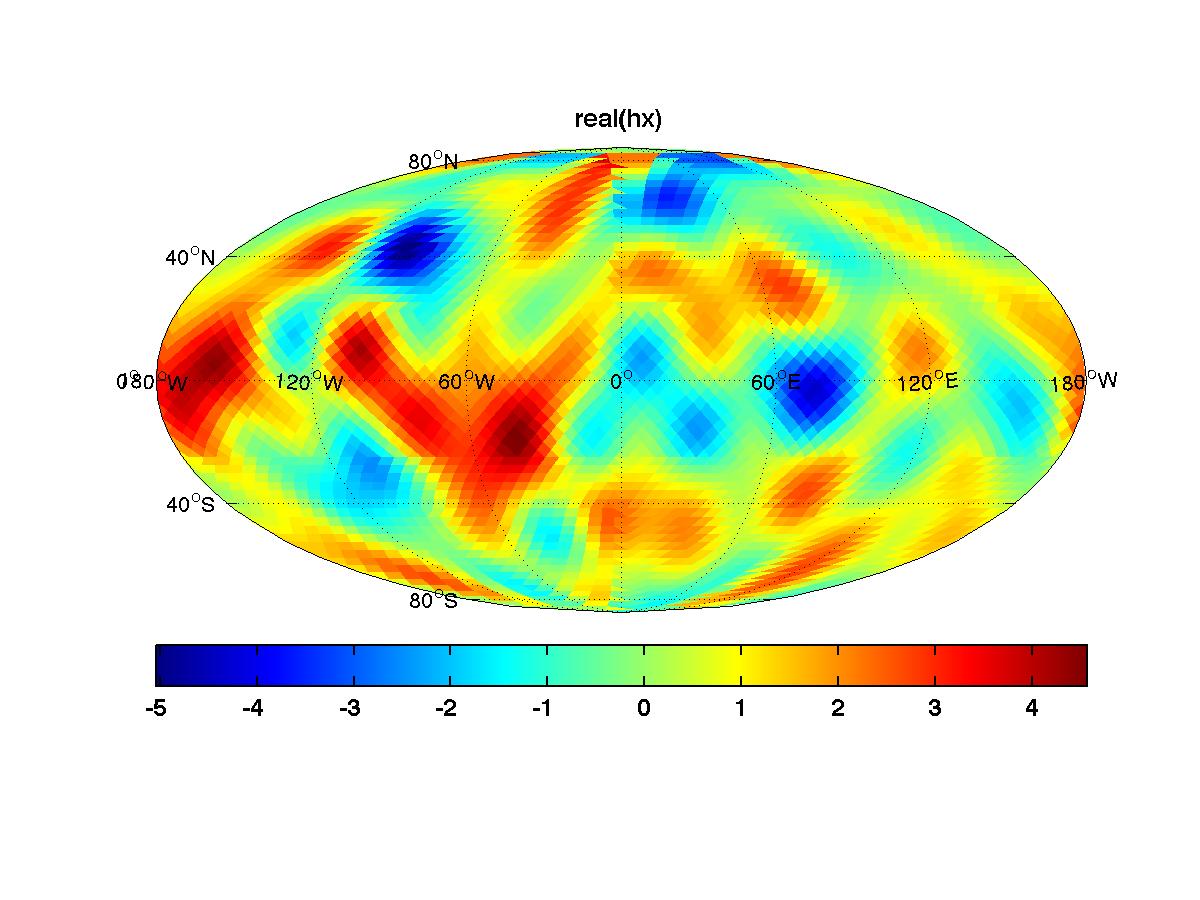}
\hspace{.05\textwidth}
\includegraphics[trim=3cm 6.5cm 3cm 3.5cm, clip=true, angle=0, width=0.4\textwidth]{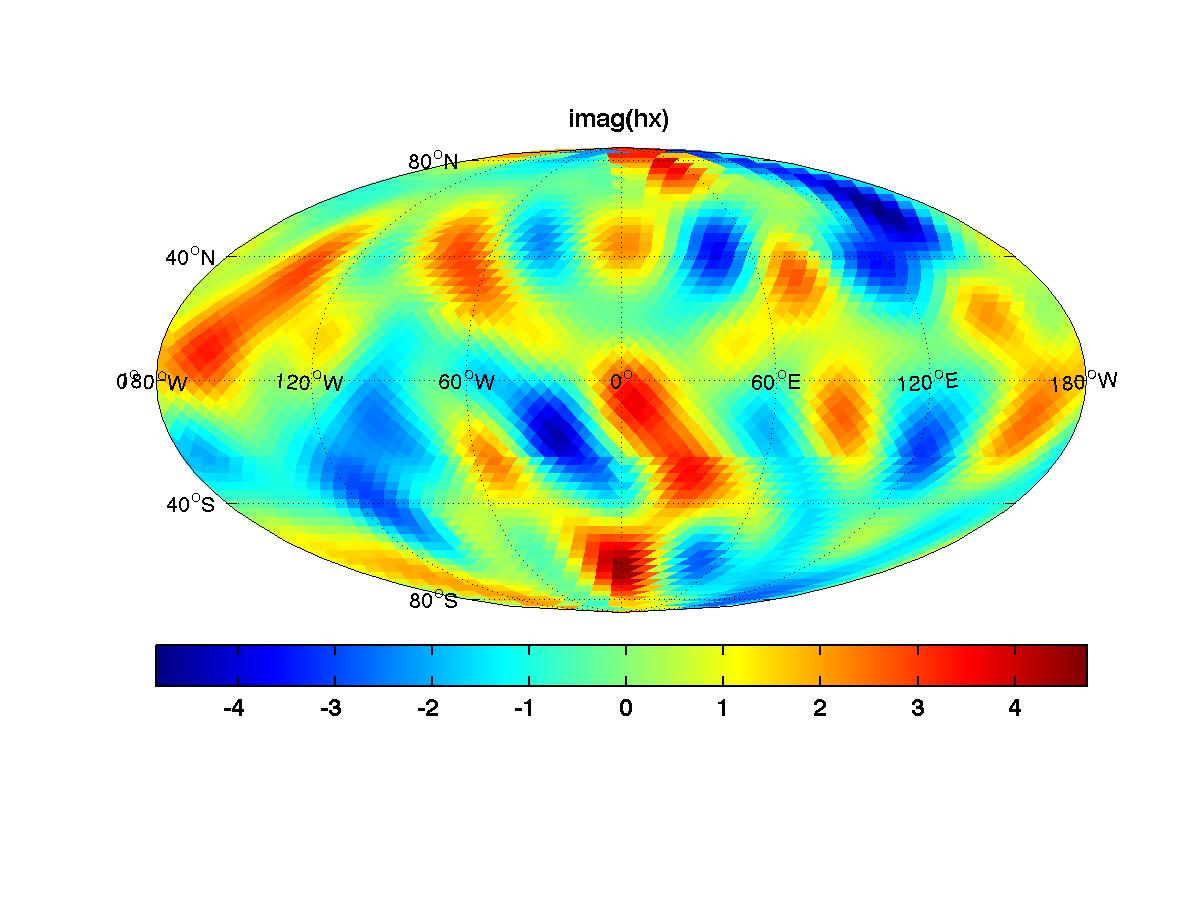}
\includegraphics[trim=3cm 6.5cm 3cm 3.5cm, clip=true, angle=0, width=0.4\textwidth]{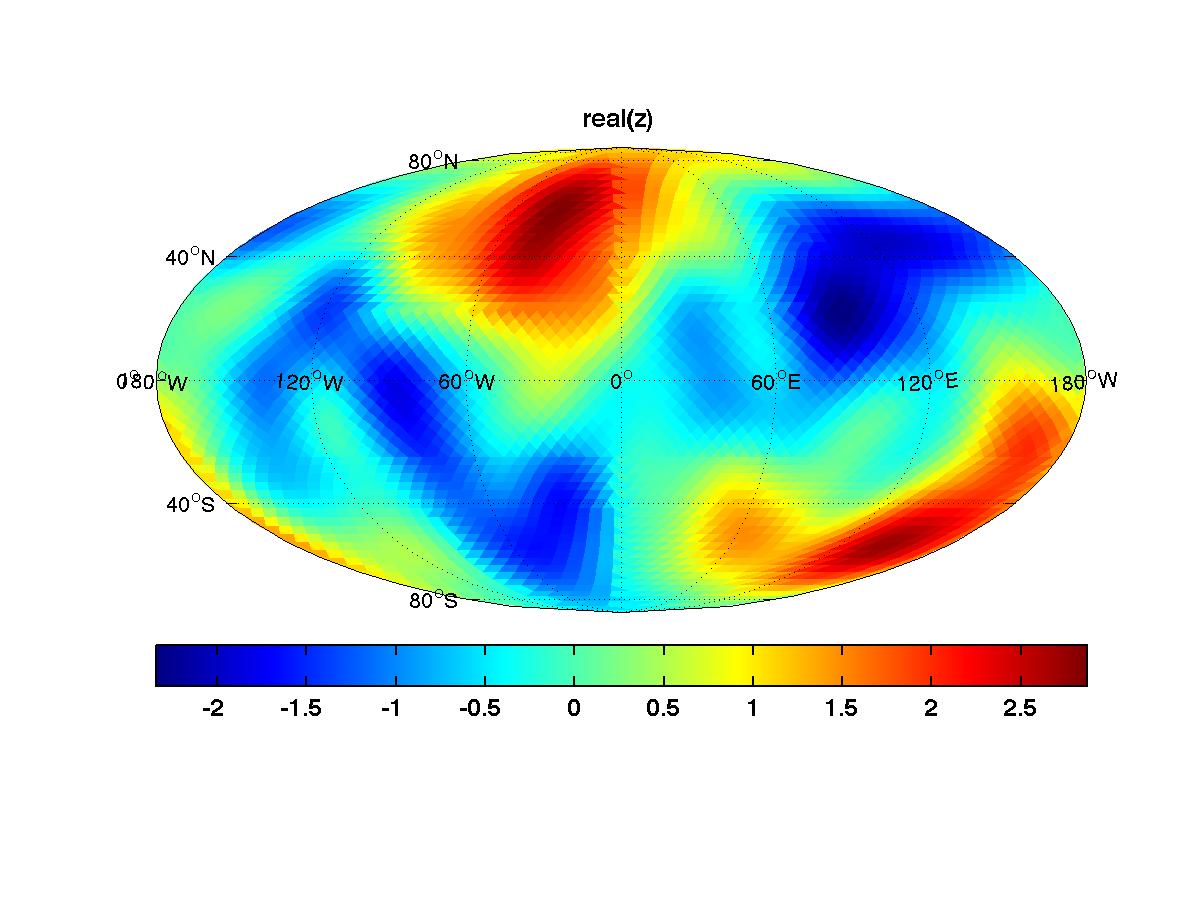}
\hspace{.05\textwidth}
\includegraphics[trim=3cm 6.5cm 3cm 3.5cm, clip=true, angle=0, width=0.4\textwidth]{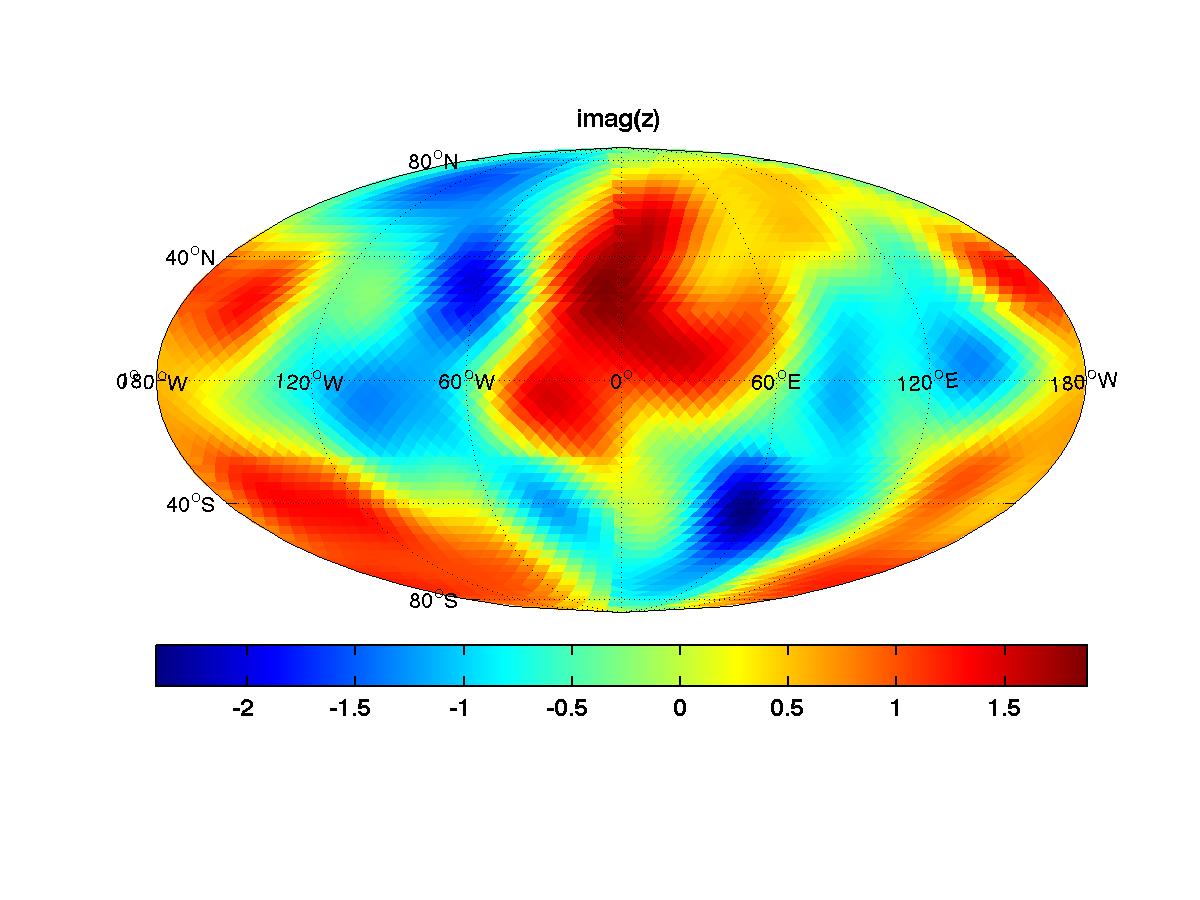}
\caption{Mollweide projections of the real and imaginary parts 
of $h_+(\hat k)$, $h_\times(\hat k)$ for a simulated gravitational-wave background 
(top two rows) and the corresponding pulsar response map (bottom row).}
\label{f:responseMap_background}
\end{center}
\end{figure*}

Note that the Fourier transform of the timing 
residual response $\tilde r(f)$ is related to the Fourier
transform of the redshift response $\tilde z(f)$ by the 
frequency-dependent scaling 
$\tilde r(f)=\tilde z(f)/(i2\pi f)$,
which follows from Eq.~(\ref{e:r(t)}).
So it is a simple matter to go back-and-forth between these two types of response.

\subsection{Overlap reduction function}
\label{sec:overlap}

Consider two pulsars, labeled by $1$ and $2$, and let $\hat u_1$,
$\hat u_2$ be unit vectors pointing from Earth to each pulsar.  
To calculate the expected value of the correlation between the 
signals from these two pulsars due to a statistically isotropic 
gravitational-wave background, we need to evaluate the integrals 
(\ref{e:RGC}) 
that appear in Eq.~(\ref{eq:gamma_12,l}) for the components of 
the overlap reduction function.
This can be done in any reference frame.  Here we follow the 
standard approach of \cite{Mingarelli:2013, TaylorGair:2013}, 
and work in the so-called ``computational" frame in which pulsar 1 is
located on the $z$-axis and pulsar $2$ is located in the $xz$-plane,
making an angle $\zeta$ with respect to the $z$-axis:
\be
\begin{aligned}
\hat u_1 &= (0, 0, 1)\,,
\\
\hat u_2 &= (\sin\zeta, 0, \cos\zeta)\,.
\end{aligned}
\ee
We will also assume that the detector locations are at the 
origin (the SSB):
\be
\vec x_1\approx \vec x_2 \approx\vec 0\,.
\ee
Furthermore, we will do the calculation in terms of the traditional 
$+$ and $\times$ detector response functions, making the approximation 
$R^A_I(f,\hat k)=(i2\pi f)^{-1}F^A_I(\hat k)$ (for $I=1,2$) 
discussed above.
(In Sec.~\ref{s:alternative_derivation}, we will give an alternative
derivation of the overlap reduction function, doing the calculation
in different reference frames for each pulsar, and working more directly
with the gradient and curl response functions.)

In this computational frame, it is easy to show that
\begin{align}
F^+_1(\hat{k})
=\frac{1}{2}\frac{u^a_1 u^b_1}{1+\hat k\cdot\hat u_1} e^+_{ab}(\hat k)
=\frac{1}{2}(1-\cos\theta)\,,
\quad
F^\times_1(\hat{k})=\frac{1}{2}\frac{u^a_1 u^b_1}{1+\hat k\cdot\hat u_1} 
e^\times_{ab}(\hat k)=0\,.
\end{align}
Since $W_{(lm)}(\hat{k})$ and $X_{(lm)}(\hat{k})$ are both
proportional to $e^{im\phi}$ and $F^+_1(\hat{k})$ is independent of
$\phi$, the integral over $\phi$ gives zero for all $m\neq 0$. We note
also that $X_{(l0)} = 0$ for all $l$, which follows from Eq.~\eqref{e:Xlm}. 
Thus, Eq.~(\ref{eq:gamma_12,l}) simplifies to
\be
\Gamma_{12,l}(f)
=
\frac{1}{(2\pi f)^2}
\frac{(N_l)^2}{4}
\left[ \int {\rm d}^2\Omega_{\hat k}\> F_1^+(\hat k) W_{(l0)}(\hat{k}) \right] \,\, 
\left[ \int {\rm d}^2\Omega_{\hat k'}\> F_2^+{}^*(\hat k') W^*_{(l0)}(\hat{k}') \right]\,,
\label{eq:ptagamma_12,l}
\ee
where the factor of $(2\pi f)^{-2}$ comes from the frequency-dependent
factors of $(i2\pi f)^{-1}$ in Eqs.~(\ref{e:RA_pulsars}) and (\ref{e:RP_pulsars}). Note that 
the frequency dependence in Eq.~\eqref{eq:ptagamma_12,l} is 
usually absorbed into $H(f)$ in Eq.~\eqref{e:r1r2corr}, see e.g.,~\cite{TaylorGair:2013}.
We will denote the frequency-independent part of the overlap reduction 
functions with an overbar, e.g., $\bar\Gamma_{12}$ and $\bar \Gamma_{12,l}$.

We will consider the integral over $\hat{k}$ first, making use of 
Eq.~(\ref{e:Wlm}) written in terms of $x=\cos\theta$:
\begin{equation}
\int {\rm d}^2\Omega_{\hat k}\> F_1^+(\hat k) W_{(l0)}(\hat{k}) 
=  \sqrt{\frac{(2l+1)\pi}{4}} \int_{-1}^{1} {\rm d}x\>
(1-x) \left(2(1-x^2) \frac{{\rm d}^2}{{\rm d}x^2} 
- 2x\frac{{\rm d}}{{\rm d}x} + l(l+1)\right) P_l(x)\,,
\end{equation}
where $P_l(x)$ is a Legendre polynomial.  This last expression in
brackets can be simplified further, noting that
\begin{equation}
\left(2(1-x^2) \frac{{\rm d}^2}{{\rm d}x^2} 
- 2x\frac{{\rm d}}{{\rm d}x} + l(l+1)\right) P_l(x) 
= (1-x^2) \frac{{\rm d}^2}{{\rm d}x^2} P_l(x)\,.
\end{equation}
Thus,
\begin{equation}
\int {\rm d}^2\Omega_{\hat k}\> F_1^+(\hat k) W_{(l0)}(\hat{k}) =
\sqrt{\frac{(2l+1)\pi}{4}} \int_{-1}^{1}{\rm d}x\>
(1-x)(1-x^2) \frac{{\rm d}^2}{{\rm d}x^2} P_l(x)\,.
\end{equation}
If we now integrate by parts twice, and use the orthogonality of the
Legendre polynomials, Eq.~(\ref{eq:PlmOrth}) 
(noting that $l \geq 2$ for this tensor harmonic basis), we find that 
only the second boundary term contributes, yielding
\begin{equation}
\int {\rm d}^2\Omega_{\hat k}\> F_1^+(\hat k) W_{(l0)}(\hat{k}) 
= 2\sqrt{(2l+1)\pi} (-1)^l\,.
\label{e:F+W_simple}
\end{equation}

To evaluate the integral over $\hat{k}'$, we note that
\be
\begin{aligned}
F_2^+{}^*(\hat k')
&=\frac{1}{2}\frac{u_2^a u_2^b}{1+\hat k'\cdot\hat u_2}e^+_{ab}(\hat k')
\\
&=\frac{1}{2}
\frac{(\sin\zeta\cos\theta'\cos\phi' -\cos\zeta\sin\theta')^2- (-\sin\zeta\sin\phi')^2}
{1+\cos\zeta\cos\theta'+\sin\zeta\sin\theta'\cos\phi'}
\\
&=\frac{1}{2}\left[
(1-\cos\zeta\cos\theta'-\sin\zeta\sin\theta'\cos\phi')
-\frac{2\sin^2\zeta\sin^2\phi'}{1+\cos\zeta\cos\theta'+\sin\zeta\sin\theta'\cos\phi'}
\right]\,.
\end{aligned}
\ee
(We do not need to calculate $F_2^\times(\hat k')$, since it does not
enter the expression for the overlap reduction function in our chosen
reference frame.)  The integral can be written following
\cite{Mingarelli:2013} as a sum of two parts:
\begin{equation}
\int {\rm d}^2\Omega_{\hat k'}\> F_2^+{}^*(\hat k') W^*_{(l0)}(\hat{k}') 
= Q_l + R_l\,,
\end{equation}
where
\begin{align}
Q_l &= \sqrt{\frac{(2l+1)\pi}{4}}
\int_{-1}^1 {\rm d}x\>
(1-x\cos\zeta) 
(1-x^2) \frac{{\rm d}^2}{{\rm d}x^2} P_l(x)\,,
\\
R_l &=   -\sqrt{\frac{(2l+1)}{4\pi}}\sin^2\zeta\
\int_{-1}^{1}{\rm d}x\>
(1-x^2) \frac{{\rm d}^2}{{\rm d}x^2} P_l(x)
\int_0^{2\pi} {\rm d}\phi'\>
\frac{\sin^2\phi'}{1+x\cos\zeta +\sqrt{1-x^2}\sin\zeta\cos\phi'}\,.
\label{e:pulsar2int}
\end{align}
The $Q_l$ integral is similar in form to the integral over $\hat{k}$
given above and can be evaluated in the same way:
\begin{equation}
Q_l =  \sqrt{(2l+1)\pi} \left[(1-\cos\zeta)+(1+\cos\zeta)(-1)^l\right]\,.
\end{equation}
For $R_l$, the integral over $\phi'$ is familiar from the standard
computation of the Hellings and Downs curve. $R_l$ can be reduced 
to the form
\be
\begin{aligned}
R_l &= -\sqrt{(2l+1)\pi}\left[
\int_{-1}^{-\cos\zeta}{\rm d}x\,\, \frac{(1+\cos\zeta)}{(1-x)} 
(1-x^2) \frac{{\rm d}^2}{{\rm d}x^2} P_l(x)
+ \int_{-\cos\zeta}^1{\rm d}x \,\, \frac{(1-\cos\zeta)}{(1+x)}  
(1-x^2) \frac{{\rm d}^2}{{\rm d}x^2} P_l(x)
\right]
\\
&= -\sqrt{(2l+1)\pi} \left[ 
(1+\cos\zeta) \int_{-1}^{-\cos\zeta} {\rm d}x\>(1+x) 
\frac{{\rm d}^2}{{\rm d}x^2} P_l(x) 
+(1-\cos\zeta) \int_{-1}^{-\cos\zeta} {\rm d}x\>(1-x) 
\frac{{\rm d}^2}{{\rm d}x^2} P_l(x)\right]\,.
\end{aligned}
\ee
Integrating each term by parts we obtain 
\begin{equation}
R_l =  -\sqrt{(2l+1)\pi} \left[ (1-\cos\zeta) 
+ (1+\cos\zeta) (-1)^l - 2(-1)^l P_l(\cos\zeta) \right]\,.
\end{equation}
Thus, the $\hat k'$ integral is simply
\begin{equation}
\int {\rm d}^2\Omega_{\hat k'} F_2^+{}^*(\hat k') W^*_{(l0)}(\hat{k}') 
= 2\sqrt{(2l+1)\pi}(-1)^lP_l(\cos\zeta)
\end{equation}
Putting the above results together, we obtain
\begin{equation}
\Gamma_{12,l}(f) = 
\frac{1}{(2\pi f)^2}
(N_l)^2 (2l+1) \pi P_l(\cos\zeta)\,. 
\label{e:overlap_l,statiso}
\end{equation}
Thus, for a statistically isotropic background, the overlap reduction
functions for the individual $l$ modes are proportional to Legendre
polynomials.  The full overlap reduction function, assuming
$C_l^{GG}=C_l^{CC}\equiv C_l$ and $C_l^{GC}=0=C_l^{CG}$, 
is given by the sum
\begin{equation}
\Gamma_{12}(f) = 
\frac{1}{(2\pi f)^2}
\sum_{l=2}^\infty
C_l (N_l)^2 (2l+1) \pi P_l(\cos\zeta)\,.
\label{e:overlap,statiso}
\end{equation}
%

\subsection{Alternative derivation of the overlap reduction function}
\label{s:alternative_derivation}

The above result for $\Gamma_{12,l}(f)$ is surprisingly simple, considering
the somewhat involved calculation needed to derive it.  This raises
the question as to whether there is an alternative approach that would
lead more directly to the final result.  The answer
to this question is yes, and it is based on the observation that
Eq.~(\ref{eq:gamma_12,l}) for $\Gamma_{12,l}(f)$ involves a sum of
{\em products} of two integrals, each of the form given by 
Eq.~(\ref{e:RP_pulsars_exact}).
This means that the integrals for pulsars 1 and 2 need not be 
evaluated in the same reference frame as we did in the
previous section, but can be evaluated in different
reference frames, appropriately chosen to simplify the calculation of
the integral for each pulsar separately.  In particular, we can rotate
coordinates so that, for each pulsar, $\hat u_I$ is directed along the
transformed $z$-axis.

To be more explicit, consider a particular pulsar $I$ located in direction
$\hat u_I$.  In the ``cosmic" reference frame, where the angular
dependence of the gravitational-wave background is to be described,
let the angular coordinates of the pulsar be $(\zeta_I,\chi_I)$, so that
\be
u^a_I = (\sin\zeta_I\cos\chi_I, \sin\zeta_I\sin\chi_I, \cos\zeta_I)\,.
\ee
A rotation 
\be
\mb{R}(\alpha,\beta,\gamma)
=\mb{R}_z(\gamma)\mb{R}_y(\beta)\mb{R}_z(\alpha)
=
\left(
\begin{array}{ccc}
\cos\gamma & \sin\gamma & 0 \\
-\sin\gamma & \cos\gamma & 0 \\
0 & 0 & 1
\end{array}
\right)
\left(
\begin{array}{ccc}
\cos\beta & 0 & -\sin\beta \\
0 & 1 & 0 \\
\sin\beta & 0 & \cos\gamma 
\end{array}
\right)
\left(
\begin{array}{ccc}
\cos\alpha & \sin\alpha & 0 \\
-\sin\alpha & \cos\alpha & 0 \\
0 & 0 & 1
\end{array}
\right)
\ee
defined by the Euler angles 
\be
(\alpha,\beta,\gamma)=(\chi_I,\zeta_I,0)
\ee
will rotate the cosmic frame (with coordinates $x^a$)
to the computational frame for pulsar $I$ 
(with coordinates $x_I^{\bar a}=\mb{R}(\chi_I,\zeta_I,0)^{\bar a}{}_a x^a$), 
so that 
\be
u_I^{\bar a}=(0,0,1)\,.
\ee

To evaluate the response functions in the computational frame, 
we need to know how the integrands 
transform under a rotation.  Ignoring the pulsar term and making
the approximation $\vec x_I\approx \vec 0$, these integrals 
simplify to
\begin{align}
R^P_{I(lm)} (f) = (i2\pi f)^{-1} F^P_{I(lm)}\,,
\quad{\rm where}\quad
F^P_{I(lm)} \equiv \int_{S^2} {\rm d}^2\Omega_{\hat k}\>
\frac{1}{2}
\frac{u_I^a u_I^b}{1+\hat k\cdot\hat u_I}\,Y^P_{(lm)ab}(\hat k)
\label{e:IPlm_pulsars}
\end{align}
as mentioned in the text following Eqs.~(\ref{e:RA_pulsars})
and (\ref{e:RP_pulsars}).
The frequency-dependent factor $(i2\pi f)^{-1}$ is a scalar 
with respect to rotations, so we need only determine the 
transformation properties of $F^P_{I(lm)}$.
Since the gradient and curl spherical harmonics transform 
like tensors with respect to the $ab$ indices and like 
ordinary spherical harmonics with respect to the $lm$ indices 
\cite{Challinor:2000}, we have:
\be
Y^{P}_{(lm)ab}(\theta, \phi)=
\sum_{m'=-l}^l
\left[D^l{}_{mm'}(\chi,\zeta,0)\right]^*\,
Y^{P}_{(lm')\bar a\bar b}(\bar\theta_I,\bar\phi_I)
\mb{R}(\chi_I,\zeta_I,0)^{\bar a}{}_a \mb{R}(\chi_I,\zeta_I,0)^{\bar b}{}_b\,.
\label{e:harmtrans}
\ee
Here $D^l{}_{m'm}(\chi_I,\zeta_I,0)$ is the Wigner-D matrix 
associated with the rotation $\mb{R}(\chi_I,\zeta_I,0)$, and
$(\bar\theta_I,\bar\phi_I)$ are the angular coordinates of 
the direction $\hat k$ in the computational frame for pulsar $I$.
Thus, for the gradient response:
\be
\begin{aligned}
F^G_{I(lm)} 
&=\int_0^{2\pi}{\rm d}\phi\int_{-1}^1{\rm d}\cos\theta\>
\frac{1}{2}
\frac{u_I^a u_I^b}{1+\hat k\cdot\hat u_I}\,Y^G_{(lm)ab}(\theta,\phi)
\\
&=\sum_{m'=-l}^l
\left[D^l{}_{mm'}(\chi_I,\zeta_I,0)\right]^*
\int_0^{2\pi}{\rm d}\bar\phi_I\int_{-1}^1{\rm d}\cos\bar\theta_I\>
\frac{1}{2}
\frac{u_I^{\bar a} u_I^{\bar b}}{1+\hat k\cdot\hat u_I}\,
Y^G_{(lm')\bar a\bar b}(\bar\theta_I,\bar\phi_I)
\\
&=\sum_{m'=-l}^l
\left[D^l{}_{mm'}(\chi_I,\zeta_I,0)\right]^*
\int_0^{2\pi}{\rm d}\bar\phi_I\int_{-1}^1{\rm d}\cos\bar\theta_I\>
\frac{N_l}{2}\left[
F^+_I(\bar\theta,\bar\phi)W_{(lm')}(\bar\theta_I,\bar\phi_I)+
F^\times_I(\bar\theta_I,\bar\phi_I)X_{(lm')}(\bar\theta_I,\bar\phi_I)\right]\,,
\end{aligned}
\ee
where we used Eq.~(\ref{e:RGC}) with the $R$'s replaced by $F$'s 
to get the last line.
Since in the computational frame
\be
F_I^+(\bar\theta_I,\bar\phi_I) = \frac{1}{2}(1-\cos\bar\theta_I)\,,
\qquad
F_I^\times(\bar\theta_I,\bar\phi_I) = 0\,,
\ee
we can ignore the $F_I^\times(\bar\theta_I,\bar\phi_I)$ term.
In addition, because $F_I^+(\bar\theta_I,\bar\phi_I)$ is independent of 
$\bar\phi_I$, we only get a contribution from $m'=0$.
Thus,
\be
\begin{aligned}
F^G_{I(lm)}
&=\left[D^l{}_{m0}(\chi_I,\zeta_I,0)\right]^*
\int_0^{2\pi}{\rm d}\bar\phi_I\int_{-1}^1{\rm d}\cos\bar\theta_I\>
\frac{N_l}{2}F^+_I(\bar\theta_I,\bar\phi_I)W_{(l0)}(\bar\theta_I,\bar\phi_I)
\\
&=\left[D^l{}_{m0}(\chi_I,\zeta_I,0)\right]^*
N_l\sqrt{(2l+1)\pi}(-1)^l 
\\
&=2\pi(-1)^l N_l Y_{(lm)}(\zeta_I,\chi_I)\,,
\label{e:calI_finalresult}
\end{aligned}
\ee
where we used Eq.~(\ref{e:F+W_simple}) for 
the $W_{(l0)}(\bar\theta_I,\bar\phi_I)$ integration and 
Eq.~(\ref{e:WignerD_CC}) for the Wigner-D matrix.

Proceeding in exactly the same manner for the curl response, 
we find
\be
\begin{aligned}
F^C_{I(lm)} 
&=\int_0^{2\pi}{\rm d}\phi\int_{-1}^1{\rm d}\cos\theta\>
\frac{1}{2}\frac{u_I^a u_I^b}{1+\hat k\cdot\hat u_I}\,Y^C_{(lm)ab}(\theta,\phi)
\\
&=\sum_{m'=-l}^l
\left[D^l{}_{mm'}(\chi_I,\zeta_I,0)\right]^*
\int_0^{2\pi}{\rm d}\bar\phi_I\int_{-1}^1{\rm d}\cos\bar\theta_I\>
\frac{1}{2}\frac{u_I^{\bar a} u_I^{\bar b}}{1+\hat k\cdot\hat u_I}\,
Y^C_{(lm')\bar a\bar b}(\bar\theta_I,\bar\phi_I)
\\
&=\sum_{m'=-l}^l
\left[D^l{}_{mm'}(\chi_I,\zeta_I,0)\right]^*
\int_0^{2\pi}{\rm d}\bar\phi_I\int_{-1}^1{\rm d}\cos\bar\theta_I\>
\frac{N_l}{2}
\left[
F^\times_I(\bar\theta_I,\bar\phi_I)W_{(lm')}(\bar\theta_I,\bar\phi_I)-
F^+_I(\bar\theta_I,\bar\phi_I)X_{(lm')}(\bar\theta_I,\bar\phi_I)\right]
\\
&=-\left[D^l{}_{m0}(\chi_I,\zeta_I,0)\right]^*
\int_0^{2\pi}{\rm d}\bar\phi_I\int_{-1}^1{\rm d}\cos\bar\theta_I\>
\frac{N_l}{2}F^+_I(\bar\theta_I,\bar\phi_I)X_{(l0)}(\bar\theta_I,\bar\phi_I)
\\
&=0\,,
\label{e:calJ_finalresult}
\end{aligned}
\ee
where the last equality follows from 
$X_{l0}(\bar\theta_I,\bar\phi_I)=0$ by Eq.~(\ref{e:Xlm}). 
The significance of this result will be discussed in detail
in Sec.~\ref{sec:overlap_general}.

Returning now to Eq.~(\ref{eq:gamma_12,l}), and including the
factors of $(i2\pi f)^{-1}$, we have
\be
\begin{aligned}
\Gamma_{12,l}(f)
&= \sum_{m=-l}^l
\sum_P R^P_{1(lm)}(f)R^{P*}_{2(lm)}(f)
\\
&=
\frac{1}{(2\pi f)^2}
4\pi^2 (N_l)^2\sum_{m=-l}^l 
Y_{(lm)}(\zeta_1,\chi_1)Y_{(lm)}^*(\zeta_2,\chi_2) \label{e:gam12lalt}
\\
&=
\frac{1}{(2\pi f)^2}
(N_l)^2(2l+1)\pi P_l(\hat u_1\cdot\hat u_2)\,,
\end{aligned}
\ee
where we used the addition theorem for (ordinary)
spherical harmonics, Eq.~(\ref{e:addition_theorem_0Y}),
to eliminate the summation over $m$.
Note that this reproduces the result from the previous
subsection, Eq.~(\ref{e:overlap_l,statiso}),
with $\cos\zeta = \hat u_1\cdot\hat u_2$, where $\zeta$ is the
angle between the directions to the two pulsars.

\subsection{Recovery of the Hellings and Downs curve}
\label{s:recovery_HD}

For an isotropic, unpolarised and uncorrelated gravitational-wave background, 
we expect to recover the Hellings and Downs curve from the
frequency-independent part of the overlap reduction function
\be
\bar\Gamma_{12} \equiv
\sum_{l=2}^\infty C_l(N_l)^2(2l+1)\pi P_l(\hat u_1\cdot\hat u_2)\,.
\label{eq:LegendreExpansion}
\ee
For such a background, we saw in Sec.~\ref{s:statprop} that
\begin{equation} 
C^{GG}_l=C^{CC}_l=C_l=1\,,
\quad
C^{GC}_l=0=C_l^{CG}\,,
\label{eq:Cisotropic}
\end{equation}
for $l\ge 2$.
To show that these are indeed the coefficients that recover the
Hellings and Downs curve, we decompose the Hellings and Downs curve as
a superposition of Legendre polynomials
\begin{equation}
\frac{1}{2}\left(
1 + \frac{3}{2} \left(1-\cos\zeta\right)
\left[\ln\left(\frac{1-\cos\zeta}{2}\right) -\frac{1}{6}\right]\right)
= \sum_{l=0}^{\infty} a_l P_l(\cos\zeta)\,.
\end{equation}
The coefficients are given in the usual way
\begin{equation}
a_l = \frac{2l+1}{2} \int_{-1}^{1} {\rm d}x\>
\frac{1}{2}\left( 1 
+\frac{3}{2} \left(1-x\right)\left[\ln\left(\frac{1-x}{2}\right) 
-\frac{1}{6}\right] \right) P_l(x)\,,
\end{equation}
which follows from 
\begin{equation}
\int_{-1}^{1} {\rm d}x\> P_l(x) P_{l'}(x) = 
\frac{2}{2l+1}\delta_{ll'}\,.
\end{equation}
Considering $a_0$ and $a_1$ first:
\be
\begin{aligned}
4 a_0 
&=  \int_{-1}^{1} {\rm d}x\>
\left( 1 + \frac{3}{2} \left(1-x\right)
\left[\ln\left(\frac{1-x}{2}\right) -\frac{1}{6}\right] \right)
\\
&=  \left[x - \frac{3}{4} (1-x)^2 \ln\left(\frac{1-x}{2}\right) 
+ \frac{(1-x)^2}{2} \right]_{-1}^{1} = 1-(-1+2) =0\,,
\end{aligned}
\ee
and
\be
\begin{aligned}
\frac{4}{3}a_1 
&= \int_{-1}^{1} {\rm d}x\>
\left( x + \frac{3}{2} x \left(1-x\right)
\left[\ln\left(\frac{1-x}{2}\right) -\frac{1}{6}\right] \right)
\\
&= \left[ \frac{3}{8}x^2 + \left(\frac{1}{2}(1-x)^3 -\frac{3}{4}(1-x)^2 \right)
\ln\left(\frac{1-x}{2}\right) +\frac{3}{8}(1-x)^2 - 
\frac{(1-x)^3}{6} + \frac{x^3}{12}   \right]_{-1}^1 
\\
&= \left(\frac{3}{8}+\frac{1}{12}\right) 
- \left(\frac{3}{8}+\frac{3}{2}-\frac{8}{6}-\frac{1}{12}\right) = 0\,.
\end{aligned}
\ee
The vanishing of these coefficients is to be expected, as the
expansion in Eqs.~(\ref{e:habdecomp}) and (\ref{e:gamma(f)}) starts at $l=2$. 

For $l \geq 2$, since $1-(1-x)/4 =3P_0(x)/4+P_1(x)/4$,
orthogonality of the Legendre polynomials ensures
\begin{equation}
\frac{4}{2l+1}a_l 
= \int_{-1}^1 {\rm d}x\>
\frac{3}{2} (1-x) \ln\left(\frac{1-x}{2}\right) P_l(x)
= 6 \int_0^1 {\rm d}y\> y\ln(y) P_l(1-2y)\,,
\end{equation}
where we have made the substitution $y=(1-x)/2$.  The shifted Legendre
polynomials $\tilde{P}_l(y) \equiv P_l(1-2y)$ are given by the formula
\begin{equation}
\tilde{P}_l(y) = \frac{1}{l!} \frac{{\rm d}^l}{{\rm d}y^l} 
\left[(y-y^2)^l\right]\,,
\end{equation}
and therefore we obtain
\be
\begin{aligned}
\frac{4}{2l+1}a_l 
&= \frac{6}{l!} \int_0^1{\rm d}y\> 
y \ln(y) \frac{{\rm d}^l}{{\rm d}y^l} \left[(y-y^2)^l\right]
\\
&= \frac{6}{l!} \left[ y \ln(y)  \frac{{\rm d}^{l-1}}{{\rm d}y^{l-1}} 
\left[(y-y^2)^l\right] \right]_0^1 
- \frac{6}{l!} \int_0^1{\rm d}y\> 
\left(1+\ln(y)\right) \frac{{\rm d}^{l-1}}{{\rm d}y^{l-1}} \left[(y-y^2)^l\right]
\\
&= -\frac{6}{l!} \left[ (1+ \ln(y)) \frac{{\rm d}^{l-2}}{{\rm d}y^{l-2}} 
\left[(y-y^2)^l\right] \right]_0^1+\frac{6}{l!} 
\int_0^1 {\rm d}y\>
\frac{1}{y} \frac{{\rm d}^{l-2}}{{\rm d}y^{l-2}} \left[(y-y^2)^l\right]
\\
&= \frac{6}{l!} \int_0^1 {\rm d}y\>
\frac{1}{y^2} \frac{{\rm d}^{l-3}}{{\rm d}y^{l-3}} \left[(y-y^2)^l\right] 
= \cdots 
= \frac{6(l-2)!}{l!} \int_0^1 {\rm d}y\>\frac{(y-y^2)^l}{y^{l-1}}
=\frac{6(l-2)!}{l!} \int_0^1 {\rm d}y\>y (1-y)^l  
\\
&=\frac{6(l-2)!}{l!} \int_0^1 {\rm d}y\> (1-y)y^l
=\frac{6(l-2)!}{l!} \left[\frac{y^{l+1}}{l+1}-\frac{y^{l+2}}{l+2}\right]_{0}^1 
= \frac{6(l-2)!}{(l+2)!}
= 3 N_l^2\,,
\end{aligned}
\ee
where we have used the fact that, for $k < n$, ${\rm d}^k/{{\rm d}y^k}
\left[(y-y^2)^n\right]=0$ at $y=0,1$. 
Thus,
\begin{equation}
a_l = \frac{3}{4}(N_l)^2 (2l+1)\,.
\end{equation}
\end{widetext}
Comparing these coefficients with the
$C_l$ given in Eq.~(\ref{eq:LegendreExpansion}) for two pulsars 
separated by the angle $\zeta$, we see that the
$l$-dependence of the $a_l$ agrees with 
$C_l={\rm const}~(=3/4\pi)$ for $l\ge 2$, precisely as found above. 
We note that, since the correlation function for an 
isotropic background must depend only on the pulsar separation, we would 
expect to be able to expand it as a sum of Legendre polynomials. 
This property was exploited in a search for isotropic 
gravitational-wave backgrounds with non-general-relativistic polarisations 
in~\cite{lee2008}, where a Legendre polynomial expansion 
up to $l=5$ was used as a fit to numerically-computed 
correlation curves for longitudinal polarisation modes. 

We now explore the utility of this decomposition by generating and analysing a set of
pulsar TOAs.  The dataset was constructed using the {\bf GWbkrgd}
plugin within the pulsar-timing software package
\textsc{Tempo2}~\cite{tempo2-1,tempo2-2,tempo2-3}, 
which injects an isotropic,
unpolarised and uncorrelated gravitational-wave background 
into a set of specified
pulsar observations.  We employ an array of $10$ pulsars spread
uniformly over the sky, and observed fortnightly for $5$ years. The
injected background power-spectrum is flat, creating a correlated
white-noise influence in pulsars separated across the sky.

\begin{figure*}[htbp]
\begin{center}
\includegraphics[angle=0, width=0.6\textwidth]{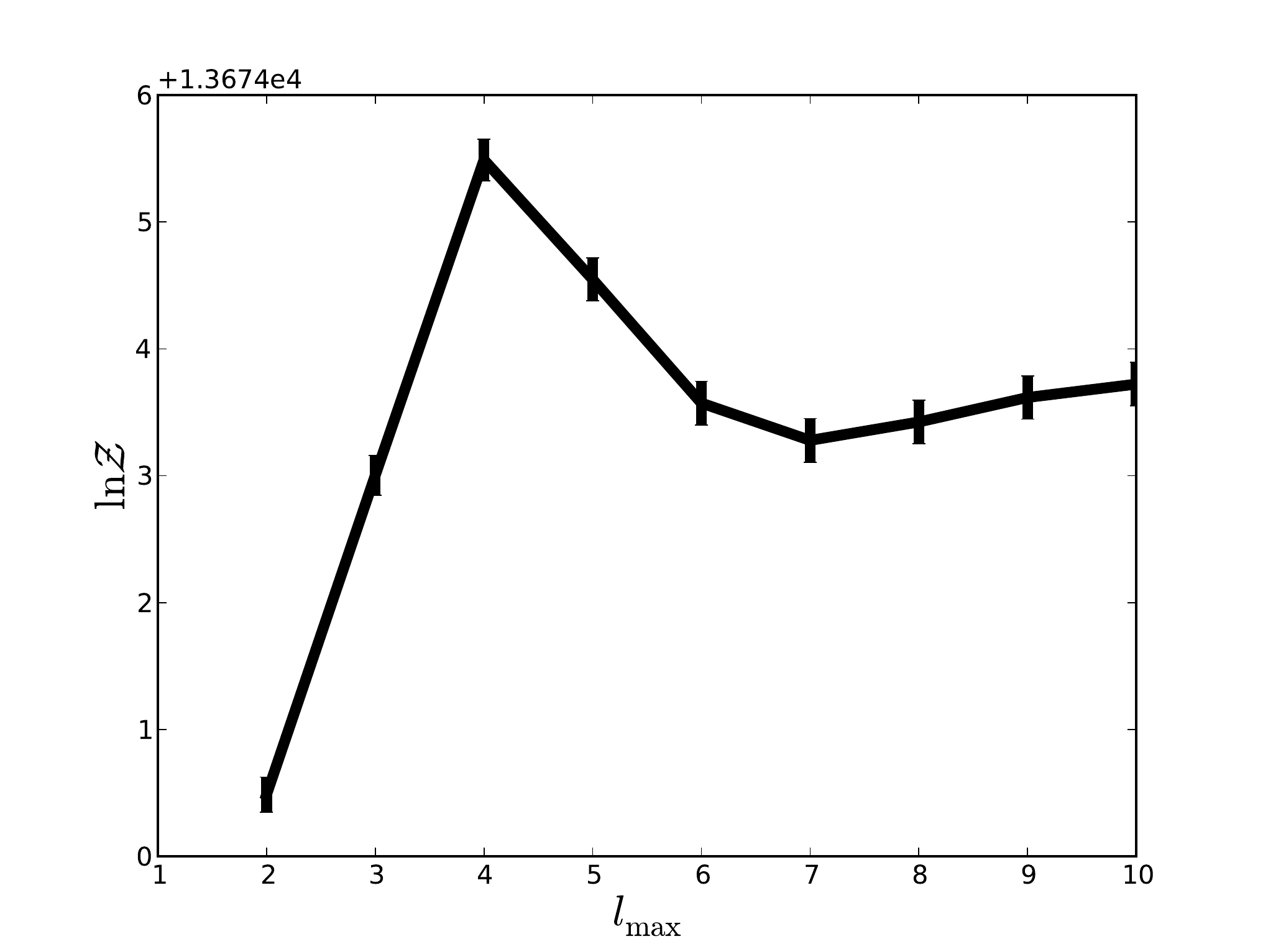}
\caption{Bayesian evidence for models of the overlap reduction 
  function truncated at varying $l_{\rm max}$.
  To produce this plot, 
  we injected an isotropic, unpolarised and uncorrelated gravitational-wave
  background with a white-noise spectrum into a set of
  realistic-format pulsar TOAs.  Testing truncated expansions of the
  form in Eq.~(\ref{eq:LegendreExpansion}), and recovering the
  Bayesian evidence, we find that an expansion up to and including
  $l=4$ is sufficient to recover the Hellings and Downs curve.}
\label{f:UncorrLegendreEvidence}
\end{center}
\end{figure*}

Using a Bayesian time-domain likelihood formalism
\cite{van-haasteren-2009,van-haasteren-levin-2012}, we test models of
the overlap reduction function which truncate the expansion in
Eq.~(\ref{eq:LegendreExpansion}) at varying $l_{\rm max}$.  Bayesian
statistics uses the volume under the prior-weighted likelihood (the
Bayesian {\it evidence}) as a model-comparison statistic. However,
this typically involves the evaluation of an expensive
multi-dimensional integral, which can sometimes be prohibitive for
costly likelihood computations or high-dimensional parameter spaces.
Fortunately, mature nested-sampling \cite{skilling2004} algorithms,
such as \textsc{MultiNest}
\cite{feroz2008,feroz2009,importanceMultiNest2013}, now exist to tackle
this problem, and as such, we employ \textsc{MultiNest} in all the following
parameter-estimation and evidence recovery calculations.

\begin{figure*}[htbp]
\begin{center}
\subfigure[]{\label{f:LegendreCorrCurve}\includegraphics[angle=0, width=0.49\textwidth]{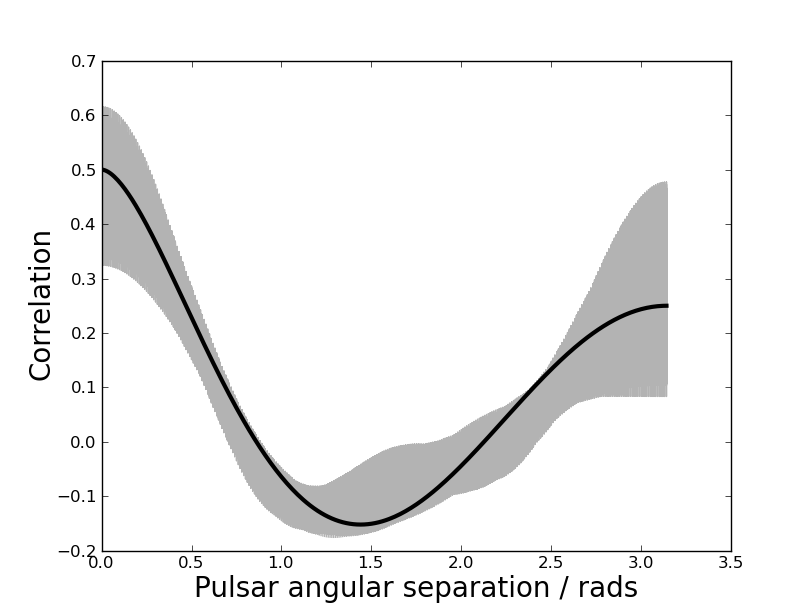}}
\subfigure[]{\label{f:LegendreExpansionCl}\includegraphics[angle=0, width=0.49\textwidth]{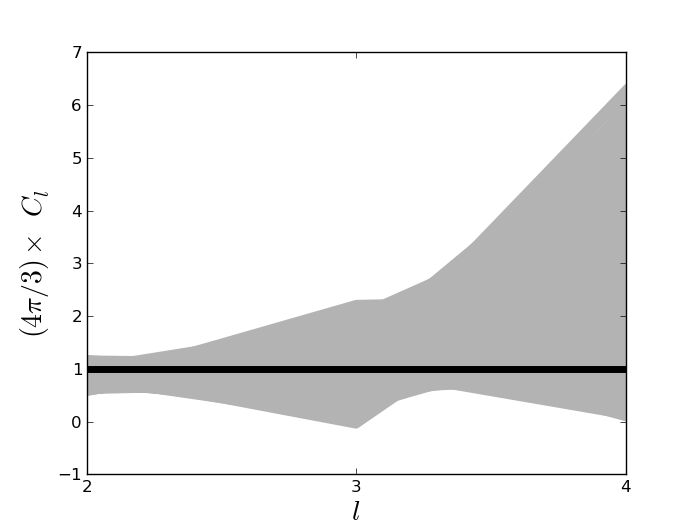}}
\caption{In (a) we explicitly show the envelope of overlap reduction
  functions corresponding to expansion-coefficients lying in the
  $95\%$ credible interval of the Bayesian analysis of an isotropic, unpolarised and
  uncorrelated gravitational-wave background. With an expansion up to and including
  $l=4$, our reconstruction is sufficiently consistent with the
  injected Hellings and Downs curve. In (b) we demonstrate that
  our numerical analysis is consistent with the analytic hypothesis
  that $C_l={\rm const},\;\forall\; l\geq 2$.}
\label{f:LegendreReconstruction}
\end{center}
\end{figure*}

Our results are shown in Fig.~\ref{f:UncorrLegendreEvidence}, where
we see that an expansion up to and including $l=4$ is sufficient to
recover the shape of the Hellings and Downs curve.  We can show this
explicitly by drawing the envelope of overlap reduction functions
corresponding to expansion-coefficients lying in the $95\%$ credible
interval of the Bayesian analysis. The result is shown in
Fig.\ \ref{f:LegendreCorrCurve}, along with the injected Hellings and
Downs curve. We also show in Fig.\ \ref{f:LegendreExpansionCl} that
the coefficients of the expansion are consistent with the analytic
result of $C_l={\rm const},\;\forall\; l\geq 2$.

\begin{figure*}[htbp]
\begin{center}
\subfigure[]{\includegraphics[angle=0, width=0.4\textwidth]{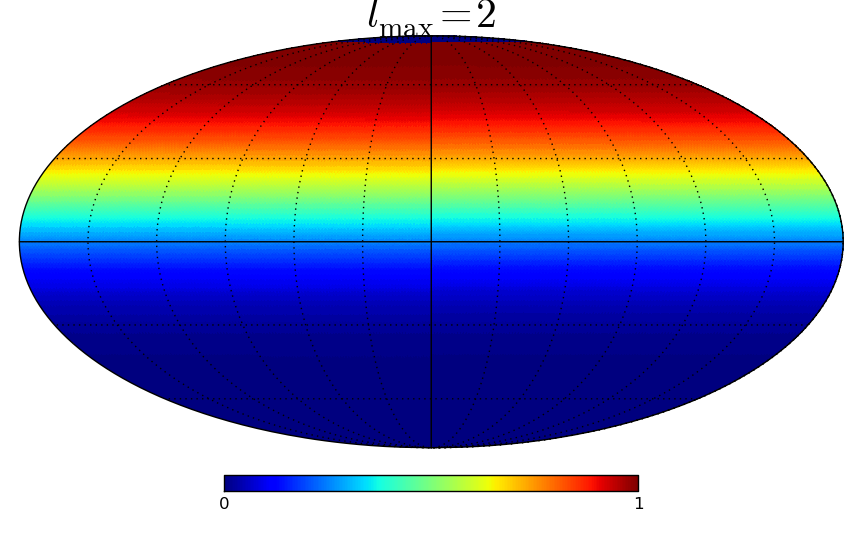}}
\hspace{.05\textwidth}
\subfigure[]{\includegraphics[angle=0, width=0.4\textwidth]{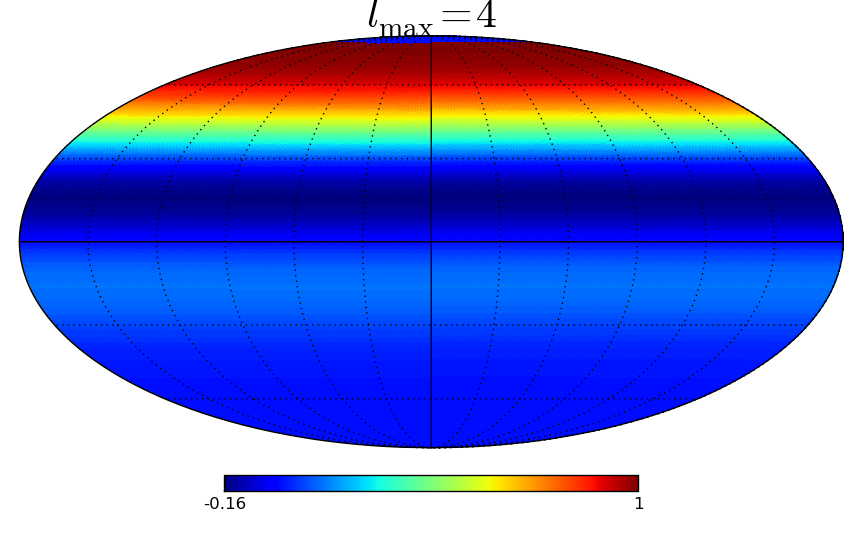}}\\
\subfigure[]{\includegraphics[angle=0, width=0.4\textwidth]{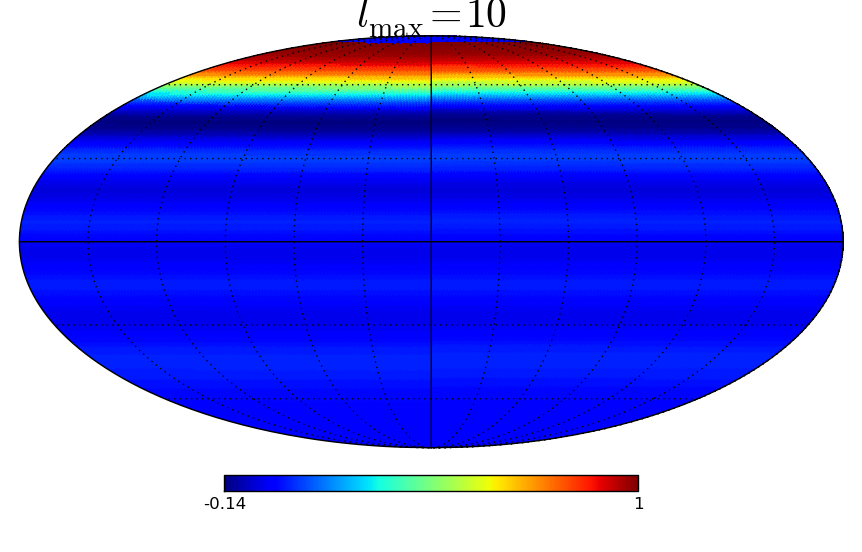}}
\hspace{.05\textwidth}
\subfigure[]{\includegraphics[angle=0, width=0.4\textwidth]{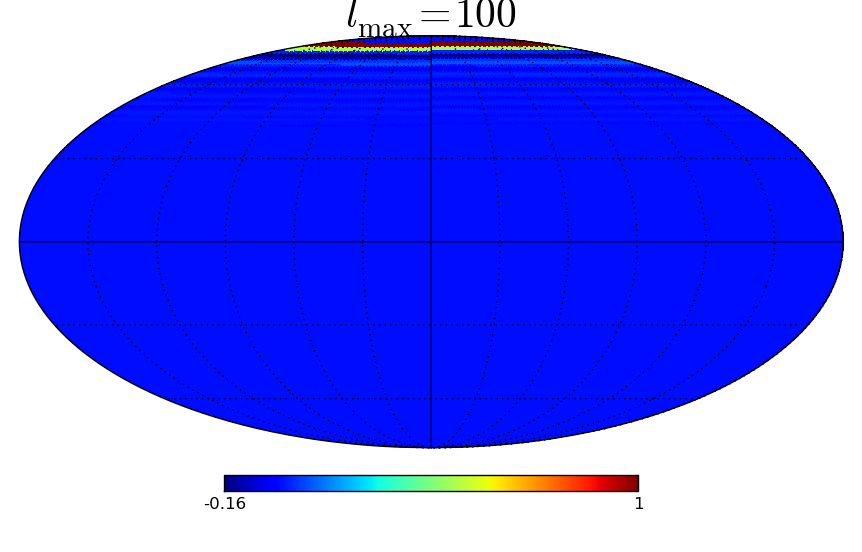}}
\caption{Theoretical correlation maps ($|\langle h_+(\hat{k})h_+^*(\hat{k}')\rangle_k|$) showing the degree to which
  Fourier modes along the $z$-axis and elsewhere on the sky are
  correlated. An uncorrelated gravitational-wave background should
  have a delta-function in sky-location for the quadratic expectation
  value of Fourier modes (see Eq.~(\ref{e:uncorrXV})). As we include
  higher multipoles in the expansion of
  Eq.~(\ref{eq:LegendreExpansion}), we converge toward this
  behaviour.}
\label{f:TheoreticalCorrMaps}
\end{center}
\end{figure*}

\begin{figure*}[htbp]
\begin{center}
\subfigure[]{\includegraphics[angle=0, width=0.4\textwidth]{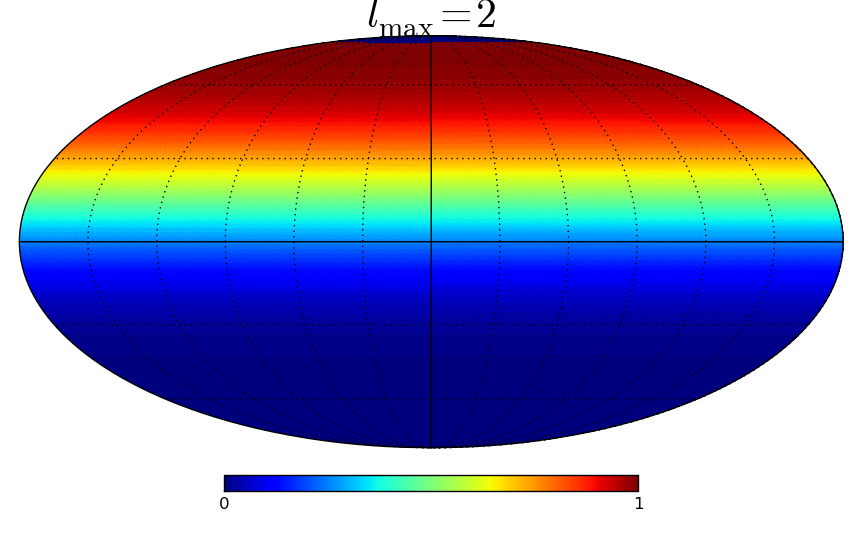}}
\hspace{.05\textwidth}
\subfigure[]{\includegraphics[angle=0, width=0.4\textwidth]{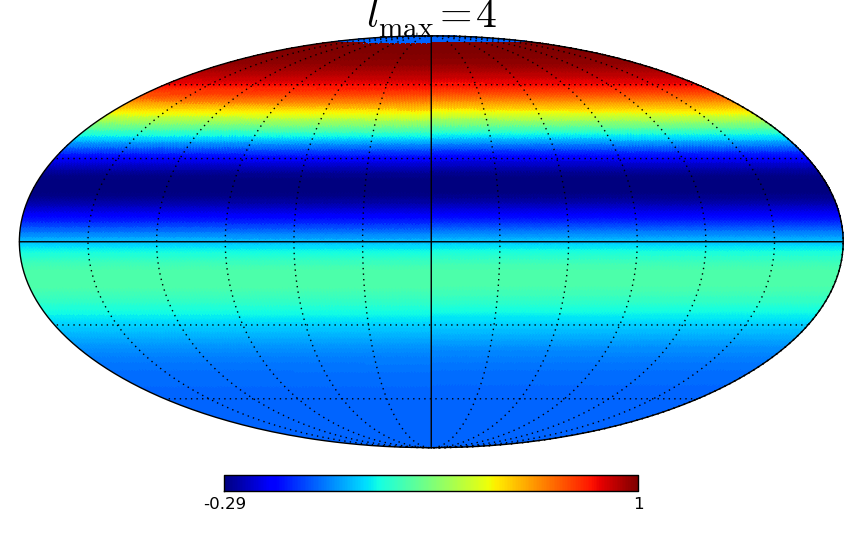}}
\caption{Correlation maps (displaying $|\langle h_+(\hat{k})h_+^*(\hat{k}')\rangle_k|$) constructed from the {\it
    maximum-a-posteriori} $C_l$ values of our Bayesian analysis of a
  pulsar TOA dataset containing an injected isotropic, unpolarised and 
  uncorrelated gravitational-wave background. 
  We see that the truncated expansion
  of Eq.~(\ref{eq:LegendreExpansion}) which gives the highest Bayesian
  evidence (corresponding to $l_{\rm max}=4$) adequately recovers the
  restricted sky-correlation that is characteristic of an
  uncorrelated gravitational-wave background.}
\label{f:RecoveredCorrMaps}
\end{center}
\end{figure*}

Furthermore, by expanding to higher multipoles we converge toward the
injected case of an uncorrelated background. From
Eq.\ (\ref{e:uncorrXV}) we see that, by definition, an uncorrelated
gravitational-wave background should have a delta-function in
sky-location for the quadratic expectation value of the Fourier
amplitudes. We show this explicitly in
Fig.\ \ref{f:TheoreticalCorrMaps}, where the correlation between
Fourier modes along the $z$-axis and elsewhere on the sky are
colour-coded to indicate strength of correlation. For $l_{\rm
  max}=100$ the correlation between the Fourier mode along the
$z$-axis with other sky-locations is almost negligible, effectively
giving the delta-function characteristic of an uncorrelated
background. From our analysis of the pulsar TOA dataset containing an
injected uncorrelated gravitational-wave background, we can produce {\it
  maximum-a-posteriori} correlation maps from the recovered $C_l$
values. In Fig.~\ref{f:RecoveredCorrMaps}, we see that the expansion
with highest Bayesian evidence (corresponding to $l_{\rm max}=4$)
adequately replicates the restricted sky-correlation of Fourier modes
that is characteristic of an uncorrelated background. 

Note that these
skymaps of the Fourier-mode correlation are not directly measurable in
a PTA analysis -- the maps in Fig.~\ref{f:TheoreticalCorrMaps} were
constructed from the theoretical expectation of the correlation of 
the grad/curl-expanded ``plus''/``cross'' GW amplitudes for an
isotropic background (see Eq.~(\ref{e:acorr}) and the discussion in 
Sec.~\ref{s:implications}). Likewise the maps in Fig.~\ref{f:RecoveredCorrMaps}
are not directly measurable, but are constructed by inserting the measured 
coefficients of the overlap reduction function expansion 
from Eq.~(\ref{eq:LegendreExpansion}) into the
expressions for the expectation of the Fourier-mode correlations.

\section{Correlation functions for arbitrary backgrounds}
\label{s:arbback}

\subsection{Characterising anisotropic backgrounds}
\label{sec:overlap_general}

The most general form of Eq.~(\ref{e:acorr}) that we can write 
down for the correlation between different modes of the background, 
while still assuming that different frequency components are 
uncorrelated and that the frequency and spatial dependence can be 
factorised is
\be
\langle a^{P}_{(lm)}(f) a^{P'*}_{(l'm')}(f') \rangle 
= C^{PP'}_{lml'm'}(f) \, \delta(f-f')\,,
\label{e:aniso_expectation_values}
\ee
where
\be
C^{PP'}_{lml'm'}(f) = C^{PP'}_{lml'm'} H(f)\,.
\ee
If the background is of cosmological origin we would impose 
additional restrictions on the correlation functions. 
It would be normal to impose statistical isotropy (i.e., 
no preferred direction on the sky) and possibly requiring 
parity-invariance or no polarisation. 
This will be discussed in more detail in Section~\ref{s:implications}. 
In the following we make no further restrictions on the correlation 
functions, to allow for a completely general background.

For expectation values of the form given in Eq.~(\ref{e:aniso_expectation_values}), 
the correlation between the output of the two detectors given in 
Eqs.~(\ref{e:r1r2corr})--(\ref{eq:gamma_12,l}) generalizes to
\begin{equation}
\langle r_1(t) r_2(t') \rangle
=\int_{-\infty}^\infty {\rm d}f\>
e^{i2\pi f(t-t')}H(f){\Gamma}_{12}(f)\,,
\end{equation}
where
\begin{widetext}
\begin{align}
{\Gamma}_{12}(f)
&=\sum_{(lm)}\sum_{(l'm')}\sum_P\sum_{P'}
C^{PP'}_{lml'm'}
R^P_{1(lm)}(f)
R^{P'*}_{2(l'm')}(f)\,,
\label{e:Gamma12_general2}
\end{align}
with $R^P_{I(lm)}(f)$ given by 
Eq.~(\ref{e:RPlm}).
For pulsar timing arrays, the above expression simplifies.
As we showed in Sec.~\ref{s:alternative_derivation},
\begin{align}
R^G_{I(lm)}(f)=
(i 2\pi f)^{-1}2\pi(-1)^l N_l Y_{(lm)}(\hat u_I)\,,
\qquad
R^C_{I(lm)}(f)=0\,.
\label{e:RG=Y,RC=0}
\end{align}
Thus,
\be
\Gamma_{12}(f) =
f^{-2}\sum_{(lm)}\sum_{(l'm')}
(-1)^{l+l'}N_l N_{l'} 
C^{GG}_{lml'm'}
Y_{(lm)}(\hat u_1) Y_{(l'm')}^*(\hat u_2)\,.
\label{e:overlap_final}
\ee
\end{widetext}

A couple of remarks are in order: 
(i) The overlap reduction function does not depend on
$C^{CC}_{lml'm'}$, $C^{GC}_{lml'm'}$, or $C^{CG}_{lml'm'}$.  This
means that a pulsar timing array is insensitive to the curl modes of
the radiation field, no matter how great the amplitude of the curl
modes may be. This is a surprising result but arises from the fact
that all pulsar lines of sight share a common end, at the SSB, 
which we have taken to be the origin. We can rotate
the axis of the coordinate system used to compute $R^P_{(lm)}(f)$ for any
given pulsar so that the pulsar is in the $z$-direction and then the
response has no dependence on the azimuthal coordinate. For curl
modes, in this frame $F^{\times}(\hat k)=0$, while $F^+(\hat k)$ is 
multiplied by
$X_{(lm)}(\hat k) \propto m\,e^{i m \phi}$. The integral over $\phi$ therefore
vanishes for any function that has no azimuthal dependence. A
ground-based interferometer such as LIGO that is static also has zero
response to curl modes (see Appendix~\ref{s:LIGO_response} for a
derivation of the response to both grad and curl modes). The reason
that we have no sensitivity to curl modes is that the metric
perturbation for these modes vanishes at the origin of
coordinates. This is somewhat analogous to a separation between even
and odd modes. If we have waves propagating in opposite directions in
one dimension, a measurement at the origin can only determine half of
the parameters characterising the waves since odd modes are always
zero there. If we have a pointable detector we can recover all the 
modes by pointing the detector first to the left and then to the right, 
and we can also do this by adding a measurement at a second point. 
(CMB temperature and polarisation measurements, for example, are
made with detectors that are sensitive to only a small fraction of the
sky.)  Gravitational-wave detectors are not pointable, 
but if we had a network of spatially-separated
detectors we should have sensitivity to these modes as the origin can
be taken to be at the location of one detector only. Mathematically,
shifting the SSB to the point $\vec{x}_0$ introduces an extra term
$\exp[-i 2 \pi f \hat{k}\cdot\vec{x}_0/c]$ in the response, which
breaks the azimuthal symmetry. Physically, although curl modes are
transformed to curl modes under rotations, a translation mixes curl
and grad modes, which leads to this non-zero response in the shifted
reference frame.

The symmetry is also broken by detector motion. Ground-based
interferometers move due to both the rotation of Earth and the
orbital motion of Earth around the Sun. Pulsar-timing arrays 
use radio telescopes based on Earth which also move. 
However, there are two fundamental
differences between PTAs and ground-based interferometers. First,
the frequency ranges are very different. The light-travel time from
Earth to the Sun is $\sim\!500~{\rm s}$ while the maximum frequency a PTA
is sensitive to is $\sim\!10^{-6}~{\rm Hz}$ (assuming observations
approximately every two weeks). The value of $f \hat{k} \cdot \vec{x}/c$
therefore varies by less than $0.0005$ for a PTA. For LIGO the maximum frequency is $\sim\!1~{\rm kHz}$ and over a year $f\hat{k} \cdot \vec{x}/c$ varies by $5\times10^5$ due to the motion of Earth around the Sun, and by an additional $\sim\!20$ on a daily basis due to Earth's rotation. Even for the lowest frequencies ($f\sim\!10~{\rm Hz}$) that LIGO can detect, the daily variation from Earth's orbital motion is $\gtrsim 10$. The second difference is in the nature of the data. For a PTA the raw data are the arrival times of pulses from given pulsars at the detector. This is typically (and
straightforwardly) converted into an arrival time at the SSB, which is
a fixed origin. For ground-based interferometers, the measurement is
of a path-length difference between two arms with a particular
instantaneous orientation, which cannot be readily converted into the
response of a fixed-orientation detector at some other point. We
conclude that PTAs really are static point detectors with no response
to curl modes, while ground-based interferometers will be able to
measure these modes in practice as a result of the motion of the
detectors. The energy density in the gravitational-wave background 
is given by the sum of the squares of the grad and curl amplitudes, 
so the curl component is a real physical part of the background 
to which PTAs are insensitive. (Note that including the pulsar term 
does not change this conclusion as it is also axisymmetric in the 
frame in which the pulsar position is along the $z$-axis.)

(ii) We can recover Eq.~(\ref{e:overlap,statiso}) 
for the overlap reduction function for a statistically 
isotropic background with
$C_l^{GG}=C_l^{CC}\equiv C_l$ and $C_l^{GC}=0=C_l^{CG}$
by simply setting 
$C^{GG}_{lml'm'} = C_l\,\delta_{ll'}\delta_{mm'}$.
The insensitivity of a PTA to curl modes is irrelevant 
when searching for such a background, since all the 
information about the correlation structure of the 
background is contained in $C^{GG}_l$.
(This will be discussed in more detail in Sec.~\ref{s:limited-sensitivity}.)
Making the above substitution for $C^{GG}_{lml'm'}$
into Eq.~(\ref{e:overlap_final}) 
and using the addition theorem for (ordinary) spherical harmonics, 
we obtain Eq.~(\ref{e:overlap,statiso}) with $\cos\zeta=\hat u_1\cdot\hat u_2$.

\begin{widetext}
\subsection{Representation of anisotropic uncorrelated backgrounds}

Equations~(\ref{e:isocorr1}) and (\ref{e:isocorr23}) 
give the correlation functions that have to be imposed on the 
$a^P_{(lm)}(f)$'s in order to recover an isotropic, unpolarised and 
uncorrelated background.  
If instead we want to represent an unpolarised, uncorrelated but 
{\em anisotropic} background, then we require
\be
\begin{aligned}
&\langle h_+(f,\hat{k}) h_+^*(f',\hat{k}') \rangle 
= \langle h_\times(f,\hat{k}) h_\times^*(f',\hat{k}') \rangle 
= \frac{1}{2}P(\hat{k})H(f)\delta^2(\hat{k},\hat{k}')\delta(f-f')\,,
\\
&\langle h_+(f,\hat{k}) h_\times^* (f',\hat{k}') \rangle 
= \langle h_\times(\hat{k}) h_+^* (\hat{k}') \rangle 
= 0\,,
\label{e:uncorranisoXV}
\end{aligned}
\ee
where $P(\hat{k})$ defines the anisotropic distribution of
gravitational-wave power on the sky.  
(We are assuming here that the spectral and angular 
dependence of the background factorise as $P(\hat k)H(f)$.)
Since an arbitrary scalar function
$P(\hat{k})$ can be written as a linear combination of spherical
harmonics, we consider a single term in this sum, $Y_{LM}(\hat{k})$.
Using Eq.~(\ref{e:aG+iaC}) to solve for $a^G_{(lm)}(f)$ and 
$a^C_{(lm)}(f)$, we find
\be
\begin{aligned}
\langle a^{G}_{(lm)}(f) a^{G*}_{(l'm')}(f') \rangle 
&= \langle a^{C}_{(lm)}(f) a^{C*}_{(l'm')}(f') \rangle 
\\
&=\frac{1}{2}\int {\rm d}^2\Omega_{\hat{k}}\> Y_{LM}(\hat{k}) 
\left( {}_{2}Y_{lm}^*(\hat{k})  {}_{2}Y_{l'm'}(\hat{k})  +  
{}_{-2}Y_{lm}^*(\hat{k}) {}_{-2}Y_{l'm'}(\hat{k}) \right)
H(f)\delta(f-f')\,,
\\
\langle a^{G}_{(lm)}(f) a^{C*}_{(l'm')}(f') \rangle 
&= -\langle a^{C}_{(lm)}(f) a^{G*}_{(l'm')}(f') \rangle
\\
&=\frac{i}{2}\int {\rm d}^2\Omega_{\hat{k}}\> Y_{LM}(\hat{k}) 
\left( {}_{2}Y_{lm}^*(\hat{k})  {}_{2}Y_{l'm'}(\hat{k})  -  
{}_{-2}Y_{lm}^*(\hat{k}) {}_{-2}Y_{l'm'}(\hat{k}) \right)
H(f)\delta(f-f')\,.
\label{e:anisocorr}
\end{aligned}
\ee
The integrals can be written in terms of Wigner 3-$j$ symbols 
(see for example~\cite{Wigner:1959}, \cite{Messiah:1962}):
\be
\begin{aligned}
&\langle a^{G}_{(lm)}(f) a^{G*}_{(l'm')}(f') \rangle 
= \langle a^{C}_{(lm)}(f) a^{C*}_{(l'm')}(f') \rangle 
\\
&\hspace{.5in}
=\frac{(-1)^m}{2}\sqrt{\frac{(2L+1)(2l+1)(2l'+1)}{4\pi}} 
\left( \begin{array}{ccc}L&l&l'\\M&-m&m' \end{array} \right) 
\left[\left( \begin{array}{ccc}L&l&l'\\0&2&-2 \end{array} \right) + 
\left( \begin{array}{ccc}L&l&l'\\0&-2&2 \end{array} \right) \right] 
H(f)\delta(f-f')\,,
\\
&\langle a^{G}_{(lm)}(f) a^{C*}_{(l'm')}(f') \rangle 
=-\langle a^{C}_{(lm)}(f) a^{G*}_{(l'm')}(f') \rangle
\\
&\hspace{.5in}
=\frac{i(-1)^m}{2}\sqrt{\frac{(2L+1)(2l+1)(2l'+1)}{4\pi}} 
\left( \begin{array}{ccc}L&l&l'\\M&-m&m' \end{array} \right) 
\left[\left( \begin{array}{ccc}L&l&l'\\0&2&-2 \end{array} \right) - 
\left( \begin{array}{ccc}L&l&l'\\0&-2&2 \end{array} \right) \right]
H(f)\delta(f-f')\,.
\end{aligned}
\ee
Explicit expressions for the Wigner 3-$j$ symbols are given in 
Appendix~\ref{s:spinweightedY}.

It is clear from the above that the representation of an anisotropic
background using this formalism is quite complicated. However, that 
is also the case for the standard approach~\cite{Mingarelli:2013,TaylorGair:2013}
to searching for an anisotropic uncorrelated background.
The spherical harmonic components of the frequency-independent
part of the overlap reduction function 
for such a background are given by \cite{Mingarelli:2013}:
\be
\bar\Gamma_{12,lm} =\int_{S^2}{\rm d}^2\Omega_{\hat k}\>
Y_{lm}(\hat k)\left[F_1^+(\hat k)F_2^+(\hat k)+
F_1^\times(\hat k)F_2^\times(\hat k)\right]\,,
\label{e:Gamma12_anisotropic_text}
\ee
where $l=0,1,2,\dots$.
These integrals can be evaluated using techniques similar to 
those used in Sec.~\ref{sec:overlap}.
This is described in detail in Appendix~\ref{app:UncorrAniso}. 
There we derive analytic expressions for $\bar\Gamma_{12,lm}$
for {\em all} values of $l$ and $m$, extending the analytical results of 
\cite{Mingarelli:2013}.
Figure~\ref{f:gammaLM} shows plots of $\bar\Gamma_{12,lm}$ for
$l=0,1,\cdots,5$, calculated using formulas from 
Appendix~\ref{app:UncorrAniso}.
The plots for $l=0,1,2,3$ are identical to those in Figure~2 of 
\cite{Mingarelli:2013}, as expected.
\begin{figure*}[htbp]
\begin{center}
\subfigure[]{\includegraphics[width=.49\textwidth]{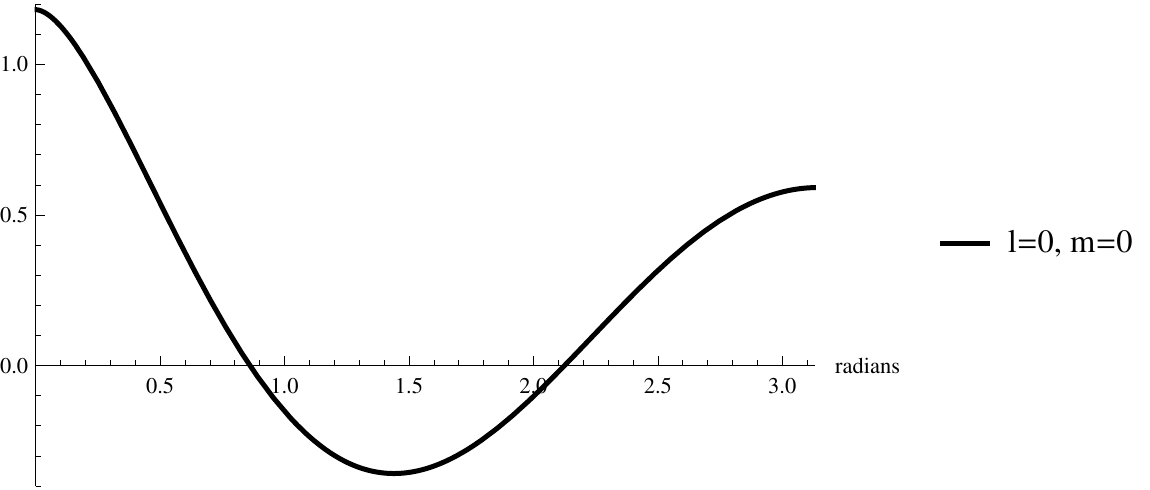}}
\subfigure[]{\includegraphics[width=.49\textwidth]{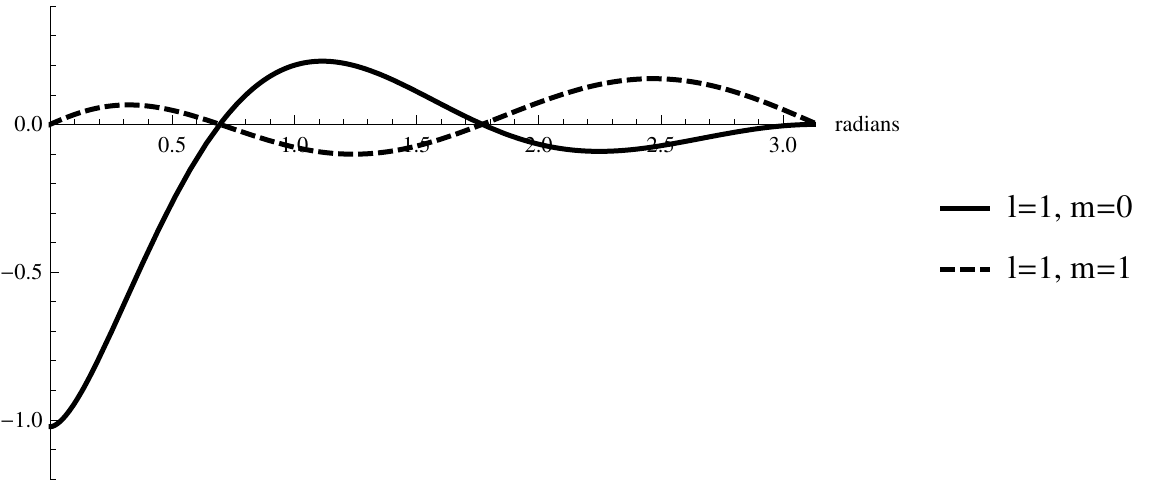}}
\subfigure[]{\includegraphics[width=.49\textwidth]{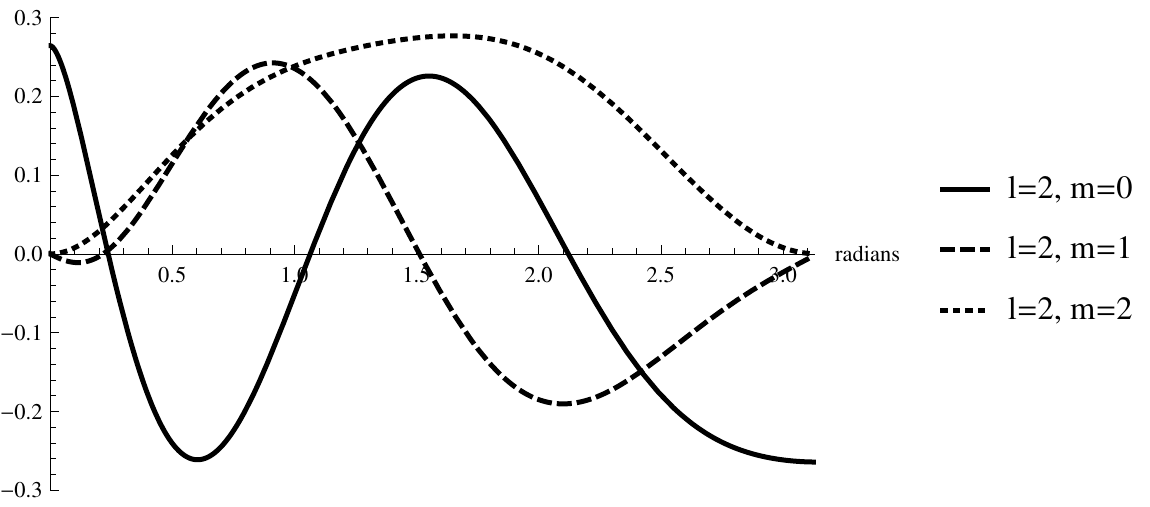}}
\subfigure[]{\includegraphics[width=.49\textwidth]{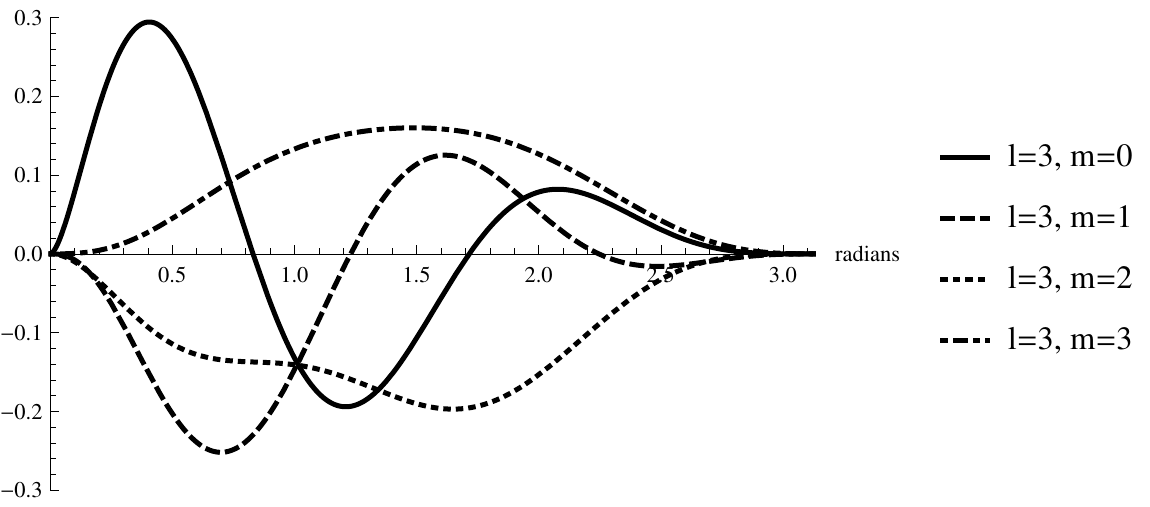}}
\subfigure[]{\includegraphics[width=.49\textwidth]{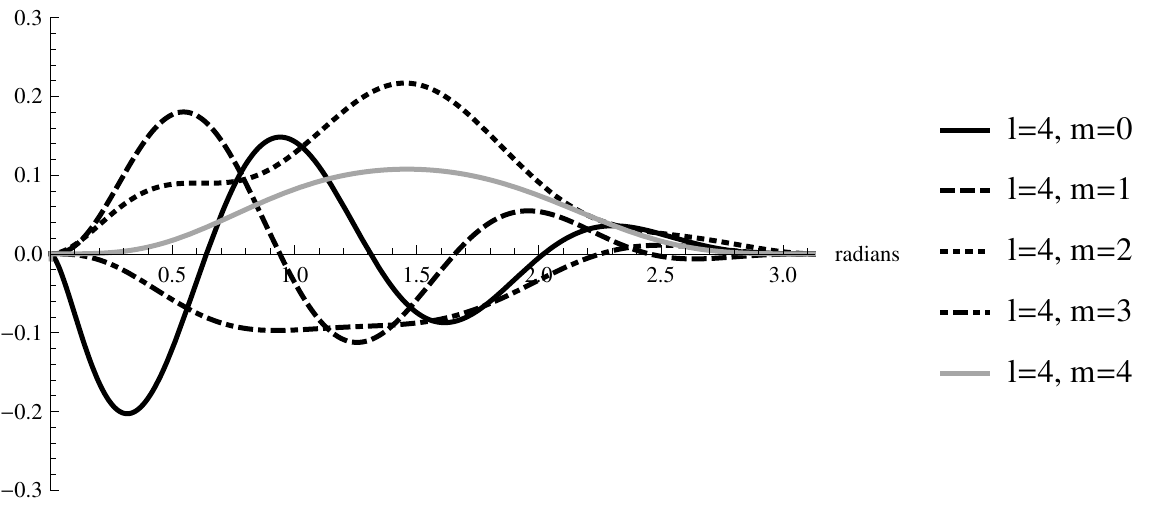}}
\subfigure[]{\includegraphics[width=.49\textwidth]{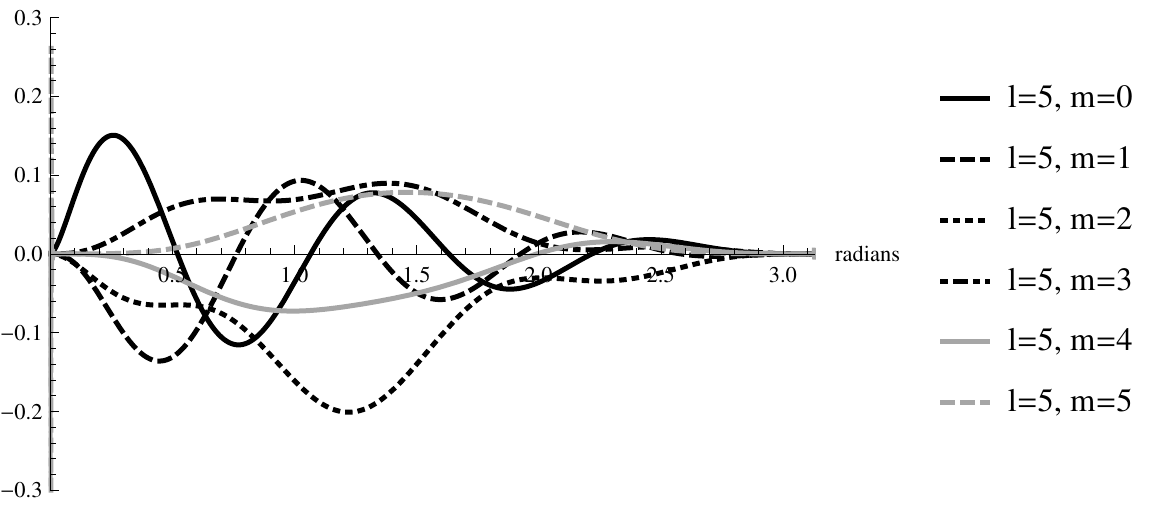}}
\caption{Plots of $\bar\Gamma_{12,lm}$ for $l=0,1,\cdots,5$ as a function of
  the angle between the two pulsars for an anisotropic, unpolarised
  and uncorrelated background.}
\label{f:gammaLM}
\end{center}
\end{figure*}
We will not consider
anisotropic uncorrelated backgrounds further in this paper, but refer
the reader to~\cite{Mingarelli:2013,TaylorGair:2013} where these are
considered in depth.

\end{widetext}

\section{Mapping the gravitational-wave background}
\label{s:mapping}

In the preceding sections we showed how this formalism can be used to
describe two specific models for the gravitational-wave 
background. However, the advantage of this approach is that it allows
us to characterise the background and produce a map of it, without
making any assumptions about its underlying nature. As described
below, we will be able to produce a map of both the amplitude and the
phase of the gravitational-wave background as a function of position
on the sky. This map will have a resolution limited by the number of
pulsars in the array and in general we will measure the small-$l$
(large angular scale) components more accurately, so we expect the
resolution of the map to increase as more pulsars are added. This map
encodes all the information that it is possible to extract from our data
about the gravitational-wave background. The power distribution on the
sky will identify hot spots where there is enhanced gravitational-wave
emission, which might correspond to nearby individual sources or even
clusters containing multiple sources. In addition, the variation of
the phase across the sky will indicate whether there is any
correlation between the emission at different sky locations. The
amount of information that can be extracted will of course depend on
the resolution of the map that we can measure. The expectation is that
the background will be isotropic, unpolarised and uncorrelated, and we saw above
that we would need to measure components up to $l_{\rm max}=4$ to recover
the Hellings and Downs curve characteristic of that case. Our ability
to identify departures from this model will depend on the size of
those departures. If the departures are large they will become
apparent much more quickly, perhaps even $l_{\rm max}=3$ would be
sufficient, but for small departures we would require many more
components. To identify hot spots we would need an angular resolution
comparable to the size of the hot spot. Taking the latter to be the
typical resolution of an individual source on the sky can be used to
set requirements on $l_{\rm max}$ and hence the number of pulsars. We
will discuss all of these considerations in this section.

\subsection{Likelihood function for the $a^G_{(lm)}$'s}

By measuring the coefficients $a^P_{(lm)}(f)$ using a pulsar timing
array we can reconstruct a map of the gravitational-wave sky, or at
least that part of it, spanned by the gradient modes $a^G_{(lm)}(f)$, 
which is visible to a PTA. A typical PTA will consist of measured 
timing residuals $\delta t_I(t)$ for a set of $N$ pulsars, 
labelled by $I$. The timing
residuals will be a combination of a gravitational-wave signal component and 
a noise component 
\be
\delta t_I(t) = r_I(t) + n_I(t)\,, 
\quad 
\tilde{\delta t}_I(f) = \tilde r_I(f) + \tilde n_I(f)\,, 
\ee
where $I=1,2,\cdots, N$ and tilde denotes Fourier transform. 
Usually the stochastic
background is taken to be part of the noise, but in the approach
described here we are using the decomposition into grad and curl
modes to construct a template for the gravitational-wave
background; therefore it is the signal component of the timing
residuals. In the Fourier domain we have (cf.~Eqs.~(\ref{e:habRP}) and
(\ref{e:RPlm})):
\begin{align}
\tilde r_I(f)
\equiv \tilde r_I(f;\vec a)
=\sum_{(lm)}
R^G_{I(lm)}(f)
a^G_{(lm)}(f) \, ,
\end{align}
where $\vec a\equiv \{a^{G}_{(lm)}(f)\}$ is shorthand notation for the
collection of parameters describing the gravitational-wave background. 
Assuming stationary and Gaussian noise, but allowing for possible
correlated noise between different pulsars, we have
\be
\langle \tilde n_I(f)\tilde n^*_J(f')\rangle 
= C_{n\,IJ}(f)\delta(f-f')\,,
\ee
and the likelihood of measuring the data $\tilde{\delta t}_I(f)$ is then
\begin{widetext}
\be
p(\delta t|C_n,\vec a) 
\propto
\exp\left\{-\frac{1}{2} \int_{-\infty}^\infty {\rm d}f\>
\left[ (\tilde\delta t_I^*(f)-\tilde r_I^*(f;\vec a)) C^{-1}_{n\,IJ}(f)
(\tilde\delta t_J(f)-\tilde r_J(f;\vec a)) \right] \right\}\,,
\label{e:likefdom}
\ee
\end{widetext}
where summation over the pulsar indices $I, J$ is assumed. The values
of the coefficients $a^G_{(lm)}(f)$ can then be recovered either by
using Bayesian inference to construct their posterior distributions
given prior probability distributions for the signal and noise
parameters, or by computing the frequentist maximum-likelihood
estimators of the parameters, which we will illustrate below.

As written above there are an uncountable number of model parameters,
since we not only have a sum over different $(lm)$ modes of the
background, but we are allowing the coefficients $a^G_{(lm)}(f)$ to be
arbitrary functions of frequency. To make progress there are several
possible approaches: (i) we assume that the $a^G_{(lm)}(f)$ are independent
of frequency, for instance by filtering the data so that we consider 
only a narrow range of frequencies for which this would be a reasonable
approximation; (ii) we assume that the frequency and spatial
dependence of the coefficients factorise, $a^G_{(lm)}(f) =
\sqrt{H(f)}\,a^G_{(lm)}$; (iii) we assume the background consists of a
finite number of components only, with frequencies $f_j$. In approach
(iii) the included frequency components could be the entire set of
discrete frequencies to which we are sensitive with a finite data set,
or a smaller number of components. It has been shown in the context of
the International Pulsar Timing Array (IPTA) 
mock data challenge that an isotropic stochastic background
can be well represented as a superposition of a small number of
components in this way~\cite{LentatiHyperEff}. All three of these
approaches are mathematically equivalent in the sense that they reduce
the number of $a^G_{(lm)}(f)$ coefficients to a finite value 
for a given $(lm)$,
but we will use approach (iii) to illustrate how to obtain the 
maximum-likelihood estimators of the $a^G_{(lm)}(f)$'s in the following.

If we make this assumption and also assume the data set is finite so
that the integral over frequency in Eq.~(\ref{e:likefdom}) reduces to
a sum, then we can introduce a new signal vector which combines the
response of all the pulsars 
\be
\vec{\delta t} \equiv \{\delta t_i\}^T \equiv
\{ \tilde{\delta t}_1(f_1),\tilde{\delta t}_1(f_2), \cdots, 
\tilde{\delta t}_N(f_{n})\}^T\,,
\ee
where $i=1,2,\cdots, nN$ labels a component of $\vec{\delta t}$,
and $j=1,2,\cdots, n$ labels a frequency component $f_j$ 
to which we are sensitive.
The (symmetric) $nN\times nN$ correlation matrix for the whole 
signal then takes the form 
\be 
F_{ii'} =C_{n\,II'}(f_j)\,\delta_{jj'}\,,
\ee
where $j = {\rm mod}(i,n)$ labels a discrete frequency 
and $I = \lceil i/n\rceil$ labels a pulsar.
(Here $\lceil X \rceil$ denotes the smallest integer greater than $X$.)
Similarly, the response of the pulsar timing array to the 
background can be written as $H\vec a$, where
\be
H_{ik} = R^G_{I(lm)_k}(f_{j_k})\,\delta_{jj_k}
\ee
and
\be
\vec a\equiv \{a_k\}^T \equiv\{a^G_{(lm)_k}(f_{j_k})\}^T\,.
\ee
Here $k$ labels a particular component of the signal parameter 
vector $\vec a$, which is specified by spherical harmonic indices
$(lm)_k$ and discrete frequency index ${j_k}$.
Note that each column of $H$ corresponds to a single component of $\vec a$, 
and describes the frequency-domain response of the PTA network to
the corresponding mode of the background. Using this notation the likelihood takes
the general form
\be p(\delta t|F,\vec a) \propto
\exp\left[-\frac{1}{2} \left( \vec{\delta t} - H \vec{a} \right)^\dag F^{-1} \left(
  \vec{\delta t} - H \vec{a} \right) \right]\,.
\label{e:discretelike}
\ee 

The reason for introducing this notation is that, in practice, PTA
data are measured in the time-domain and is not evenly sampled, which
makes analysing the data in the frequency-domain as described above
quite challenging. In addition, a timing model is fitted out for each
pulsar in the array to account for various astrophysical effects
including the proper motion of the pulsar, its spin-down and possible
binary parameters~\cite{LorimerKramer:2012}. This timing-model fit can also affect the
gravitational-wave signal in the data. Accounting for this by
introducing timing-model parameters into the likelihood and then
marginalising over them results in a modified likelihood that takes
exactly the same form as Eq.~(\ref{e:discretelike}). The measured data
$\vec{\delta t}$ are again the concatenation of the measured residuals from
the pulsars in the array, the model matrix, $H$, now contains the time-domain 
response of the PTA to the modes of the background, and the 
correlation matrix takes the form $F^{-1} = G \left(G^T C_n
G\right)^{-1} G^T$, where $C_n$ is the time-domain correlation matrix
for the noise, and $G$ is constructed from the left null-space of the
timing-model design matrix, where it effectively projects the
timing-residuals into a space orthogonal to the timing-model
\citep{van-haasteren-levin-2012}. 

\subsection{Maximum-likelihood estimates of the gravitational-wave sky}
\label{s:MLestimates}

Starting from a likelihood of the form given in Eq.~(\ref{e:discretelike}), 
it is relatively simple to compute the frequentist maximum-likelihood
estimators of the signal parameters since they enter the model
linearly. We find
\be
\vec{a}\big|_{\rm ML} = \left( H^\dag F^{-1} H\right)^{-1} H^\dag F^{-1} 
\vec{\delta t}\,,
\ee
assuming that the vector $\vec{\delta t}$ and matrix $F$ are both real. Once
these maximum-likelihood estimators or posterior distributions for the
$\vec{a}$ parameters have been obtained, a sky map of the
gravitational-wave background at each frequency $f_j$ can be
constructed using Eq.~(\ref{e:h+x}) and, in the Bayesian case,
integrating over the posterior distribution of $\vec{a}$. The matrix
$H^\dag F^{-1} H$ that enters the above expression is the {\em Fisher
  information matrix} and we have assumed it is invertible, but this
will not necessarily be the case. Although we have reduced the number
of $a^G_{(lm)}(f)$ coefficients by assuming a finite number of
frequency components, there are still an infinite number of
coefficients as we vary over $l$ and $m$. In practice we won't be able
to measure all of the coefficients as we have only a finite amount of
noisy data and, most importantly, a finite number of pulsars. Each
pulsar will allow us to measure the amplitude of a sine and a cosine
quadrature at a particular frequency. As the array is static we would
therefore expect to be able to measure only $2N$ real components of the
background at any given frequency.

That this is the number of modes that can be resolved can be
understood mathematically by considering the structure of the matrix
$H$. Up until now we have allowed $H$ and $\vec{a}$ to be arbitrary
complex quantities, but the measured data in the time-domain must be
real. As we saw in Eq.~(\ref{e:acoeffsymmetry}), the 
negative-frequency components of $a^G_{(lm)}(f)$ must satisfy 
the constraint
\be
a^G_{(lm)}(-f) = (-1)^m \, a^{G*}_{(l,-m)}(f)\,.
\ee
In addition,
Eqs.~(\ref{e:RG=Y,RC=0}) and (\ref{e:conj(Y)}) imply
that $R^G_{(lm)}(-f) = (-1)^m R^{G*}_{(l, -m)}(f)$ and therefore this
constraint is sufficient to make $H\vec{a}$ a real time-series. 
Including only positive-frequency components in both $H$ and
$\vec{a}$, the likelihood takes the same form, but with $H\vec{a}$
replaced by $(H\vec{a} + H^* \vec{a}^*)$. Alternatively, we can
replace $\vec{a}$ by a real vector with twice as many components, with
alternate entries being the real and imaginary parts of the complex
$a^G_{(lm)}(f)$ components, and similarly double the number of columns of
$H$, making it into a real matrix.

If we suppose that we have included $N_m$ modes in $\vec a$, 
and consider a single frequency component $f_j$ for simplicity, then,
in the time domain, the matrix $H$ has the form 
\be
H_{ik} = \sum_{I=1}^N x_{Ii} R^G_{I(lm)_k}(f_j)\,,
\ee
in which $\{x_{Ii}\}$ is the (complex) time series of pulsar $I$ 
for the given frequency component $f_j$ (i.e., $\exp[i2 \pi f_j t]$ 
evaluated at the appropriate discrete times for that pulsar) and 
has zeros in all components that  correspond to samples from the other pulsars.
(Note that for the time-domain analysis, the index 
$i=1,2,\cdots, M$ now labels the times for all the pulsars 
in the array, and $k$, as before, labels a 
particular component of the signal parameter vector $\vec a$, 
having spherical harmonic indices $(lm)_k$ and 
frequency $f_{j_k}=f_j$ for this particular case.)
It is clear that the non-trivial left eigenspace of $H$ (the
range of $H$) is spanned by the $N$ complex vectors $\vec x_I\equiv\{ x_{Ii} \}$
(equivalently, the columns of $H$ are linear combinations of these $N$
complex vectors) and that there are therefore $N_m-N$ complex left
null eigenvectors. If we do a singular-value decomposition
\be H= U \Sigma V^{\dag}\,,\label{e:svd}
\ee
then the $M\times N_m$ rectangular
matrix $\Sigma$ will have at most $N$ non-zero elements 
on its diagonal, $\sigma_i\equiv \Sigma_{ii}$. 
(For this form of singular-value decomposition, 
$U$ and $V$ are unitary matrices having 
dimensions $M\times M$ and $N_m\times N_m$, respectively.)
This means that we will have sensitivity to $N$ complex combinations 
of the $\vec a$ coefficients, and that there will be an
additional $N_m-N$ combinations which we cannot detect. In the real
representation, $H$ has twice as many columns and there will be twice
as many non-zero singular values, essentially corresponding to the
real and imaginary parts of the columns of $U$. For a given column
$\vec{v}_i$ of $V$ and $\vec{u}_i$ of $U$, we have $H\vec{v}_i =
\sigma_i \vec{u}_i$. The vectors $\vec{v}_i$ and $-i \vec{v}_i$ give
the two corresponding singular vectors of $(H\vec{a} + H^*
\vec{a}^*)$, which map onto real vectors that are the real and
imaginary parts of the complex vector $\sigma_i\vec{u}_i$.

We can split the columns of the matrix $U$ as 
$U=[H_{\rm range}\,H_{\rm null}]$,
where ${H}_{\rm range}$ represents the first $N$ columns of $U$ which span
the range of $H$. We must then replace $H \vec{a}$ by $H_{\rm range} \vec{b}$ in
the likelihood, Eq.~(\ref{e:discretelike}), and can proceed as
before. Once the maximum-likelihood estimator of $\vec{b}$ is obtained
we get the maximum-likelihood value of $\vec{a}$, projected into the
space to which we have sensitivity, by computing $\vec{a} = V
\Sigma^{+} \vec{b}$, where $\Sigma^+$ is the pseudo-inverse of
$\Sigma$, obtained by replacing the non-zero elements on the diagonal
of $\Sigma$ by their reciprocals and taking the transpose of the 
resulting matrix.

As this work was nearing completion, we became aware of an ongoing
independent study by Cornish and van~Haasteren (private communication,
this work has subsequently appeared
as~\cite{Cornish-vanHaasteren:2014}) that is also concerned with PTA
mapping of the gravitational-wave background. Their approach is based
on a real-space description of the background using pixelised sky
maps, and they have found that an $N$-pulsar array is sensitive to
2$N$ different ``sky map basis'' elements.  These correspond to the
$N$ complex combinations of $a^G_{(lm)}(f)$ modes to which we find a
PTA is sensitive. The Cornish and van~Hassteren work will provide a
more detailed study of sky mapping, demonstrating the practical
application of ideas similar to those discussed in this section, while
the theoretical results described elsewhere in this paper provide a
way to interpret such results and an understanding of the physical
nature of the individual sky map basis elements. In addition, our
results indicate that there is a portion of the gravitational-wave sky
that a pulsar timing array will never be able to see. At fixed
resolution (which means a fixed pixel size in a real-space
representation or a fixed number of $l$ modes in our representation)
you would expect to be able to measure all of the background
components once you had more pulsars in the array than components in
the finite resolution representation.  However, this is not the
case. While it is possible to measure the whole grad component of the
sky at a fixed resolution once sufficiently many pulsars are included
in the array, the array is always blind to the curl component.

Increasing the number of frequency components does not provide
sensitivity to additional components of the background. When adding
another frequency, we can simply add another set of $a^G_{(lm)}$'s 
at the new frequency, in which case we will be able to determine 
$N$ of these
coefficients for each frequency component. Alternatively, we can assume
that the new $a^G_{(lm)}$'s are equal to the first set 
of $a^G_{(lm)}$'s, or equal to a frequency-dependent multiple of  
the first set.  In that case we still cannot
measure any additional combinations of $a^G_{(lm)}$'s,
but the additional frequency components 
will allow us to measure the same combinations of
coefficients with greater precision.

\subsection{Example recovery of a simulated background}
\label{s:example_recovery}

\begin{figure*}[htbp]
\begin{center}
\subfigure[\ Total map (grad+curl)]
{\includegraphics[trim=3cm 6.5cm 3cm 3.5cm, clip=true, angle=0, width=0.32\textwidth]{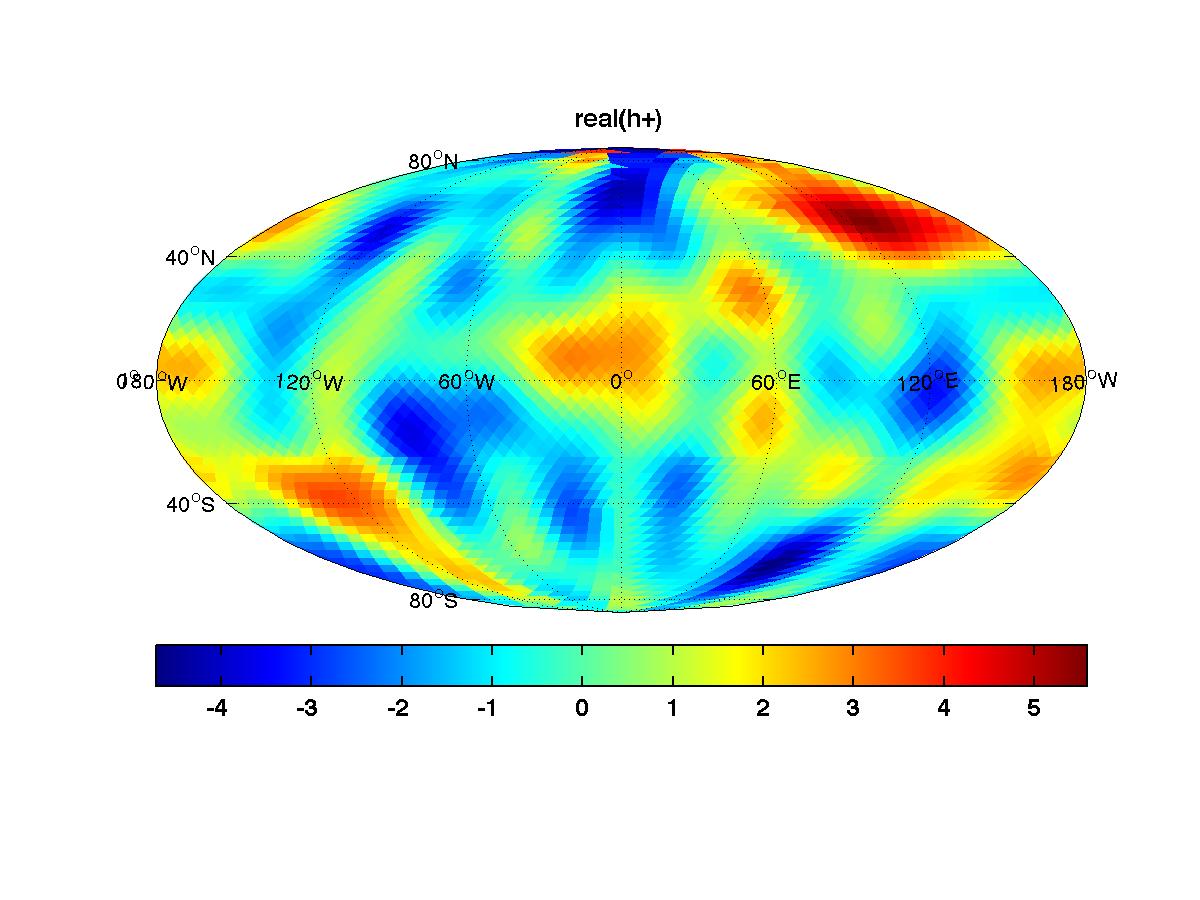}}
\subfigure[\ Gradient component]
{\includegraphics[trim=3cm 6.5cm 3cm 3.5cm, clip=true, angle=0, width=0.32\textwidth]{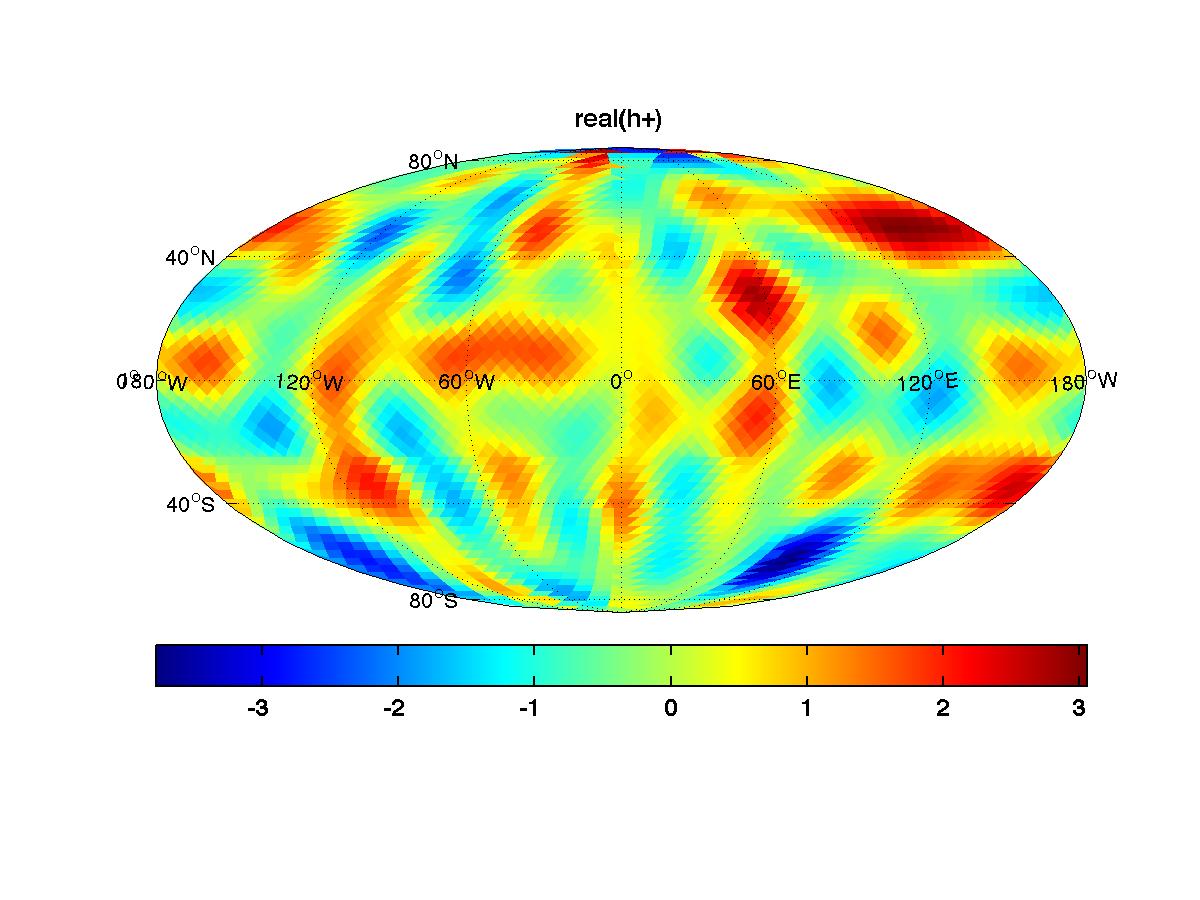}}
\subfigure[\ Curl component]
{\includegraphics[trim=3cm 6.5cm 3cm 5cm, clip=true, angle=0, width=0.32\textwidth]{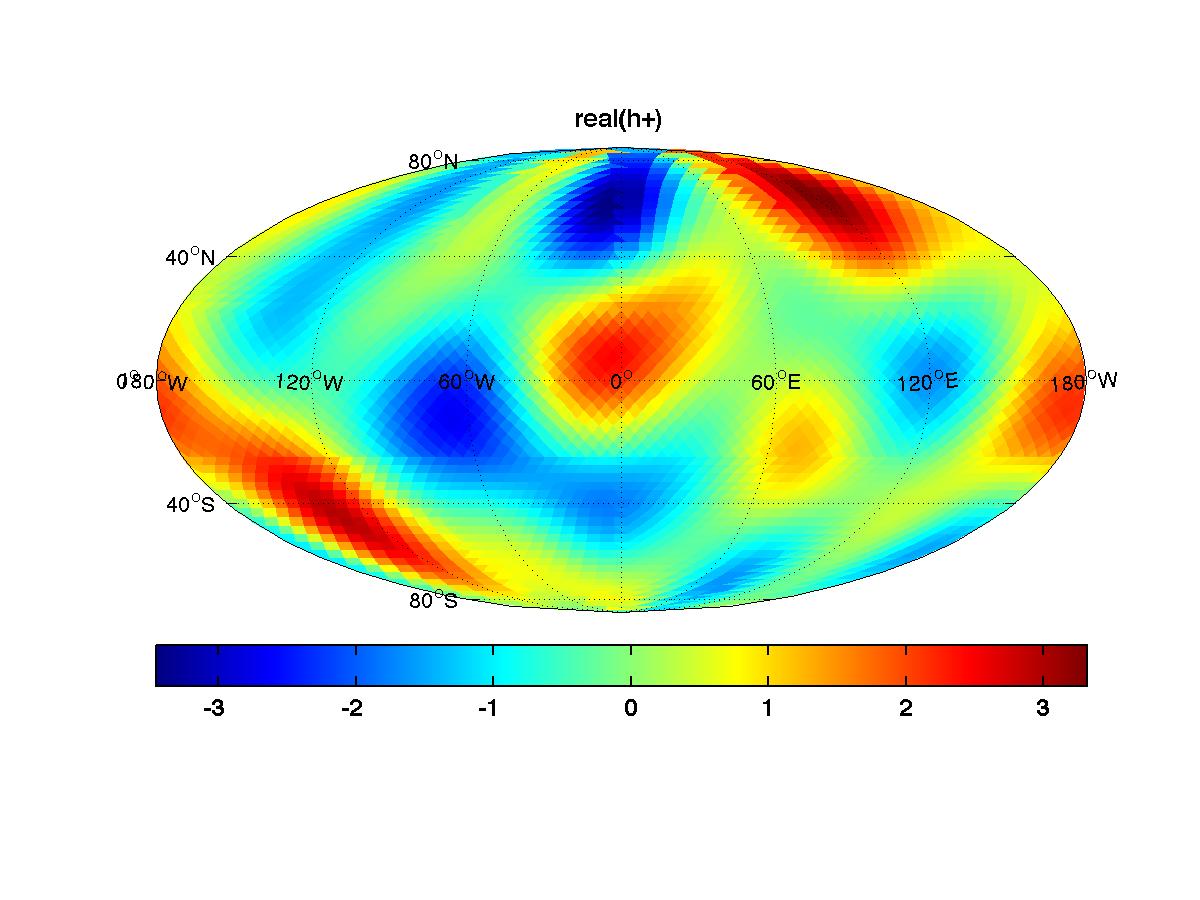}}
\caption{Mollweide projections of ${\rm Re}(h_+)$ for the simulated 
gravitational-wave background.
Panel (a) shows the total simulated background (grad+curl components);
panel (b) shows the gradient component; and
panel (c) shows the curl component.
Sky maps of ${\rm Im}(h_+)$, ${\rm Re}(h_\times)$, and ${\rm Im}(h_\times)$
are similar.}
\label{f:demo_injectedskymaps}
\end{center}
\end{figure*}

We now illustrate the mapping procedure described above by
constructing maximum-likelihood sky maps of $h_+(f,\hat k)$ and
$h_\times(f,\hat k)$ for a simulated gravitational-wave background.
For simplicity, we consider only a single frequency component and a
noise-free simulation. 
(So we will drop the $f$ from $h_{+,\times}(f,\hat k)$ and 
simply write $h_{+,\times}(\hat k)$ in what follows.)
More realistic simulations will be presented
in a detailed follow-on paper to
Ref.~\cite{Cornish-vanHaasteren:2014}.  Mollweide projections of the
real part of $h_+(\hat k)$ for the simulated background and its
gradient and curl components are shown in
Fig.~\ref{f:demo_injectedskymaps}.
The sky map for the simulated background shown in panel (a) is the
same as the real part of $h_+(\hat k)$ in the top-left panel of
Fig.~\ref{f:responseMap_background}.  (The imaginary part of $h_+(\hat
k)$ and the real and imaginary parts of $h_\times(\hat k)$ for the
simulated background are qualitatively similar and are shown in
three other panels in Fig.~\ref{f:responseMap_background}.)  
The simulated gradient component shown in 
Fig.~\ref{f:demo_injectedskymaps}, panel (b) is for a 
statistically isotropic background with 
$C_l=1$ for $l=2,3,\cdots, 10$; and the simulated curl
component shown in 
Fig.~\ref{f:demo_injectedskymaps}, panel (c) is also for a 
statistically isotropic
background, but with $C_l=4$ for $l=2,3,4,5$.  (The smaller value of
$l_{\rm max}$ for the curl component is responsible for the larger
angular-scale structure in the curl map, and the larger values for the
$C_l$'s for the curl component were chosen to make its contribution 
to the total simulated background comparable to that of the grad
component.)  The total simulated background in 
Fig.~\ref{f:demo_injectedskymaps}, panel (a) is just
the sum of the gradient and curl components.

\begin{figure*}[htbp]
\begin{center}
\subfigure{\includegraphics[trim=3cm 6.5cm 3cm 3.5cm, clip=true, angle=0, width=0.32\textwidth]{grad_realhp.jpg}}
\subfigure{\includegraphics[trim=3cm 6.5cm 3cm 3.5cm, clip=true, angle=0, width=0.32\textwidth]{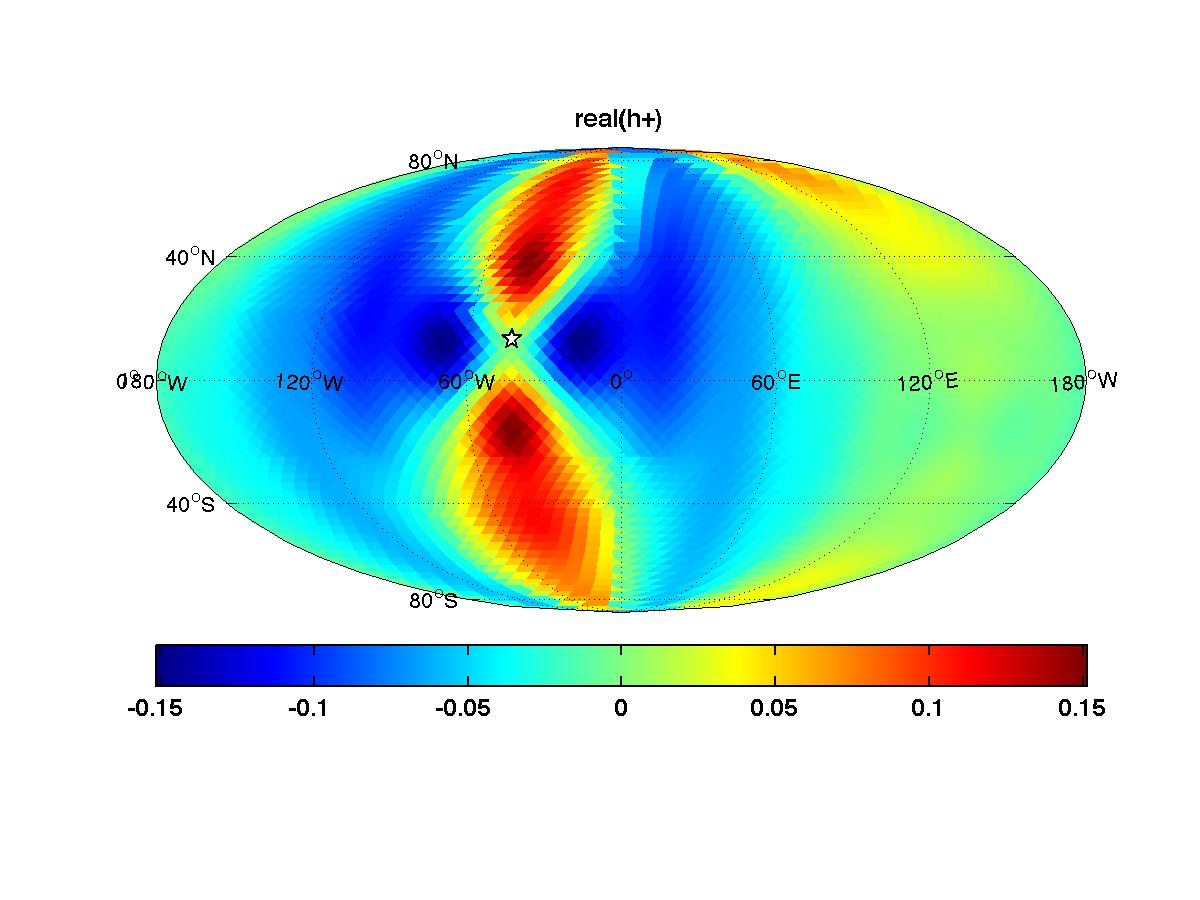}}
\subfigure{\includegraphics[trim=3cm 6.5cm 3cm 3.5cm, clip=true, angle=0, width=0.32\textwidth]{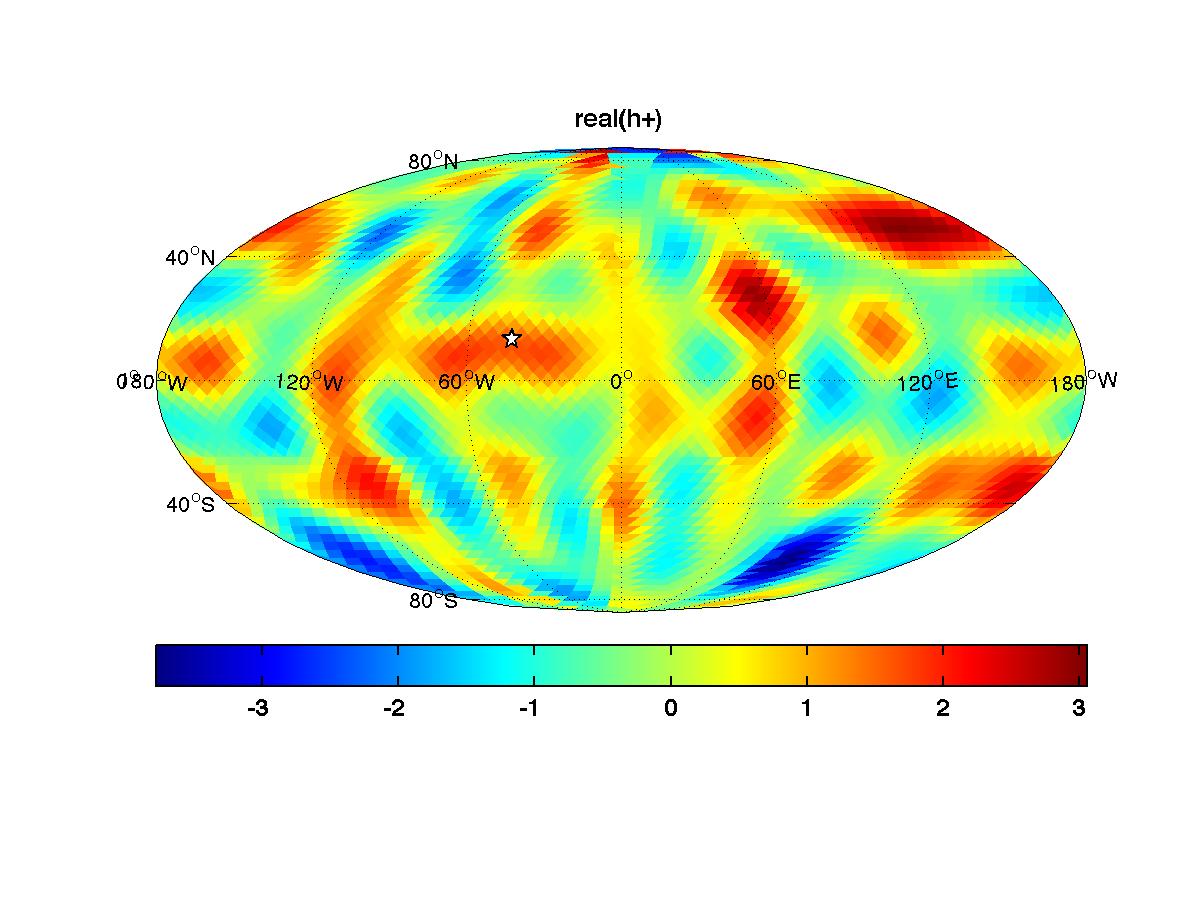}}
\subfigure{\includegraphics[trim=3cm 6.5cm 3cm 3.5cm, clip=true, angle=0, width=0.32\textwidth]{grad_realhp.jpg}}
\subfigure{\includegraphics[trim=3cm 6.5cm 3cm 3.5cm, clip=true, angle=0, width=0.32\textwidth]{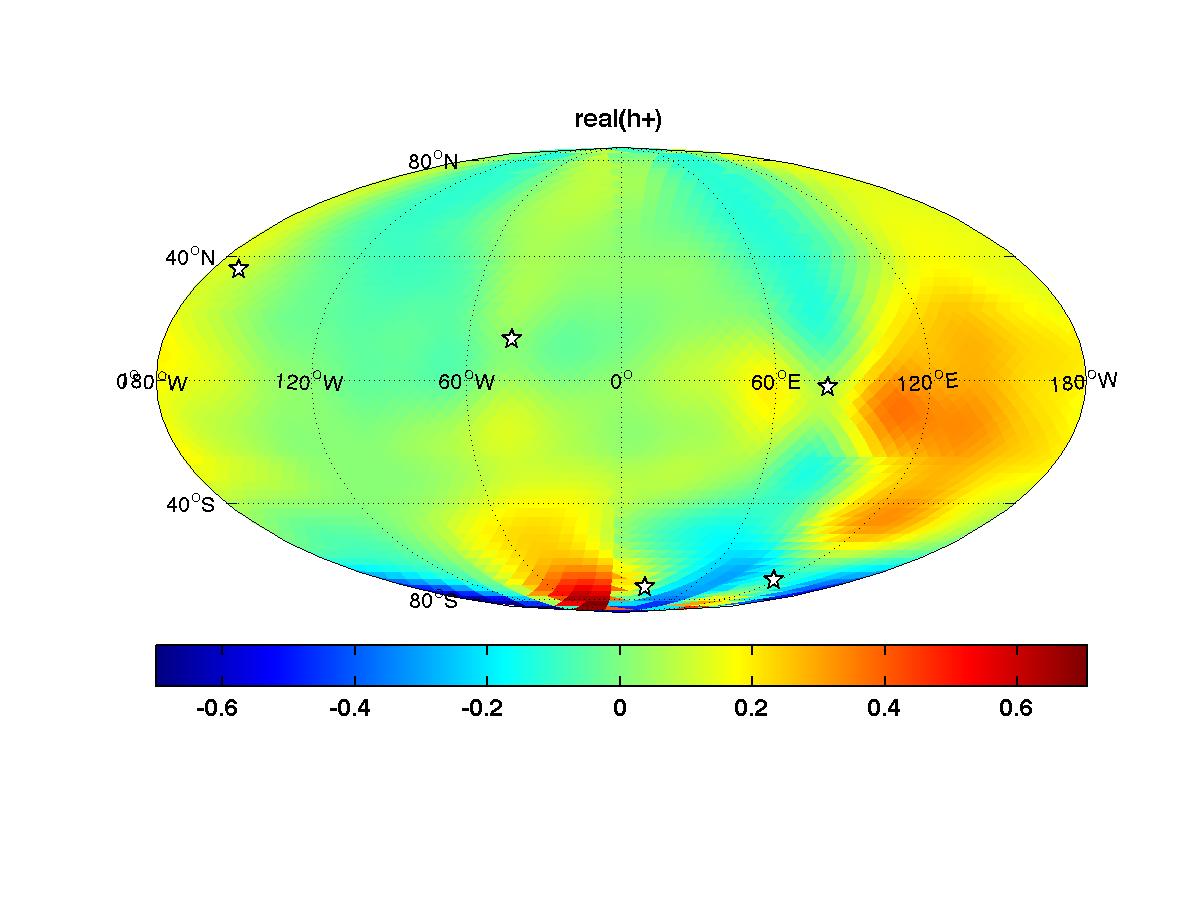}}
\subfigure{\includegraphics[trim=3cm 6.5cm 3cm 3.5cm, clip=true, angle=0, width=0.32\textwidth]{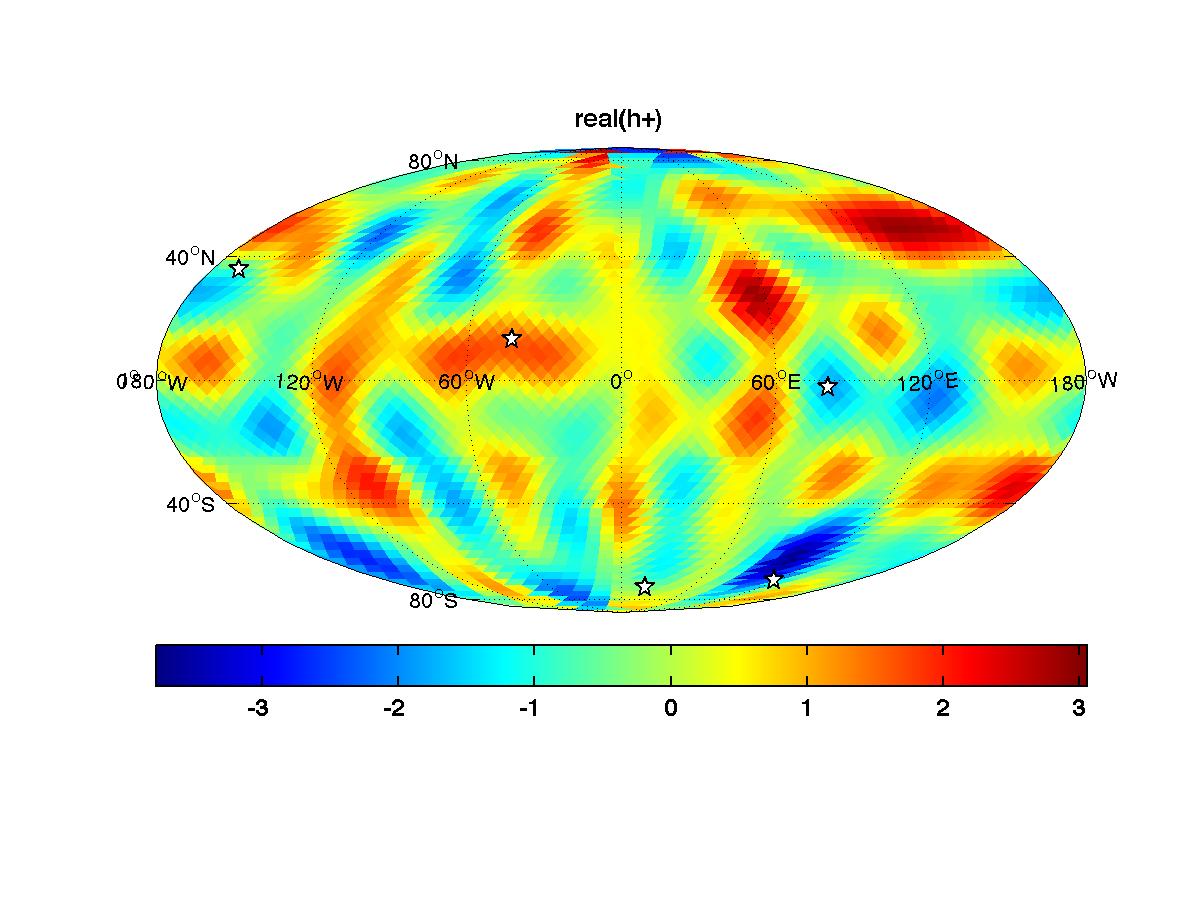}}
\subfigure{\includegraphics[trim=3cm 6.5cm 3cm 3.5cm, clip=true, angle=0, width=0.32\textwidth]{grad_realhp.jpg}}
\subfigure{\includegraphics[trim=3cm 6.5cm 3cm 3.5cm, clip=true, angle=0, width=0.32\textwidth]{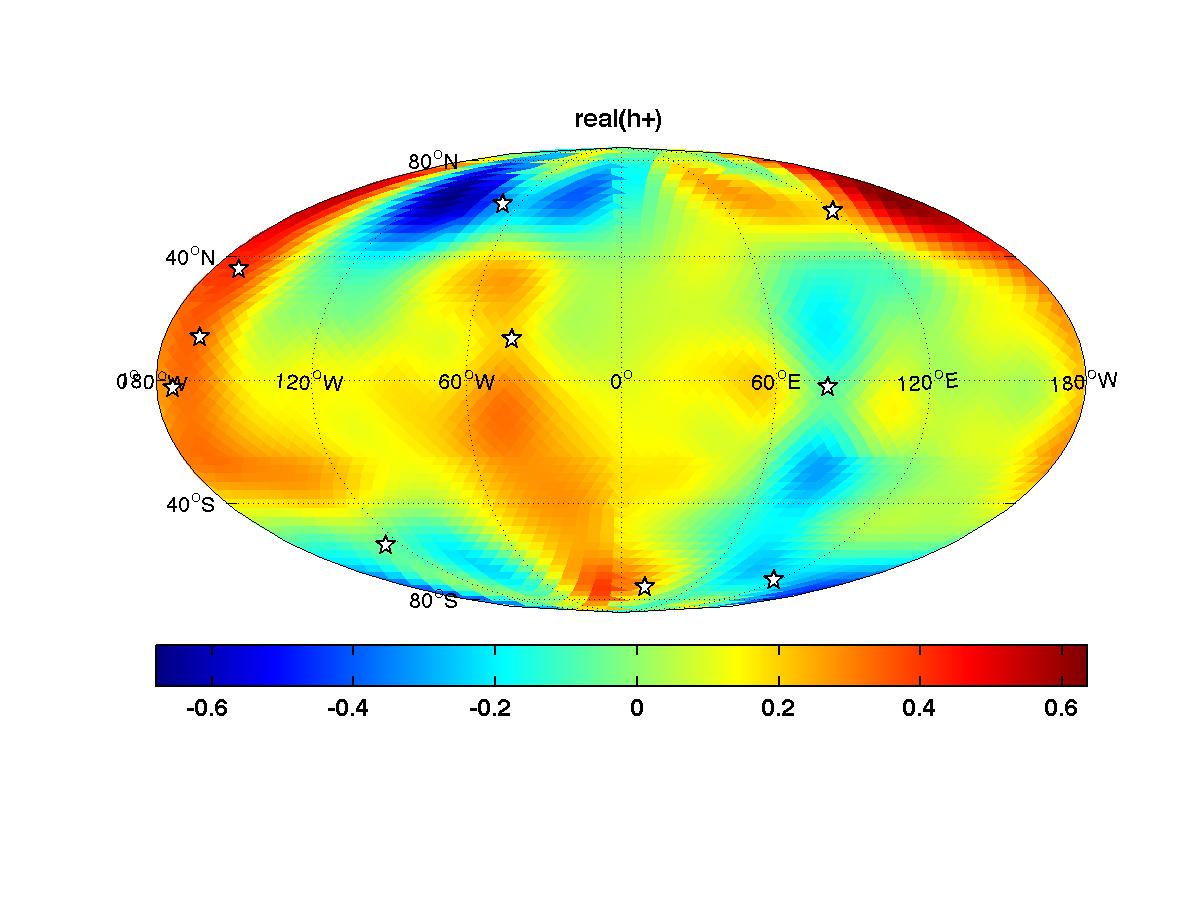}}
\subfigure{\includegraphics[trim=3cm 6.5cm 3cm 3.5cm, clip=true, angle=0, width=0.32\textwidth]{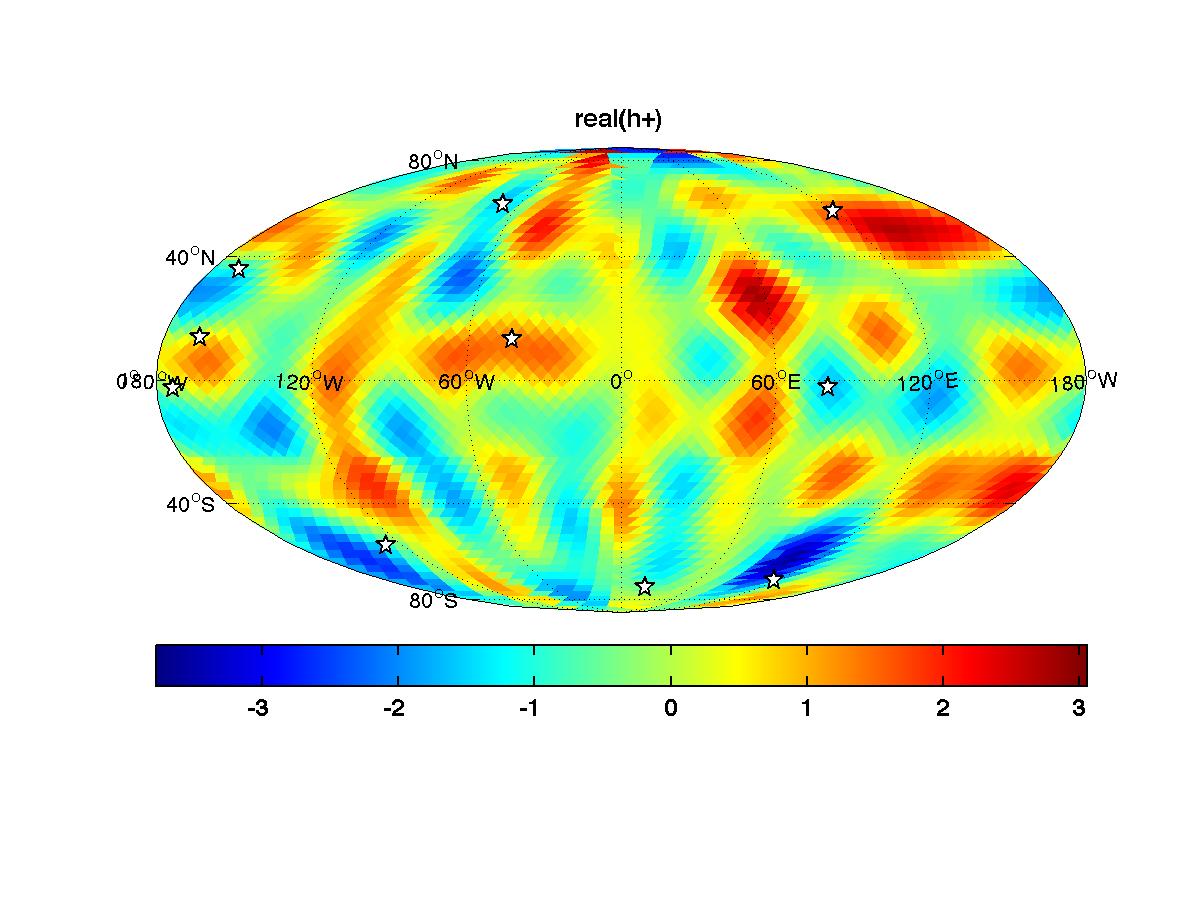}}
\subfigure{\includegraphics[trim=3cm 6.5cm 3cm 3.5cm, clip=true, angle=0, width=0.32\textwidth]{grad_realhp.jpg}}
\subfigure{\includegraphics[trim=3cm 6.5cm 3cm 3.5cm, clip=true, angle=0, width=0.32\textwidth]{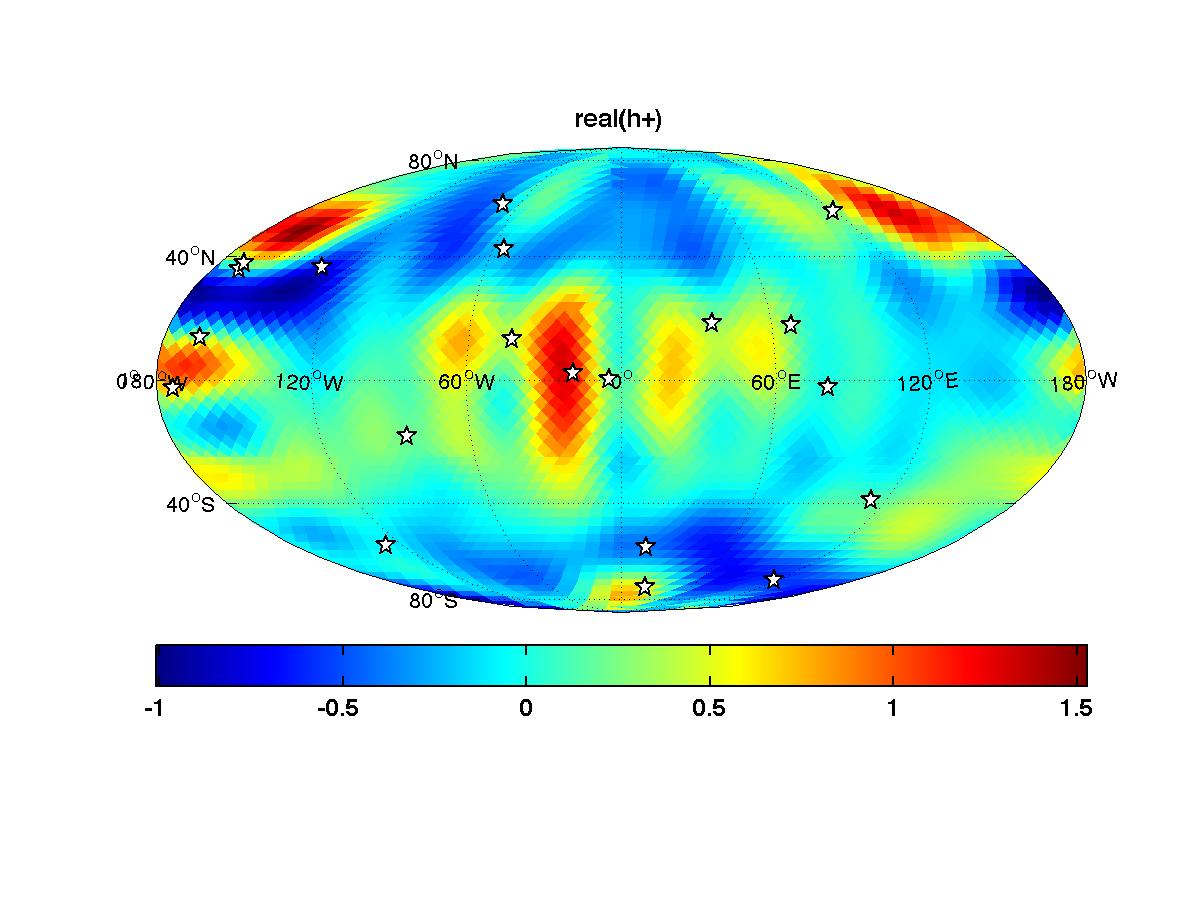}}
\subfigure{\includegraphics[trim=3cm 6.5cm 3cm 3.5cm, clip=true, angle=0, width=0.32\textwidth]{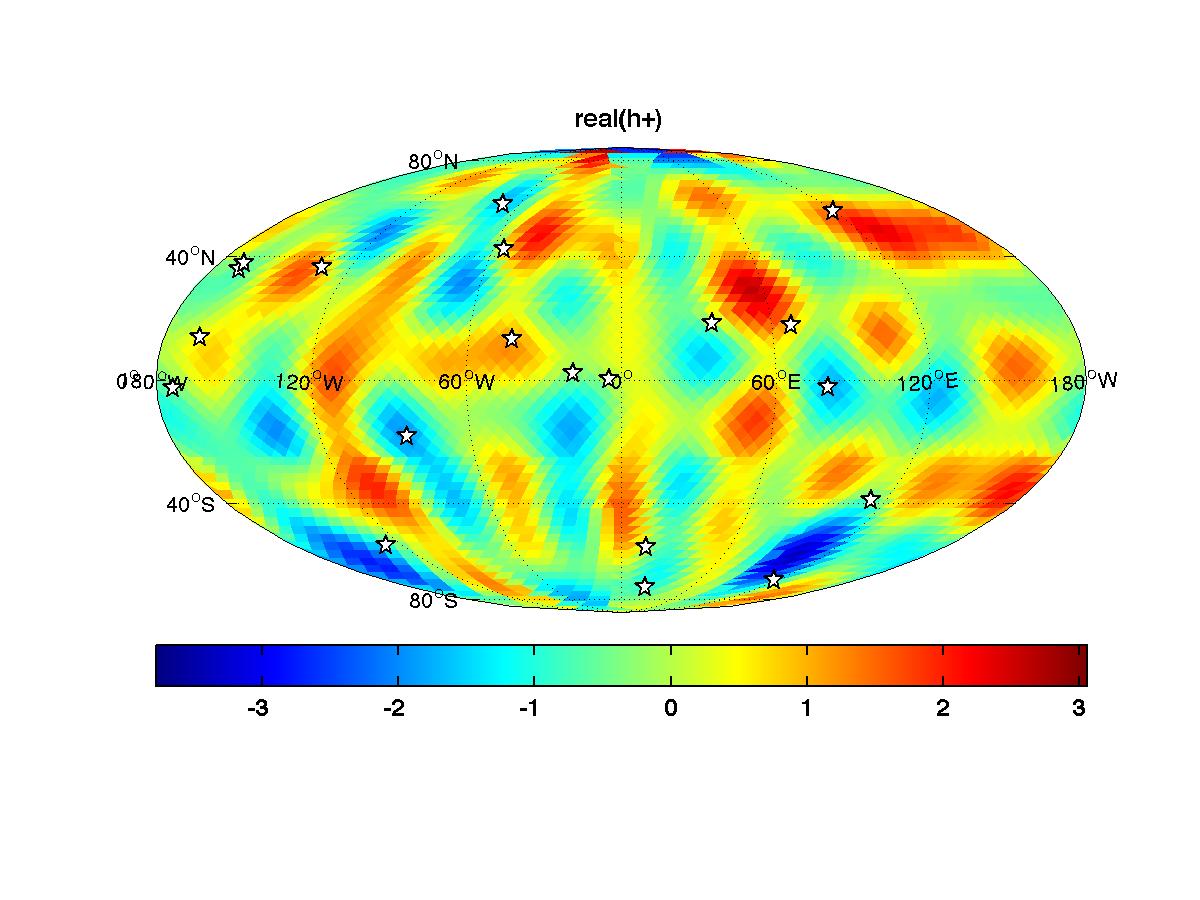}}
\subfigure{\includegraphics[trim=3cm 6.5cm 3cm 3.5cm, clip=true, angle=0, width=0.32\textwidth]{grad_realhp.jpg}}
\subfigure{\includegraphics[trim=3cm 6.5cm 3cm 3.5cm, clip=true, angle=0, width=0.32\textwidth]{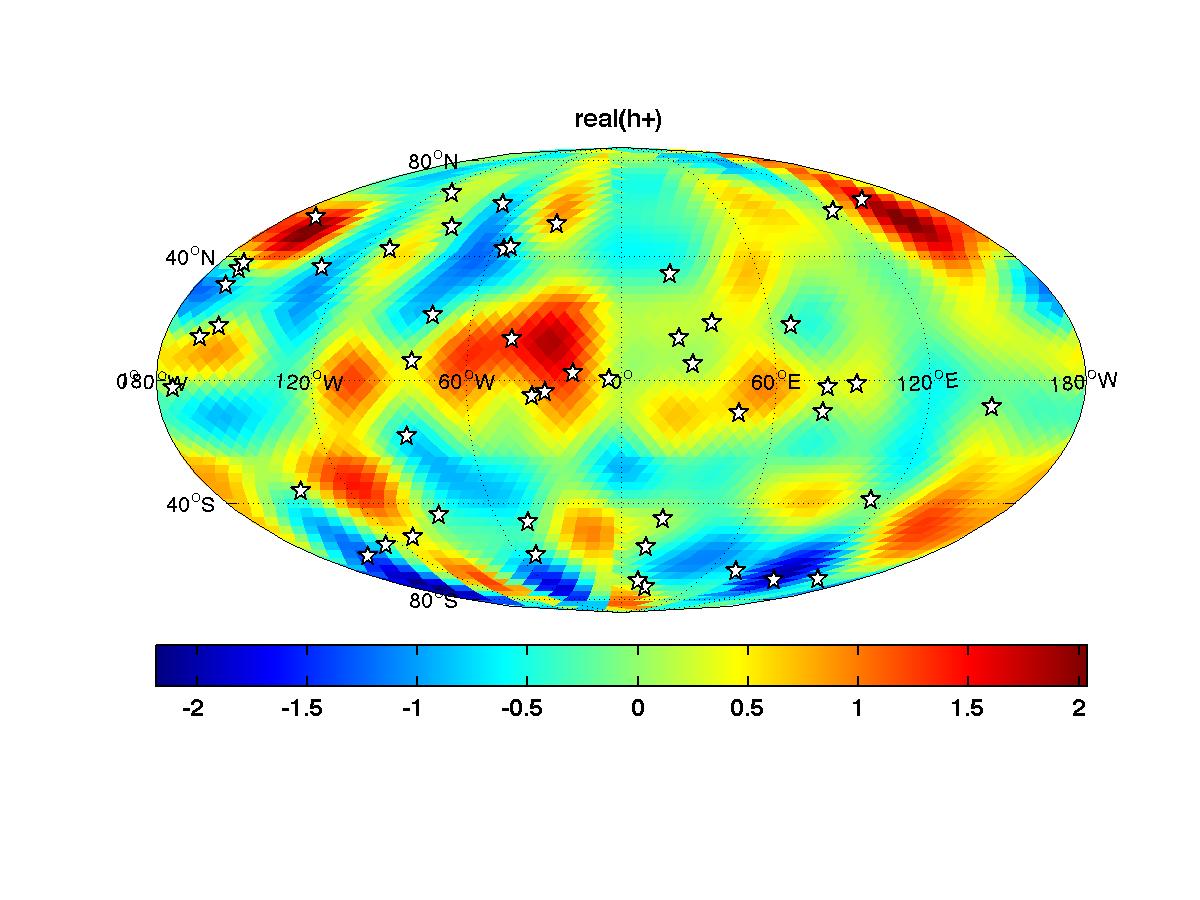}}
\subfigure{\includegraphics[trim=3cm 6.5cm 3cm 3.5cm, clip=true, angle=0, width=0.32\textwidth]{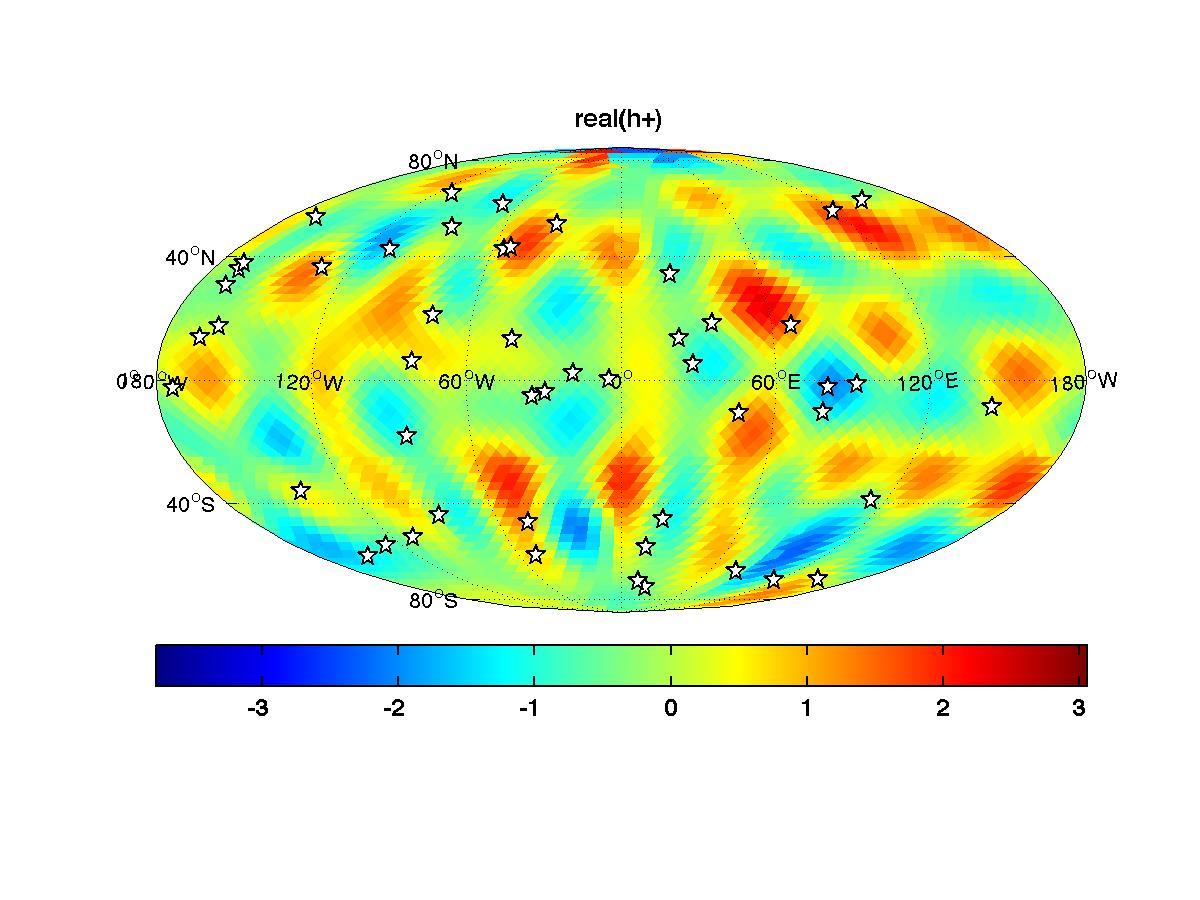}}
\subfigure{\includegraphics[trim=3cm 6.5cm 3cm 3.5cm, clip=true, angle=0, width=0.32\textwidth]{grad_realhp.jpg}}
\subfigure{\includegraphics[trim=3cm 6.5cm 3cm 3.5cm, clip=true, angle=0, width=0.32\textwidth]{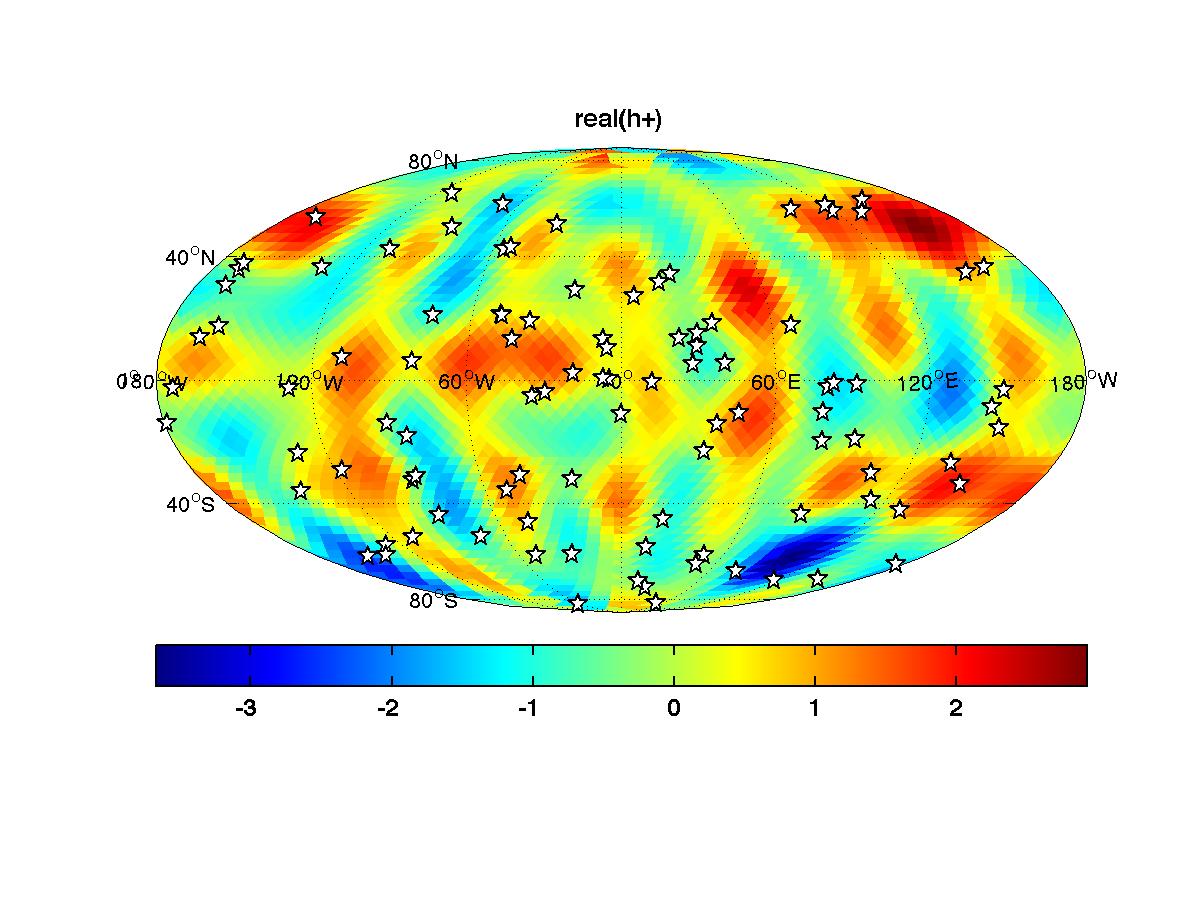}}
\subfigure{\includegraphics[trim=3cm 6.5cm 3cm 3.5cm, clip=true, angle=0, width=0.32\textwidth]{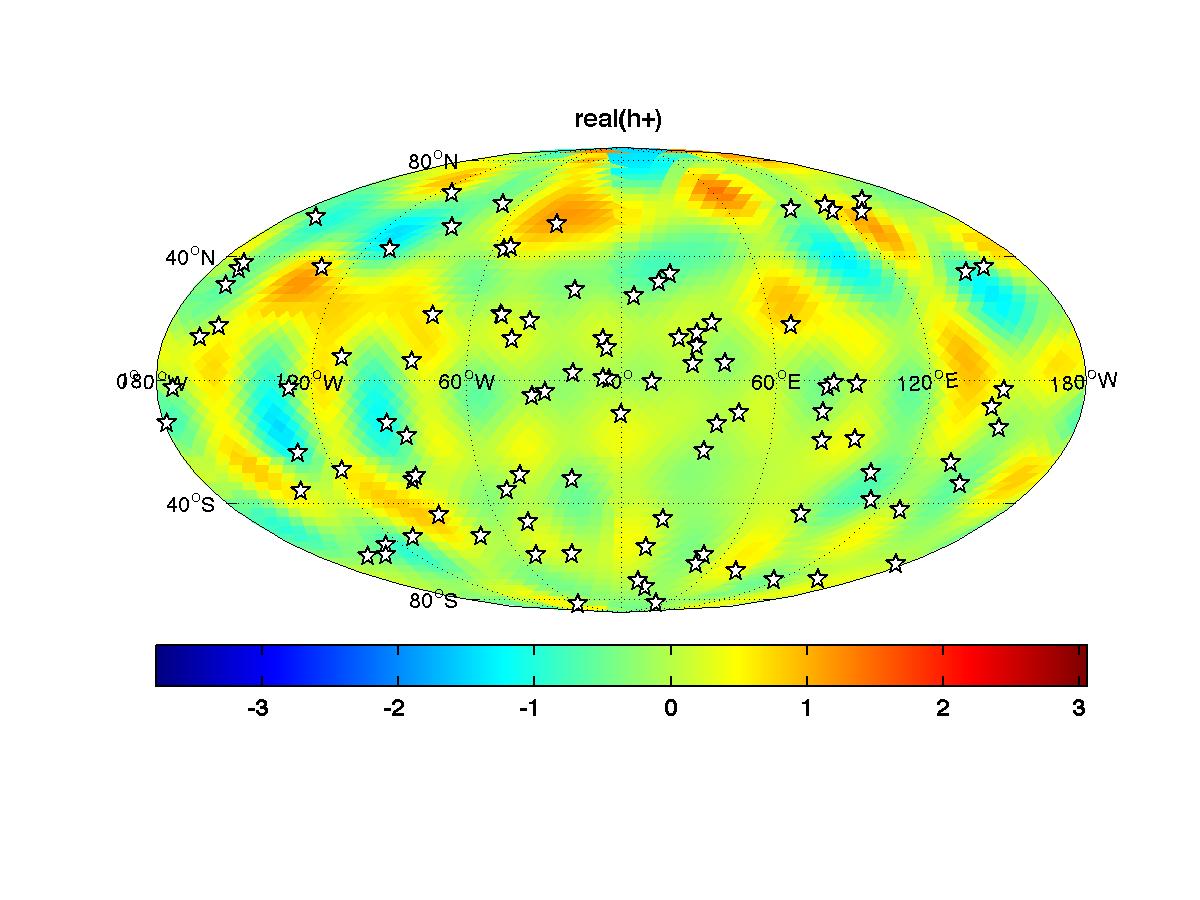}}
\caption{Mollweide projections of ${\rm Re}(h_+)$ for 
the grad-component of the simulated background (first column), 
the maximum-likelihood recovered sky maps (second column), 
and residual sky maps (third column) 
for PTAs containing different numbers of pulsars.
The residual sky maps are the difference between the grad-component
of the simulated background and the recovered maps.
Maps of ${\rm Im}(h_+)$, ${\rm Re}(h_\times)$, and ${\rm Im}(h_\times)$ are similar.
The rows correspond to PTAs containing $N=1$, 5, 10, 20, 50, and 100 pulsars, 
respectively.
The pulsar locations are shown as white stars.
We used a variable colorscale in the recovered maps to better show the angular 
structure in the small-$N$ maps, which would not have been visible if we had used the same (fixed) colorscale used to make the simulated background and residual maps.}
\label{f:demo_recoveredskymaps}
\end{center}
\end{figure*}

\begin{figure*}[htbp]
\begin{center}
\subfigure[\ Total map (grad+curl)]
{\includegraphics[trim=3cm 6.5cm 3cm 3.5cm, clip=true, angle=0, width=0.32\textwidth]{tot_realhp.jpg}}
\subfigure[\ Gradient component]
{\includegraphics[trim=3cm 6.5cm 3cm 3.5cm, clip=true, angle=0, width=0.32\textwidth]{grad_realhp.jpg}}
\subfigure[\ Curl component]
{\includegraphics[trim=3cm 6.5cm 3cm 3.5cm, clip=true, angle=0, width=0.32\textwidth]{curl_realhp.jpg}}
\subfigure[\ Grad residual map (grad$-$recovered)]
{\includegraphics[trim=3cm 6.5cm 3cm 3.5cm, clip=true, angle=0, width=0.32\textwidth]{residual_100_realhp.jpg}}
\subfigure[\ Max-likelihood recovered map]
{\includegraphics[trim=3cm 6.5cm 3cm 3.5cm, clip=true, angle=0, width=0.32\textwidth]{maxlikelihood_100_realhp.jpg}}
\subfigure[\ Total residual map (total$-$recovered)]
{\includegraphics[trim=3cm 6.5cm 3cm 3.5cm, clip=true, angle=0, width=0.32\textwidth]{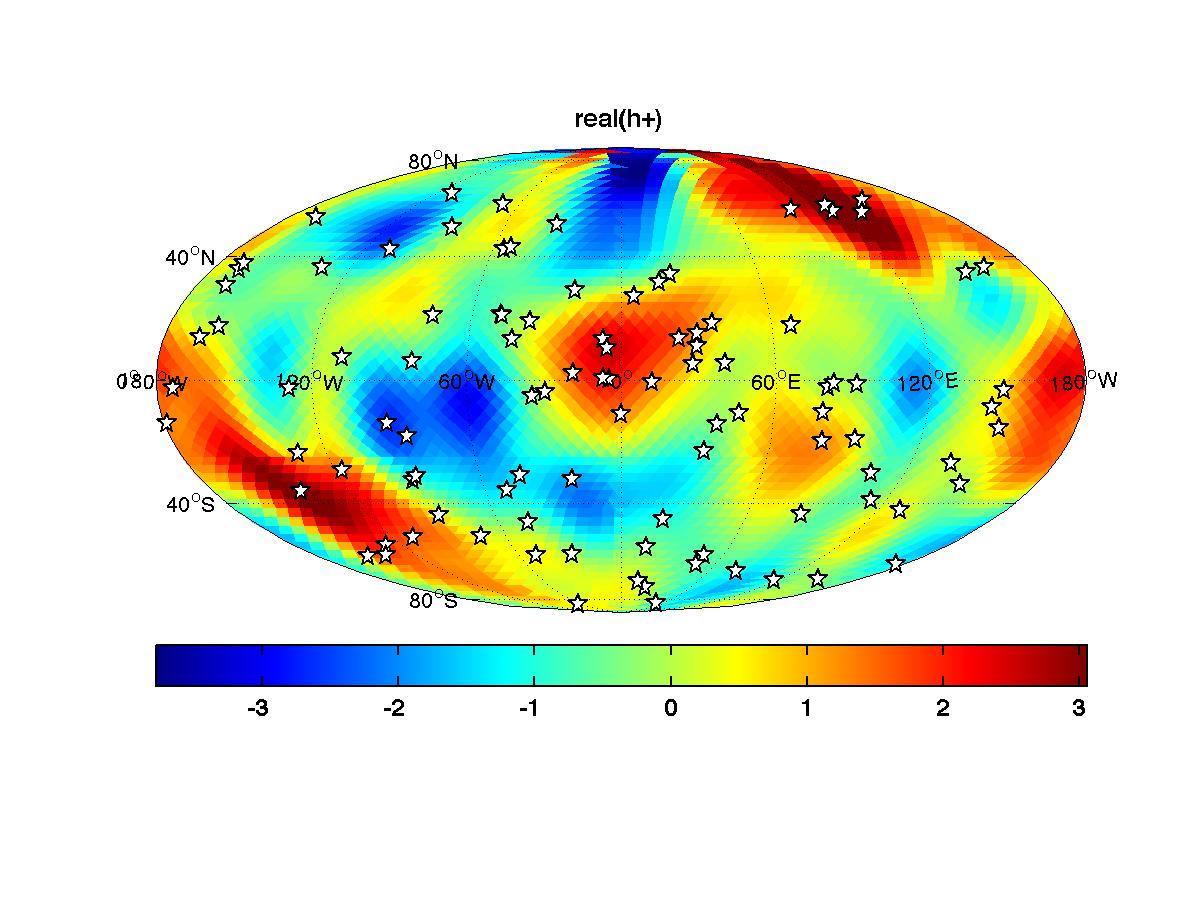}}
\caption{Mollweide projections of ${\rm Re}(h_+)$ for the different
components of the simulated background (panels a--c), 
the maximum-likelihood recovered map for a PTA with $N=100$ pulsars (panel e), 
and the corresponding residual maps for the grad-component (panel d) 
and the total simulated background (panel f).
Sky maps of ${\rm Im}(h_+)$, ${\rm Re}(h_\times)$, and ${\rm Im}(h_\times)$
are similar.
Note that the maximum-likelihood recovered map most-closely resembles the 
gradient component of the simulated background, 
since a PTA is insensitive to the curl modes of a gravitational-wave background.
The residual map for the grad-component (panel d) is cleaner than the residual
map for the total simulated background (panel f), which has angular 
structure that resembles the curl component of the background.}
\label{f:demo_comparisonskymaps}
\end{center}
\end{figure*}

Figrure~\ref{f:demo_recoveredskymaps} shows the results of the 
maximum-likelihood estimation of the simulated background for PTAs 
containing different numbers of pulsars.
Mollweide projections of the real part of $h_+(\hat k)$ are 
shown for the gradient component of the simulated background 
(first column), the maximum-likelihood recovered sky maps 
(second column), and the residual sky maps (third column).
The residual maps are the difference between the gradient component of the
simulated background and the recovered maps.  The six rows correspond
to PTAs containing $N=1$, 5, 10, 20, 50, and 100 pulsars,
respectively.  The maximum-likelihood estimates of the spherical
harmonic components are constructed out to $l_{\rm max}=10$.  The size
of the pixels used for the sky maps corresponds to an angular resolution of
$\Delta\Omega \sim 50~{\rm deg}^2$, which is much finer than that
achievable with a 100-pulsar array (see Sec.~\ref{s:limited-sensitivity}
for details).  It is apparent from the plots that the
recovered maps more closely resemble the gradient component of the
simulated background as the total number of pulsars in the array
increases.  The residual maps become ``greener" (the values are
getting closer to zero), and the angular scale of the structure in the
residual maps generally gets smaller as the number of pulsars
increases.  But as discussed in Sec.~\ref{sec:overlap_general}, we
will never be able to recover the total simulated background,
regardless of the number of pulsars in the array.  This is because a
PTA is insensitive to the curl component of the background.
Figure~\ref{f:demo_comparisonskymaps} illustrates this point by
comparing the maximum-likelihood recovered sky map for $N=100$ pulsars
with the different components of the simulated background.  The
maximum-likelihood recovered map most-closely resembles the gradient
component of the simulated background, as expected.
%

\subsection{Implications of the limited sensitivity of a PTA}
\label{s:limited-sensitivity}

A given PTA is only sensitive to a number of particular complex combinations
of the $a^G_{(lm)}$'s that is equal to the number of pulsars in the
array. But which combinations are these? Considering the response
functions, $R^G_{I(lm)}(f) = (i2\pi f)^{-1}2\pi(-1)^l N_l Y_{(lm)}(\hat{u}_I)$, we
see that for large $l$, $R^G_{I(lm)}\sim l^{-3/2}$. We therefore
expect to have more sensitivity to low-$l$ modes. In an analysis with
fewer modes than pulsars, and assuming that the noise matrix $F$ is a
multiple of the identity, the Fisher information matrix is $H^\dag H$,
the elements of which are proportional to products of
$R^G_{I(lm)}$'s. The square root of the diagonal elements of the
inverse of the Fisher matrix provide a measure of the precision of
parameter measurement, which indicates we would expect to measure the
large-$l$ components of the background with a precision no better than
$\sim\! l^{3/2}$. Considering the SVD of $H$ when there is a single pulsar
in the array, we see that the column of $V$ corresponding to the
non-zero singular value of $H$ is proportional to $R^G_{1(lm)}$ and this
confirms again that we are most sensitive to the low-$l$ modes. We
note that these are precisely the components you would get when
decomposing the transverse-traceless projection of 
$\frac{1}{2}u_1^a u_1^b/(1+\hat{k} \cdot \hat{u}_1)$ onto the 
grad and curl spherical harmonic basis. Therefore, the gravitational-wave sky map
corresponding to this single non-zero singular mode that we can
measure is proportional to the response function for that single
pulsar over the sky. This is to be expected---at fixed amplitude, an
instrument is most sensitive to a signal distribution that matches its
own relative sensitivity to different directions. 

In practice, we can use this formalism to approach sky mapping in two
different ways. Firstly, we can acknowledge the fact that we are only
sensitive to $N$ complex components of the background and limit the number of complex 
$(lm)$ modes included in the model to be less than or equal to $N$. On
average, we are equally sensitive to all $m$ modes corresponding to a
particular $l$ (although the relative sensitivity to each $m$ mode
will in practice depend on the pulsar distribution in the
PTA). Therefore we should include complete sets of modes up to some
$l_{\rm max}$, which means including $(l_{\rm max}+1)^2 - 4$ modes in total. 
(The $l$'th harmonic has $(2l+1)$ independent $m$ modes, giving
a total of $(l_{\rm max}+1)^2$ independent modes with 
$l \leq l_{\rm max}$, and we subtract $4$ because there are 
no modes with $l=0$ or $l=1$.) 
Alternatively, we can include more than $N$ modes in the
model, find the singular value decomposition of the $H$ matrix, and 
then recover the $N$ linear combinations of the $N_m$ modes to which
the array is sensitive. As discussed in the preceding paragraph, we expect the
measurements of the low-$l$ components to be more accurate than those of
high-$l$ components.

The next relevant question is how many pulsars do we need to produce a
useful map of the gravitational-wave sky? It is reasonable to require
the angular resolution of the map to be comparable to the scale on
which individual gravitational-wave sources might be
resolvable. In~\cite{SesanaVecchio:2010}, it was shown that the
angular resolution, $\Delta\Omega$, of a PTA with 50 isotropically
distributed pulsars to an individual gravitational-wave source scales
as
\be \Delta\Omega \propto 50 \left(\frac{50}{N} \right)^{1/2}
\left(\frac{10}{\mathrm{SNR }}\right)^2 \mathrm{deg}^2\,. 
\ee
This implies
\be
l \sim 25 \left(\frac{{\rm SNR}}{10}\right) 
\left( \frac{N}{50} \right)^{\frac{1}{4}}\,,
\label{e:angular_resolution}
\ee
where we have defined the angular resolution as $l \approx 180/\sqrt{\Delta
  \Omega/{\rm deg}^2}$ to be consistent with the CMB literature. The
maximum $l$ mode that can be probed with a set of $N$ pulsars scales
as $l_{\rm max} \sim \sqrt{N+4} - 1$, which grows faster than the
single-source angular resolution. Assuming an SNR of $\sim 3$ is
needed for detection, we find that we need $N \gtrsim 100$ pulsars to
reach the angular scale of individual sources, with corresponding
$l_{\rm max} \sim 10$ and angular resolution $\Delta\Omega\sim 400~{\rm deg}^2$. 
Although this is beyond the current capabilities
of PTAs, the SKA could detect several thousand millisecond
pulsars~\cite{SKApulsars}, some of which will be
sufficiently stable to contribute usefully to PTA efforts. 
We may therefore not be able to reach this limit until the SKA era.
However, we saw in
Section~\ref{s:recovery_HD} that we only need to go to $l_{\rm max}
\approx 4$ to recover an isotropic, unpolarised and uncorrelated background. 
To reach this resolution requires only $21$ pulsars, which is fewer than
ongoing PTA efforts are currently using. We should therefore be able
to produce informative maps of the gravitational-wave background and
constraints on the level of anisotropy in the near future.

The fact that we are sensitive only to a number of modes equal to the
number of pulsars in the PTA is a consequence of the fact that the
pulsar lines of sight are fixed, and so our detector does not scan
across the sky. Other gravitational-wave detectors such as LIGO or
LISA will scan the sky as Earth rotates or the spacecraft orbit the Sun,
which will give these detectors sensitivity to all of the modes of the
background, given sufficient integration time.

As described briefly in the introduction to this section, what we will
be able to learn from measurements of the gravitational-wave
background is influenced both by the resolution of the map and by the
nature of the background. It is clear from
Fig.~\ref{f:demo_recoveredskymaps} in the preceding section that
adding pulsars to the array improves the resolution of the map,
allowing us to resolve finer structures. As the number of pulsars is
increased, the size of the structures that are resolved in the map and
that are visible in the unresolved part of the background become
smaller. We have argued that we need $\sim 21$ pulsars to reach
$l_{\rm max} \approx 4$, which is required to confidently recover 
an isotropic, unpolarised and uncorrelated background, 
and we need $\sim 100$ pulsars to reach a resolution at
which we could identify hot spots of the size expected for individual
gravitational-wave sources. Once a map has been made it can be used to
constrain models. Even with a low resolution map, we can check if the
coefficients are consistent with the expectation for any particular
model. If the background is isotropic, unpolarised and uncorrelated, then with a map
of any resolution we will be able to make a statement that the
background is consistent with being isotropic, unpolarised and uncorrelated to a certain level. That level
of consistency will improve as the resolution increases. If the
background deviates from isotropic, unpolarised and uncorrelated, then we will 
most-likely not be able to identify this with a very low resolution map, 
but at some critical point the departures will start to show significantly.

It is important to note that a typical model for the background will
not predict precise values of the components of the background,
$a^G_{(lm)}$, but their statistical properties, $\langle a^G_{(lm)}
a^{G'*}_{(l'm')} \rangle = C^{GG'}_{lml'm'}$. Section~\ref{s:marglike}
below describes how to recast the detection problem as a measurement
of the $C^{GG'}_{lml'm'}$ coefficients. A given background map is a
representation of the specific values of the coefficients in our
realisation of the Universe and so even with perfect data (but finite
resolution, i.e., a finite number of pulsars) the map will not be able
to distinguish all possible models. This is the usual problem of
cosmic variance familiar from other areas of astrophysics, including
analysis of the CMB---we can observe only one Universe, which limits
our ability to determine its statistical properties. An $N$-pulsar map
can measure $N$ linear combinations of the (infinite set of)
$a^G_{(lm)}$'s, but to illustrate we will assume, for simplicity, that
these are the $N$ coefficients of lowest $l$, and we will assume that we
have measured them perfectly. A typical model can be represented by a
specification of the correlation coefficients, written as a matrix
$C$, and for an isotropic, unpolarised and uncorrelated background 
we have $C=I$. An alternative
model will either predict a specific $C$ or, more likely, make some
generic statement about its properties, e.g., the diagonal components
are significantly different from each other or the off-diagonal
components are significantly different from zero. We consider the
first case, again for simplicity. The model selection problem would ask
if the alternative model fits the measured data better than an
isotropic background. Since in this simplified problem the two models
have no free parameters, the Bayesian evidence is equal to the
likelihood ratio and, denoting the measured parameters by $\vec{a}$,
the difference in log evidence is just
\be
\Delta \ln {\cal Z} = -\frac{1}{2} 
\ln({\rm det}C) -\frac{1}{2} \vec{a}^\dag (C^{-1} - I) \vec{a}\,.
\ee
If the difference in log evidence is large enough, we will disfavour
the isotropic model. To decide if we will be able to distinguish two
models with a given $N$-pulsar array, we can compute the expectation
value, under the alternative model, of this difference in log
evidence. This expectation value is
\be
\langle \Delta \ln {\cal Z} \rangle 
= \frac{1}{2} \left[{\rm Tr}(C)- N -\ln({\rm det}C) \right]\,.
\ee
It would be possible to come up with a contrived model in which only
one of the components of the correlation matrix differs from the
identity. In such a scenario the addition of extra pulsars to the
array does not improve the distinguishability of the models. This is
not surprising, as in such a scenario only one or two of the
coefficients have different statistical properties in the alternative
model, so measuring additional coefficients by adding pulsars does not
provide any discriminating power. In more realistic alternative
models, many components of the correlation matrix will differ from
the identity. In that case, the trace 
of $C$ will grow
roughly like $N$ and so at some point the models will be
distinguishable, with the number of pulsars required depending on the
size of the differences in $C$. We will leave a complete investigation
of map making, including the effects of imperfect measurement of the
coefficients and marginalising over unknown parameters in the model of
the background, for future work, but this simple illustration serves
to demonstrate how a map might be used to constrain models and how the
ability to place constraints improves with the number of pulsars in
the array. In practice, what we will do is re-evaluate the map as new
data and new pulsars are added to the array and look for signatures of
anisotropy or correlation. As the maps contain all the measurable
information about the background, anyone will be able to check
predictions of their favourite model against the data without having
to reanalyse the entire data set.

The insensitivity of a PTA to the curl modes of the background means,
in principle, that we are losing half of the information that is
available. However, most of the information about the physics
generating the background should be encoded in both the grad and curl
components. For an unpolarised and uncorrelated background, the components of the grad and curl modes
are independent realisations from the same probability distribution,
so either of them encodes the information about that probability
distribution (which is the power distribution across the sky). For a
point source, i.e., a binary, contributing to the background, the grad
and curl components both encode the source parameters, but there are only a
small number of parameters that characterise an individual system and
so these can be extracted from the grad components alone provided there
is a sufficient number of pulsars in the array. Our ability to resolve the
parameters of multiple individual sources will be limited by the
number of pulsars in the array, but, as described above, we can only
measure $N$ complex parameters at any given frequency using $N$
pulsars, and so this limitation applies whether or not we are
sensitive to the curl modes. In principle, a correlated stochastic
background generated in the early Universe could have a pure curl
polarisation, or a curl amplitude much greater than the grad
amplitude. In that case we would never see such a background with a
pulsar timing array and would clearly have incomplete information
about the background. However, as described above, the distinction
between grad and curl modes is origin-dependent. If the background was
completely dominated by the curl component, then the Solar System
would be in a very special location in the Universe and so this
situation should be regarded as extremely unlikely on anthropic
grounds. Being unable to measure the curl background does mean,
however, that we will
not be able to compare its statistical properties to those of the grad
background, and so we will not be sensitive to statistical polarisation
of the background, i.e., that the statistical properties of the grad
and curl components are significantly different (this will be
discussed again in Sec.~\ref{s:implications}). However, the same
anthropic argument suggests that significant statistical polarisation
is unlikely. In summary, while the insensitivity of a PTA to a curl
background means that we will not be able to measure all of its
properties, the main physical properties of the background should be
evident in the grad-only part of the background and so PTA
observations still have tremendous potential to inform us about
gravitational physics.

\subsection{What combinations of timing residuals can an arbitrary background produce?}

The range of the matrix $H$ describes the only possible combinations
of timing residuals that could be induced in a given array of pulsars
by any gravitational-wave background. Cornish and van~Haasteren found
that this range is spanned by the eigenvectors of the Hellings and
Downs correlation matrix (private communication).  This follows
straightforwardly from properties of the SVD.  In the SVD,
Eq.~(\ref{e:svd}), the columns of $U$ are eigenvectors of the matrix
$H H^\dag$ and the non-zero elements of $\Sigma$ are the square-roots
of the non-zero eigenvalues of this matrix.  Writing ${\vec{r}} =
H\vec{a}$ we see that
\be
\vec{r} \vec{r}^\dag = H \vec{a} \vec{a}^\dag H^\dag.
\label{e:svdav}
\ee 
If we replace $\vec{a} \vec{a}^\dag$ by the identity matrix, the
right hand side of this equation becomes $H H^\dag$, which is the
matrix we want. This replacement can be accomplished by appropriate
averaging over the space of possible $\vec{a}$'s, for instance by
assuming these are uniformly distributed over equal ranges symmetric
about zero or by assuming they have an isotropic Gaussian distribution
with correlations $<a_k a^*_{k'}> = \delta_{kk'}$. So far, we have
used only generic properties of the SVD, but these observations are
useful only if we can efficiently compute $<\vec{r} \vec{r}^\dag>$ for
such a distribution of $\vec{a}$'s. In our case the specification
$<a_k a^*_{k'}> = \delta_{kk'}$ describes an isotropic, unpolarised and uncorrelated background (see Eqs.~(\ref{e:acorr})--(\ref{e:corrforiso})). 
We know that the expected response for such a background is the 
Hellings and Downs correlation matrix and the result follows. 
That this is true in both our formulation and that of Cornish and 
van~Haasteren is a consequence of the fact that an 
isotropic, unpolarised and uncorrelated background is described by 
$<a_k a^*_{k'}> = \delta_{kk'}$ 
in both prescriptions. This argument effectively assumes
that all possible modes are present in the vector $\vec{a}$, but a
similar argument can be used if we have only included a subset, ${\cal
  S}$, of these modes. The $U$ matrix is then different, but its
columns will be the eigenvectors of $<\vec{r} \vec{r}^\dag>_{\cal S}$,
which denotes the average response over an isotropic distribution of
the modes included in ${\cal S}$ only. Using the results in this paper
we can compute $<\vec{r} \vec{r}^\dag>_{\cal S}$ straightforwardly for
arbitrary ${\cal S}$. Specifically, we use
Eqs.~(\ref{e:gamma(f)}) and (\ref{eq:gamma_12,l}) with $C_l=1$ and the sum
over $l$ and $m$ replaced by a sum over the modes included in ${\cal
  S}$ and the response functions, $R^P_{I(lm)}(f)$, taken from
Eq.~(\ref{e:RG=Y,RC=0}).
We saw in Section~\ref{s:recovery_HD}
that including modes up to $l_{\rm max} = 4$ was sufficient to recover
the Hellings and Downs correlation curve, so we would expect that with
these $21$ modes in $\vec{a}$, the $U$ matrix would already be almost
equal to the infinite coefficient limit.

In the case that all modes are present in the background, $V$ has
infinitely many columns. However, once $U$ and $\Sigma$ have been
found it is possible to construct the first $N$ elements of the $j$'th
column of $V^\dag$ for arbitrary $j$ from 
\begin{align} 
(V^\dag)_{ij} =\Sigma^+_{iK} U^*_{IK} R^G_{Ij}\,, 
\end{align}
where $R^G_{Ij}$ is the grad
response of pulsar $I$ to the component of the background included in
the $j$'th element of $\vec{a}$.  This follows by setting $a_k =
0\ \forall \ k \neq j$ and $a_j=1$ in $H \vec{a}$, which makes
$(H\vec{a})_I = R^G_{Ij}$. From the SVD we have 
\be H\vec{a} = U\Sigma \vec V^\dag_j\,, 
\ee
where $\vec V^\dag_{j}$ denotes the $j$'th
column of $V^\dag$.  Premultiplying by the inverse of $U$, which is
$U^\dag$, and then by the pseudo-inverse of $\Sigma$ gives the final
result.  This picks out only the first $N$ elements of $\vec V^\dag_j$
due to the zeros in the $\Sigma$ matrix. However, these first $N$
elements of $\vec V^{\dag}_j$ are precisely the components of the
background to which the PTA is sensitive, and this approach allows
these to be computed to whatever resolution is required.

\subsection{Marginalized likelihood function for the $C^{GG'}_{lml'm'}$'s}
\label{s:marglike}

We finish by noting that an alternative way to proceed would be to
start with the above likelihood, but marginalize over the parameters
$\vec a$. For simplicity, we assume in the following that we are 
using the real version of the likelihood~(\ref{e:discretelike}) 
with twice as many real components in $\vec{a}$ and $H$. 
We will also assume a Gaussian distribution on $\vec{a}$ of the form
\be
p(\vec a|C) 
= \frac{1}{\sqrt{\det (2\pi C)}}
\exp\left\{-\frac{1}{2} \vec{a}^T C^{-1} \vec{a} \right\}\,,
\label{e:signal_prior}
\ee
where $C\equiv \{C^{GG}_{lml'm'}(f_k,f_{k'})\}$ 
are the model hyperparameters defined by an equation
like (\ref{e:aniso_expectation_values}).
Then by marginalizing over $\vec a$,
%
%
we obtain the likelihood 
\begin{widetext}
\be
p(\delta t|F, C) 
\propto \frac{1}{\sqrt{\det(I+C H^TF^{-1}H)}}
\exp\left\{-\frac{1}{2} \vec{\delta t}^T \left( F^{-1} - 
F^{-1} H (C^{-1}+H^T F^{-1} H)^{-1} H^T F^{-1} \right) \vec{\delta t} \right\}\,,
\ee
where we have assumed $C^T=C$ and $F^T=F$. 
By using the {\it Woodbury matrix lemma}, this can be simplified to 
\be
p(\delta t|F, C) \propto \frac{1}{\sqrt{\det(2\pi D)}}
\exp\left\{-\frac{1}{2} \vec{\delta t}{}^T D^{-1}\vec{\delta t} \right\}\,,
\quad
{\rm where\ }
D = F + HCH^T\,,
\ee
\end{widetext}
which can be seen most readily by considering the correlation matrix
directly rather than its inverse~\cite{CornishRomano:2013}.
One can use this modified likelihood to 
construct posterior distributions or estimators of the parameters
$C$ using Bayesian inference or frequentist
statistics as before. In particular, one reasonable assumption to make
in this case would be 
$C^{GG}_{lml'm'}(f_k,f_{k'}) = \delta_{kk'}C^{GG}_{lml'm'}$, 
i.e., that different frequency components are
uncorrelated and the covariance of a given component depends only on
the angular mode parameters, $l,m,l',m'$, and not on the frequency. We
note that this marginalisation is completely analogous to the
marginalisation over timing-model parameters used to derive the 
time-domain version of Eq.~(\ref{e:discretelike})
in~\citep{van-haasteren-levin-2012}. If a flat prior is used instead
of a Gaussian prior, the $C^{-1}$ term does not appear in the
likelihood and the term inside the exponential can be simplified to
the form $H_{\rm null} \left(H_{\rm null}^T F H_{\rm null}\right)^{-1}
H_{\rm null}^T$, where $H_{\rm null}$ is the matrix of left singular
eigenvectors of $H$ with zero singular values. 
The matrix $H_{\rm null}$ is given by the last $2 N_m - 2 N$ columns
of $U$ in the SVD (the factors of $2$ come
from going from a complex to a real representation). This is an
identical result to the marginalisation of the timing model parameters
described in~\citep{van-haasteren-levin-2012}.

\FloatBarrier
\section{Discussion}
\label{s:discuss}

We have described how the formalism used to characterise the
polarisation of the CMB can be applied to the analysis of
gravitational-wave backgrounds. Any symmetric transverse-traceless
tensor field on the sphere can be decomposed into a superposition of
modes that are gradients or curls of spherical harmonics. We
considered using this decomposition to analyse a gravitational-wave
background observed with a pulsar timing array. Writing a
statistically isotropic background as a superposition of these modes,
we found that the PTA overlap reduction functions for these modes were
Legendre polynomials evaluated for the pulsar angular separations.  We
showed that this was just a consequence of a pulsar having zero
response to curl modes, and a response to each gradient mode that is
proportional to the corresponding spherical harmonic evaluated at the
direction of the pulsar.  Although an infinite number of modes are
needed to precisely recover the Hellings and Downs curve for an
isotropic, unpolarised and uncorrelated background, we showed that, in
practice, the background can be recovered accurately using just the lowest three
$l$-modes, $l=2,3,4$.

The formalism can also be applied to arbitrary backgrounds and we have
given general expressions for the overlap reduction functions of all
possible modes. For PTAs, these expressions simplify considerably,
being a sum of products of ordinary spherical harmonics, and being
independent of any curl correlation coefficients.  We have also shown
how anisotropic, unpolarised and uncorrelated backgrounds can be represented in this
formalism, and how the results derived here can be used to write down
overlap reduction functions for arbitrary backgrounds of that type,
extending the analytical results in~\cite{Mingarelli:2013} to orders
above quadrupole.  Finally, we described a formalism that can be used
to reconstruct a map of the gravitational-wave sky for that part of
the background, spanned by the gradient modes, that is visible to a PTA. A PTA
containing $N$ pulsars measures two real-valued quantities 
(an amplitude and a
phase or a sine and a cosine quadrature) at each frequency with each
pulsar.  With $2N$ measurements we cannot hope to measure more than
$2N$ components of the background. These are $2N$ combinations of the components of the 
gradient modes of the background, and we will never be able to detect
curl modes of the background, no matter how many pulsars are included
in the array. This insensitivity to curl modes arises as a consequence
of the fact that the pulsar array is static and does not move or scan
across the sky. To achieve the angular resolution expected for
individually resolvable single sources, we need to probe modes of the
background up to $l_{\rm max} \sim 10$, which will require about $100$
pulsars.  This should be achievable with the SKA.

For the standard case of an isotropic, unpolarised and uncorrelated
background, the formalism described here is more complicated than
directly using the Hellings and Downs curve to model the pulsar pair
cross-correlations. However, it is only marginally more complicated;
the individual overlap reduction functions are simpler and the
sensitivity to isotropic backgrounds should be comparable. The power
of this new approach is that it provides a single unified framework to
characterise any kind of background, free from any assumptions about
its physical nature. We will obtain a map of the background that
encodes as much information as is possible to extract about the
background using the available data. It is no longer necessary to
carry out separate searches for isotropic or anisotropic backgrounds,
but the level of anisotropy can be determined by looking at the
resulting map. It is not necessary to assume that the background is
uncorrelated between different sky locations; instead the degree of
correlation will be directly measured, which will provide sensitivity
to possible new and unexpected physics. If the observed
gravitational-wave background is found to show significant
correlations between different points on the sky, it would be a very
profound discovery. A gravitational-wave background generated by a
superposition of astrophysical sources will not be correlated on the
sky as the properties of each source, including the formation time and
phase, will be independent from one another. While non-standard
scenarios of primordial background formation could in principle
generate correlations in the background, in standard scenarios the
nanohertz stochastic background is expected to be uncorrelated. An
observation of a correlated nanohertz background would therefore be
startling and necessitate serious theoretical work to develop a
plausible mechanism to explain it. Although such an outcome is perhaps
unlikely, the small additional cost of characterising the background
in this way seems very worthwhile, given the large potential reward if
something unexpected is discovered.

\subsection{Implications of a correlated background}
\label{s:implications}

We will now briefly consider the hypothetical implications of a
correlated background for the pulsar timing residuals. To define the
correlation between gravitational waves coming from two different
points on the sky we need to relate the polarisation axes at the two
points, since these define the distinction between the plus and cross
polarisations. A natural way to do this is to relate the axes by
parallel transport. There is a unique geodesic (a great circle)
linking any two points on the sky. Starting at the point with
direction $\hat{k}$ for which the polarisation axes are $\hat l$, $\hat m$ 
we can define new coordinates for the
second point, $\hat{k}'$, by taking $\hat{k}$ to be along the $z$-axis
and $\hat{l}$ to be along the $x$-axis. If in this coordinate system
$\hat{k}'$ has spherical polar coordinates $(\theta' ,\phi')$ then the
great circle connecting $\hat{k}$ to $\hat{k}'$ is the constant
azimuth line $\phi=\phi'$. It can be easily seen that parallel
transport of a vector $v^a$ on the sphere around a line of constant
azimuth keeps both $v^\theta$ and $\sin\theta\, v^\phi$ constant. The
natural polarisation axes at $\hat{k}'$ are therefore
\begin{widetext}
\begin{equation}
\hat{l}' = \left(\begin{array}{c}
\cos^2\phi'\cos\theta'+\sin^2\phi'\\ 
\cos\phi'\sin\phi'(\cos\theta'-1)\\
-\cos\phi'\sin\theta'\end{array}\right)\,, 
\qquad 
\hat{m}' = \left(\begin{array}{c}  
\cos\phi'\sin\phi'(\cos\theta'-1)\\
\sin^2\phi'\cos\theta'+\cos^2\phi'\\
-\sin\phi'\sin\theta'\end{array}\right)\,.
\end{equation}
In the CMB literature, the standard approach is to write down
correlation functions for axes aligned along the line joining the two
points, which is a great circle as above. It is easy to see that this
is equivalent to the parallel transport approach. With this definition
to relate the polarisation axes and additionally making the specific
choice that the $\hat{x}$-axis is aligned along the line joining the
two points, it is possible to compute the expected correlation between
the polarisation states of the gravitational waves at different sky
positions. Using Eq.~(\ref{e:acorr}) we find~\cite{KKS:1997}
\begin{align}
\langle h_+(f,\hat{k}) \,\, h^*_+(f',\hat{k}') \rangle_k 
&= \frac{1}{2}\sum_{l=2}^\infty \frac{2l+1}{4\pi}(N_l)^2 
\left[ C_l^{GG}(f)G_{l2}^+ (\cos\theta) + C_l^{CC}(f)G_{l2}^-(\cos\theta)\right]
\delta(f-f')\,, 
\label{e:ppcorr}
\\
\langle h_\times(f,\hat{k}) \,\, h^*_\times(f',\hat{k}') \rangle_k 
&= \frac{1}{2}\sum_{l=2}^\infty \frac{2l+1}{4\pi}(N_l)^2 
\left[ C_l^{CC}(f)G_{l2}^+ (\cos\theta) + C_l^{GG}(f)G_{l2}^-(\cos\theta)\right] 
\delta(f-f')\,,
\label{e:cccorr}
\\
\langle h_+(f,\hat{k}) \,\, h^*_\times(f',\hat{k}') \rangle_k 
&= \frac{1}{2}\sum_{l=2}^\infty \frac{2l+1}{4\pi}(N_l)^2 
\left[ C_l^{GC}(f)G_{l2}^+ (\cos\theta) - C_l^{CG}(f)G_{l2}^-(\cos\theta)\right] 
\delta(f-f')\,,
\\
\langle h_\times(f,\hat{k}) \,\, h^*_+(f',\hat{k}') \rangle_k 
&= \frac{1}{2}\sum_{l=2}^\infty \frac{2l+1}{4\pi}(N_l)^2 
\left[ C_l^{CG}(f)G_{l2}^+ (\cos\theta) - C_l^{GC}(f)G_{l2}^-(\cos\theta)\right]  
\delta(f-f')\,,
\label{e:twopointfnc}
\end{align}
where the subscript $k$ indicates we are defining the axes by parallel
transport of the polarisation axes at $\hat{k}$ to $\hat{k}'$.

If we require the background to be statistically unpolarised%
\footnote{By ``statistically unpolarised'' we mean that
the expectation values $\langle h_+(f,\hat{k}) \,\,
h^*_+(f',\hat{k}') \rangle_k$ and $\langle h_\times(f,\hat{k}) \,\,
h^*_\times(f',\hat{k}') \rangle_k $ are equal. Any particular
realisation of the background will be polarised at each point (see
Figure~\ref{f:comparison_polarization}), but that polarisation will
have no preferred direction when averaged over the sky.} 
we must
impose the constraint that the correlation functions are independent
of the choice of polarisation axes. Parallel transport is angle
preserving, so if the axes at $\hat{k}$ are rotated by an angle $\psi$,
the same is true for the parallelly-transported axes at
$\hat{k}'$. Such a rotation transforms the plus and cross
polarisations as
\begin{equation}
\bar{h}_+ = h_+\,\cos2\psi + h_\times\,\sin 2\psi, \qquad
\bar{h}_\times = -h_+\,\sin2\psi + h_\times\,\cos 2\psi\,,
\end{equation}
with the corresponding effect on the correlation function
%
%
%
\begin{multline}
\langle \bar{h}_+(f,\hat{k}) \,\, \bar{h}^*_+(f',\hat{k}') \rangle_k 
=\langle h_+(f,\hat{k}) h_+^*(f',\hat{k}') \rangle_k \cos^2 2\psi 
+\langle h_\times(f,\hat{k}) h_\times^*(f',\hat{k}') \rangle_k \sin^2 2\psi 
\\
+ \left( 
\langle h_+(f,\hat{k}) h_\times^*(f',\hat{k}') \rangle_k 
+\langle h_\times(f,\hat{k}) h_+^*(f',\hat{k}') \rangle_k 
\right) \sin 2\psi \cos 2\psi\,.
\end{multline}
Requiring this to be unchanged we must have 
\be
\begin{aligned}
\langle h_+(f,\hat{k}) \,\, h^*_+(f',\hat{k}') \rangle_k 
&= \langle h_\times(f,\hat{k}) \,\, h^*_\times(f',\hat{k}') \rangle_k\,, 
\\
\langle h_+(f,\hat{k}) \,\, h^*_\times(f',\hat{k}') \rangle_k 
&= - \langle h_\times(f,\hat{k}) \,\, h^*_+(f',\hat{k}') \rangle_k\,,
\label{pxcond}
\end{aligned}
\ee
which imposes the restrictions $C_l^{GG}(f)=C_{l}^{CC}(f)\equiv
C_l(f)$ and $C_l^{GC}(f)=-C_l^{CG}(f)$ for statistically unpolarised
backgrounds.  But since $C_l^{GC}(f) = 0 = C_l^{CG}(f)$ follows from
invariance under parity transformations for a statistically isotropic
background (as shown in Sec.~\ref{s:statiso}), the above expectation
values simplify further to:
\begin{align}
\langle h_+(f,\hat{k}) \,\, h^*_+(f',\hat{k}') \rangle_k 
&=\langle h_\times(f,\hat{k}) \,\, h^*_\times(f',\hat{k}') \rangle_k
\nonumber\\
&=\frac{1}{2}\sum_{l=2}^\infty \frac{2l+1}{4\pi}(N_l)^2 C_l(f)
\left[G_{l2}^+ (\cos\theta) + G_{l2}^-(\cos\theta)\right]
\delta(f-f')\,,
\\
\langle h_+(f,\hat{k}) \,\, h^*_\times(f',\hat{k}') \rangle_k 
&=
\langle h_\times(f,\hat{k}) \,\, h^*_+(f',\hat{k}') \rangle_k
= 0\,.
\end{align}
\end{widetext}
We note that it is also possible to have a statistically polarised but
isotropic background which would be characterised by
$C_l^{GG}(f) \neq C_l^{CC}(f)$.  The correlations computed from
Eqs.~(\ref{e:ppcorr}) and (\ref{e:cccorr}) would then no longer be
equal, but this is perfectly consistent as these correlation functions
are referred to the special choice of axes with the $\hat{x}$
direction along the line joining the two pulsars.  Since PTAs are
insensitive to the curl modes of the background we will not be able to
identify statistically polarised backgrounds using pulsar timing
arrays. As a final comment, we note that, whether statistically
polarised or not, the individual modes in this decomposition describe
a background that has non-trivial correlations in the emission from
different sky locations, which would be a startling discovery as
discussed above.

Another interesting feature of a correlated background is in the
behaviour of the pulsar term. As discussed in
Sec.~\ref{sec:detresponse}, for uncorrelated backgrounds in the limit
$fL/c\gg1$, the pulsar term averages to zero except for the pulsar
auto-correlation, for which the pulsar term contributes an equal
amount as the Earth term, increasing the total correlation by a factor
of $2$. For a correlated background, by contrast, the pulsar term
always averages to zero, even for the
auto-correlation. Mathematically, this is because for each individual
mode we evaluate the product of two integrals and this reduces to the
squared magnitude of the integral in the auto-correlation limit. The
integral over the sky of the pulsar term gives zero in the limit
$fL/c\rightarrow \infty$, so there is no contribution from this term
to the final result. Physically, the factor of 2 in the case of an
uncorrelated background arises from the fact that the correlation in the pulsar residuals
is caused by the same gravitational waves being seen by both
pulsars. When the pulsar is the same, the pulsar term in a given
gravitational wave adds coherently, while they interfere destructively
for different pulsars. In the correlated background case, the
correlation is coming from the interaction between gravitational waves
propagating in different directions, but with correlated phases. The
pulsar terms for the same pulsar but from different sky positions do
not add coherently because of the $(1 + \hat{k} \cdot \hat{u})$ factor
and therefore there are no correlated contributions from the pulsar
terms in that case.

\subsection{Distinguishing a correlated background -- 
overlap reduction functions}

\begin{figure*}[htbp]
\begin{center}
\includegraphics[width=.6\textwidth]{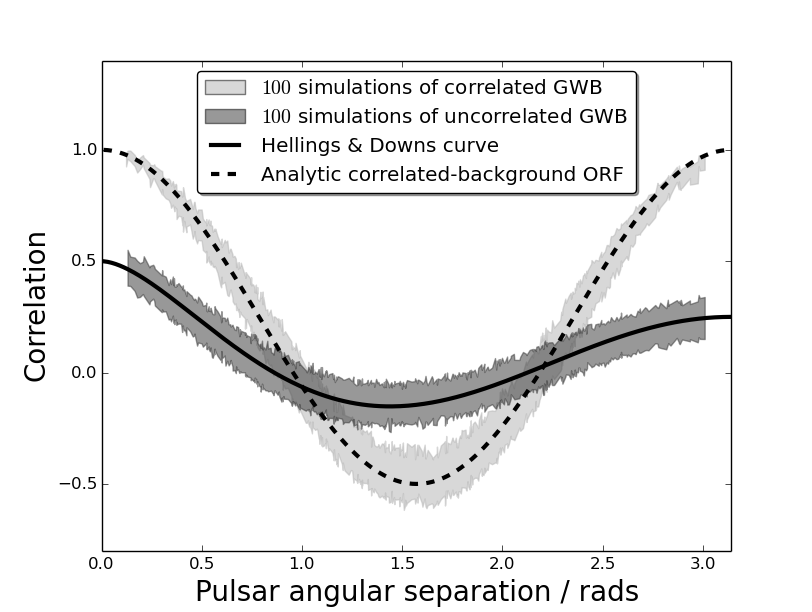}
\caption{The average correlation between pairs of pulsars (and errors), as a
  function of pulsar angular separation, for an unpolarised, 
  $l_{\rm max}=2$ correlated 
  gravitational-wave background (light-grey region) and for an
  isotropic, unpolarised and 
  uncorrelated gravitational-wave background (dark-grey region). The averages
  were computed over 100 realisations of the backgrounds. Also shown
  are the analytic correlation functions expected in the two cases:
  the former being proportional to the Legendre polynomial $P_2(\cos\zeta)$,
  and the latter being the familiar Hellings and Downs curve.}
\label{f:l2correx}
\end{center}
\end{figure*}

We have written a {\sc Tempo2} plugin to generate an arbitrary
gravitational-wave background by prescribing the gradient and curl
spherical harmonic coefficients, $\{a^G_{(lm)}, a^C_{(lm)}\}$.  The
plugin populates the sky with a large number of gravitational-wave
sources, whose complex amplitudes are chosen to be consistent with the
specified coefficients. This can be used to generate both uncorrelated
and correlated backgrounds. As a test of the plugin, in
Fig.~\ref{f:l2correx} we show the average correlation (computed over
100 realisations of the background) for distinct pulsar pairs within a
chosen ensemble of pulsars. The datasets assume an array of $32$
pulsars spread uniformly over the sky, and observed fortnightly for
$5$ years. The injected background power-spectrum is flat, creating a
correlated white-noise influence in pulsars separated across the
sky. Results are shown for both an unpolarised, $l_{\rm max}=2$
correlated background and an isotropic, unpolarised and uncorrelated background. 
The expected
analytic result is shown in both cases as well, normalised such that
the correlation for pulsars with zero angular separation is equal to
$1$. We see that the analytic results agree perfectly with the
numerical calculations, and the scatter in the two cases is
approximately the same.

The distinguishability of the two types of backgrounds can be assessed
using Bayesian evidence. We took one of the correlated-background
dataset realisations used in Fig.\ \ref{f:l2correx} (analysing a
sub-array of $10$ pulsars), and computed the Bayesian evidence within
a time-domain likelihood formalism for models which assume that the
background is correlated or uncorrelated. This gave a log
evidence-ratio of $\sim 500$ in favour of the correlated background,
showing that the two types of background could be distinguished with
high confidence. More work is needed to determine the threshold 
signal-to-noise ratio of
the background that is needed to distinguish between the different
models. The background considered here was very loud, and would have
been easily detectable regardless of the correlation signature.

\begin{figure*}[htbp]
\begin{center}
\includegraphics[angle=0, width=0.6\textwidth]{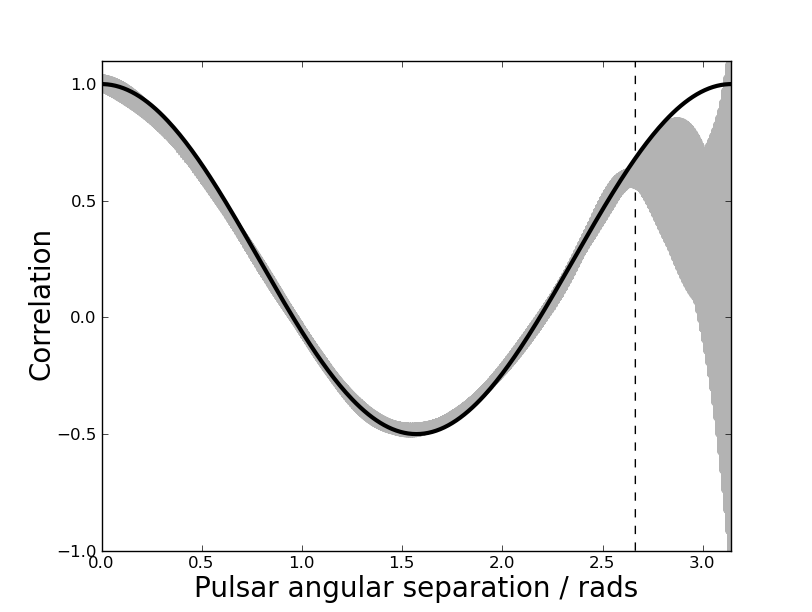}
\caption{A model-independent Bayesian reconstruction of the
correlation between pairs of pulsars, as a function of pulsar
angular separation, for a dataset with an injected unpolarised,
$l_{\rm max}=2$ correlated gravitational-wave background. 
The background spectrum is white,
with the signal injected into $10$ pulsars which are observed
fortnightly over a timespan of $5$ years. The correlation is
parametrised at $13$ distinct separations, and a cubic-spline
interpolation used to compute the correlation at all other angular
separations. The grey region shows the
envelope of splines within the $95\%$ credible interval of the full
recovered posterior distribution, while the solid black curve is the
expected analytic correlation function. The dashed black line indicates the
largest angular separation between pulsars in our chosen array.
As expected, beyond this line our reconstruction completely loses
sensitivity.}
\label{f:CorrRecon}
\end{center}
\end{figure*}

Using this same dataset, we then performed a model-independent
fully-Bayesian reconstruction of the background-induced
overlap reduction function. This technique was developed within the
context of the first IPTA Data
Challenge~\cite{IPTAmdc} to confirm that an isotropic distribution of
uncorrelated astrophysical gravitational-wave sources 
will induce the distinctive Hellings and Downs signature 
in a cross-correlation analysis~\cite{SteveIPTAchallenge2013}. 
The overlap reduction function is
parametrised at $13$ distinct pulsar angular-separations, and a
cubic-spline interpolation used to compute the correlation at all
other angular separations. The $13$ ``anchor'' values of the
overlap-reduction function are sampled, giving a posterior probability
distribution that allows us to map out an envelope of all
cubic-splines which lie within desired credible intervals. Figure
\ref{f:CorrRecon} shows this in practice, where the grey-region
defines the envelope of splines within the $95\%$ credible interval of
the full posterior probability distribution, the solid black line
shows the expected overlap reduction function, and the dashed black
line indicates the largest pulsar angular-separation in this dataset
(beyond this we lose sensitivity). We see that not only can we
distinguish between two alternative models, but we can directly
measure the correlation function in the case that the background is
correlated.

\subsection{Distinguishing a correlated background -- sky maps}

Sky maps of $h_+$ and $h_\times$ 
will also allow us to distinguish between correlated and uncorrelated 
backgrounds, provided the angular resolution of the maps is finer 
than the angular scale of the correlated background.  For coarser 
angular resolution, the correlations between neighboring sky directions 
would not be apparent in the maps, as the size of the pixels would 
be larger than the angular correlation scale of the background.
Sky maps of gravitational-wave power,
$P = |h_+|^2 + |h_\times|^2$, lack some of the discriminating 
capability of sky maps of $h_+$ and $h_\times$, 
since the phase relation between the different components of
the waves is lost in the construction of the power map.

One problem that we encounter when making sky maps is that we measure 
only a single realisation of the background, and thus do not have 
access to a statistical ensemble of such backgrounds.  This means 
that even if the background is statistically isotropic, any particular 
realisation will be anisotropic, having random hot and cold 
patches at different locations on the sky.
Although one can calculate a set of $C_l$'s for a correlated
background that is consistent with the observed power on the sky (as
they do for the CMB), one could equally-well regard the power
distribution as fundamental and decompose it in terms of ordinary
(i.e., scalar) spherical harmonics, effectively assuming it to be an
anisotropic uncorrelated background, as in \cite{Mingarelli:2013}.
Although the latter approach is a way to describe any power
distribution on the sky, the approach we adopt in this paper is more
generic since it also incorporates the phase information of the waves
at different points on the sky.

\begin{figure*}
\begin{center}
\includegraphics[trim=3cm 6.5cm 3cm 3.5cm, clip=true, angle=0, width=0.4\textwidth]{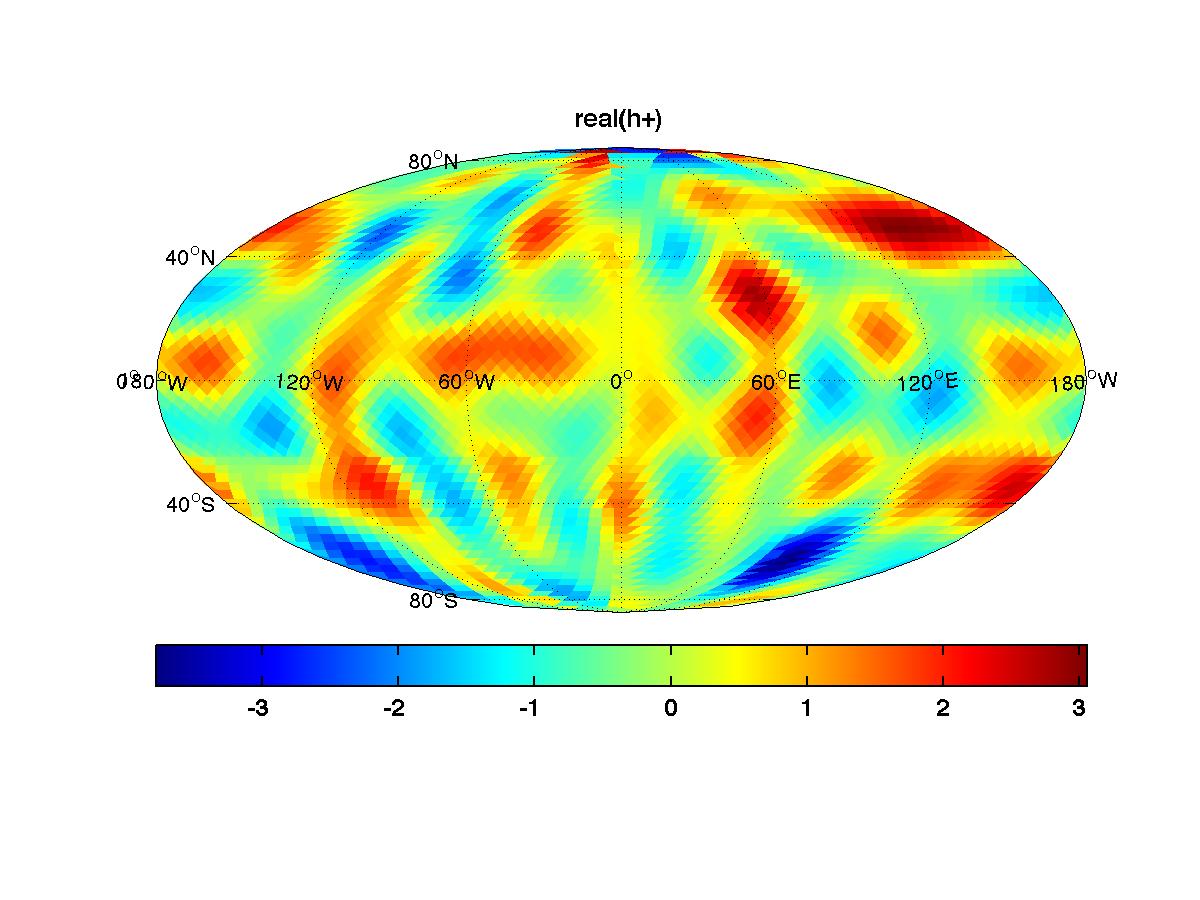}
\hspace{0.05\textwidth}
\includegraphics[trim=3cm 6.5cm 3cm 3.5cm, clip=true, angle=0, width=0.4\textwidth]{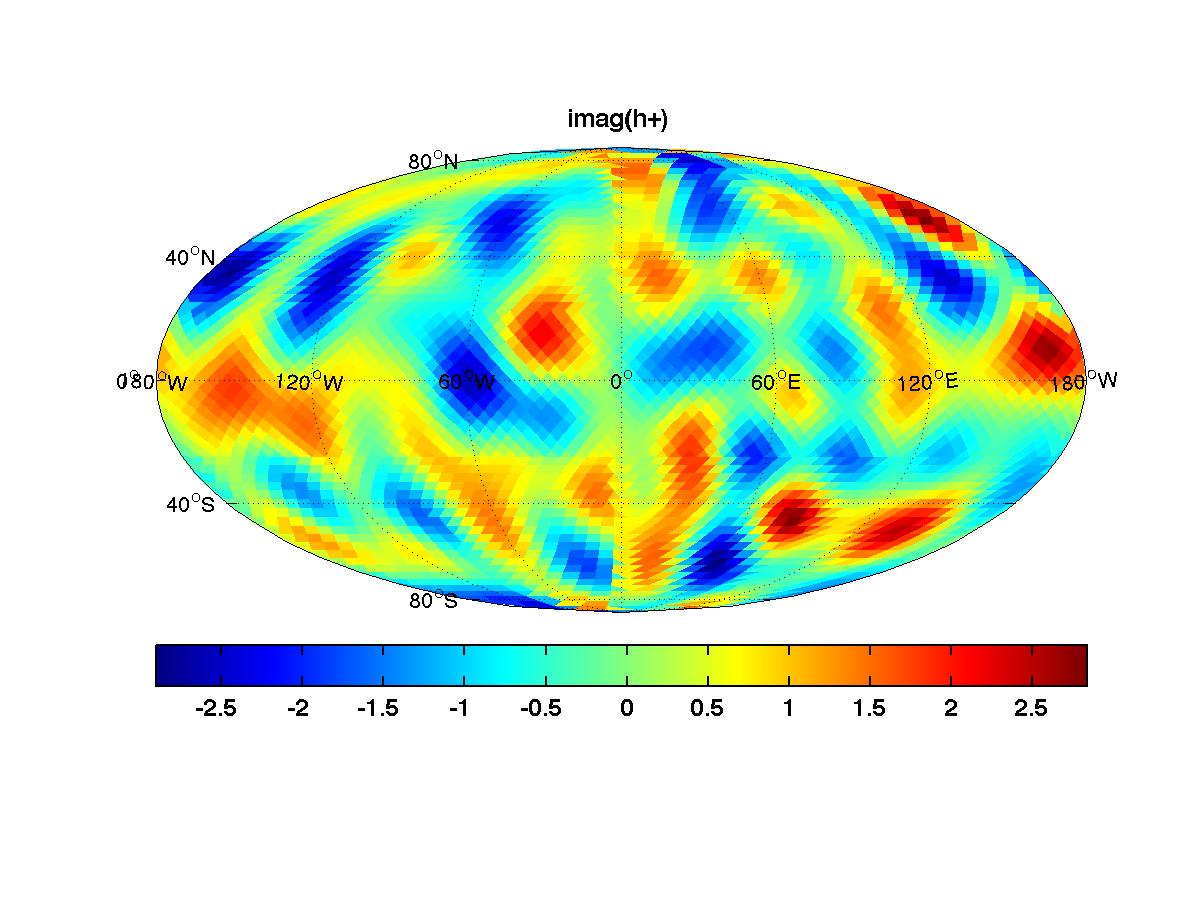}
\includegraphics[trim=3cm 6.5cm 3cm 3.5cm, clip=true, angle=0, width=0.4\textwidth]{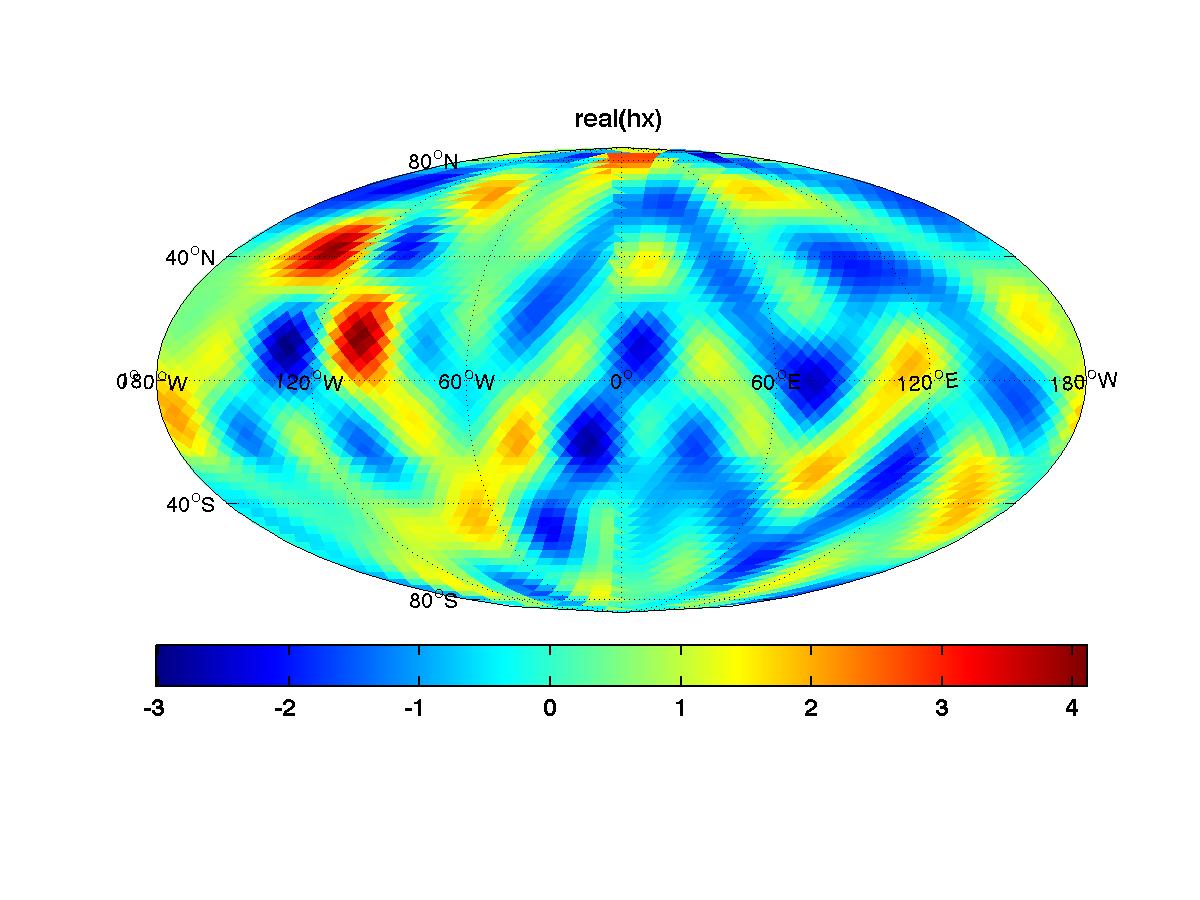}
\hspace{0.05\textwidth}
\includegraphics[trim=3cm 6.5cm 3cm 3.5cm, clip=true, angle=0, width=0.4\textwidth]{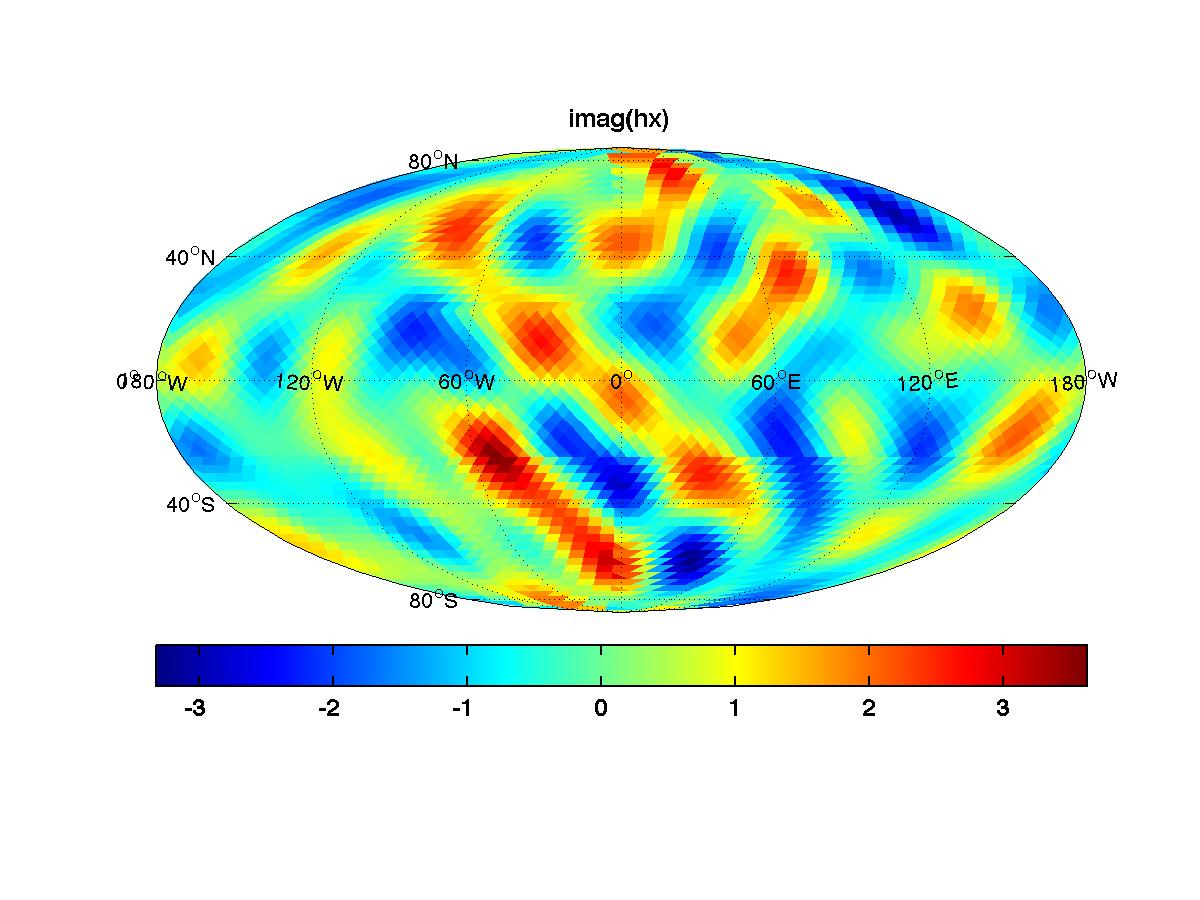}
\includegraphics[trim=3cm 6.5cm 3cm 3.5cm, clip=true, angle=0, width=0.4\textwidth]{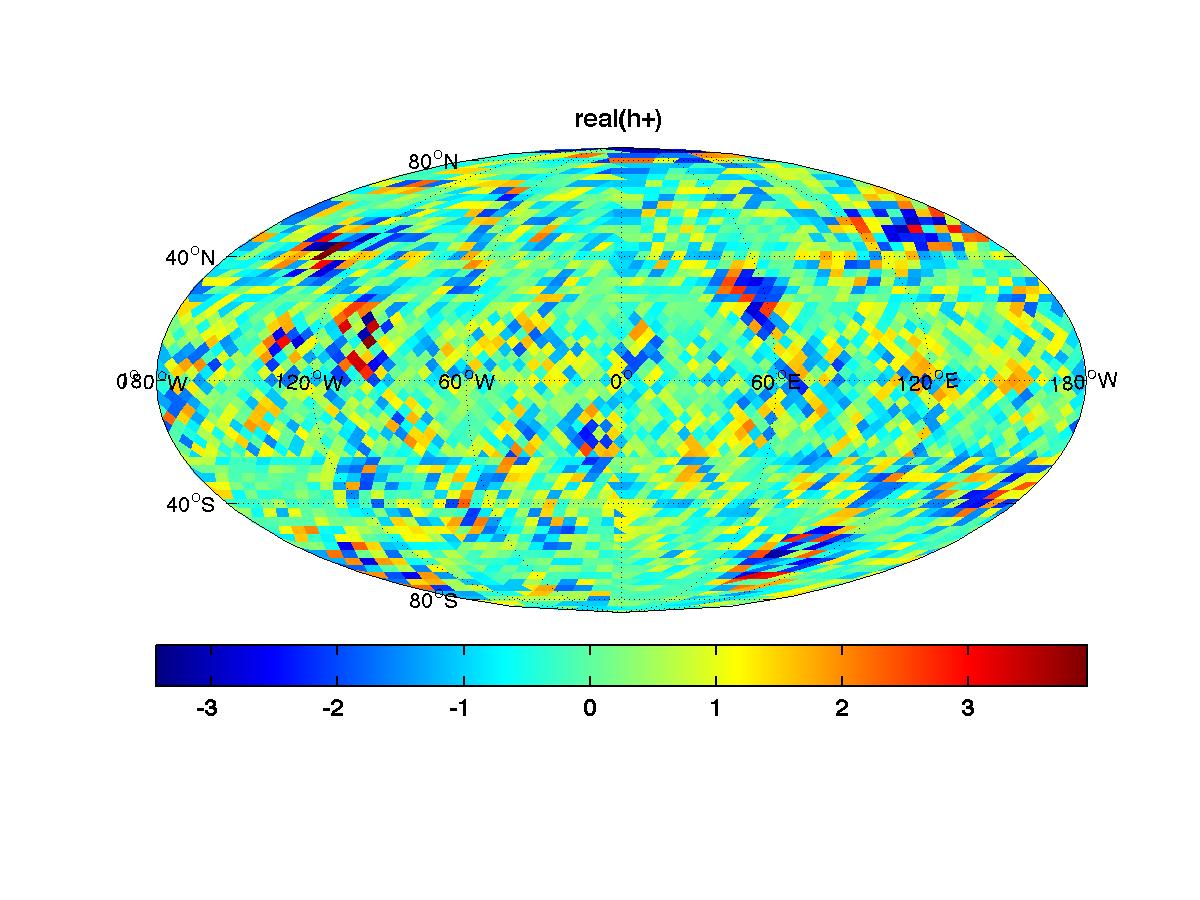}
\hspace{0.05\textwidth}
\includegraphics[trim=3cm 6.5cm 3cm 3.5cm, clip=true, angle=0, width=0.4\textwidth]{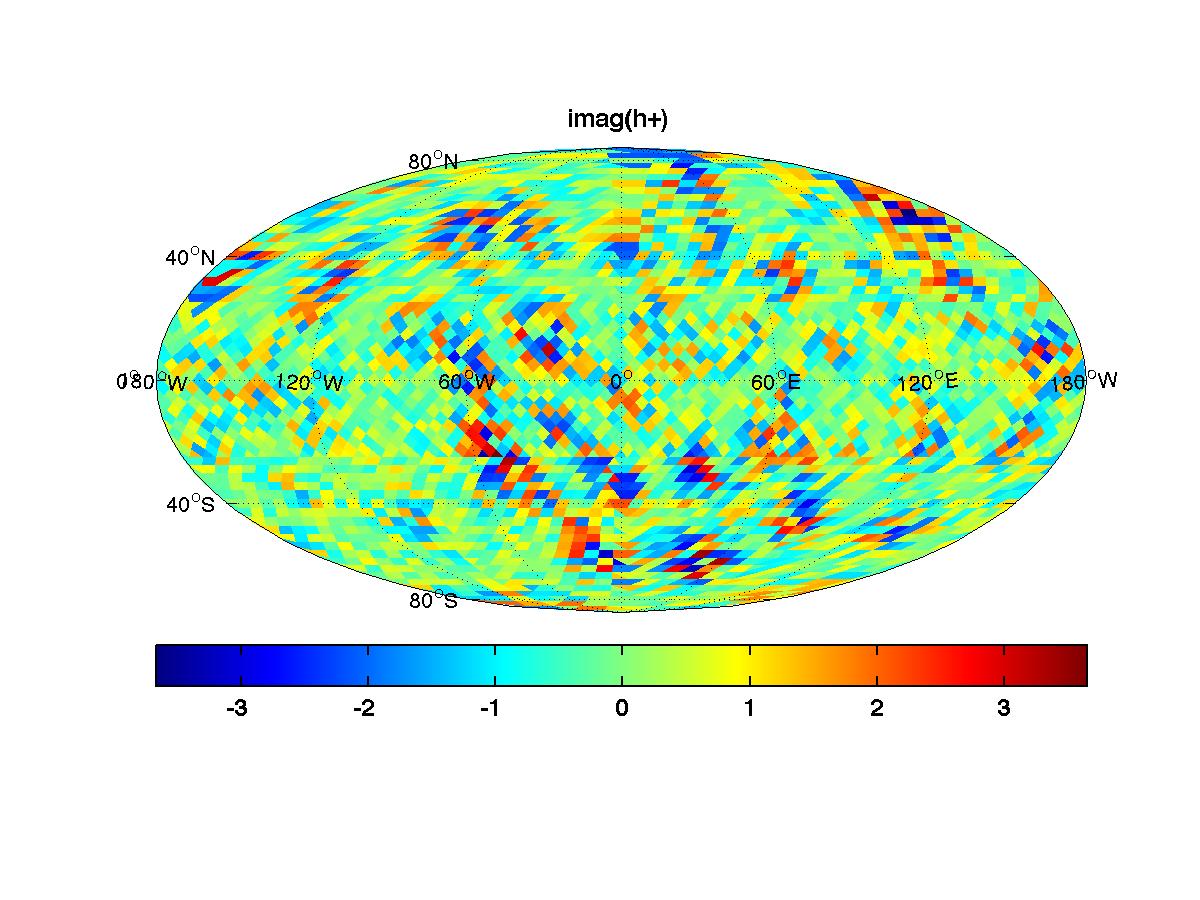}
\includegraphics[trim=3cm 6.5cm 3cm 3.5cm, clip=true, angle=0, width=0.4\textwidth]{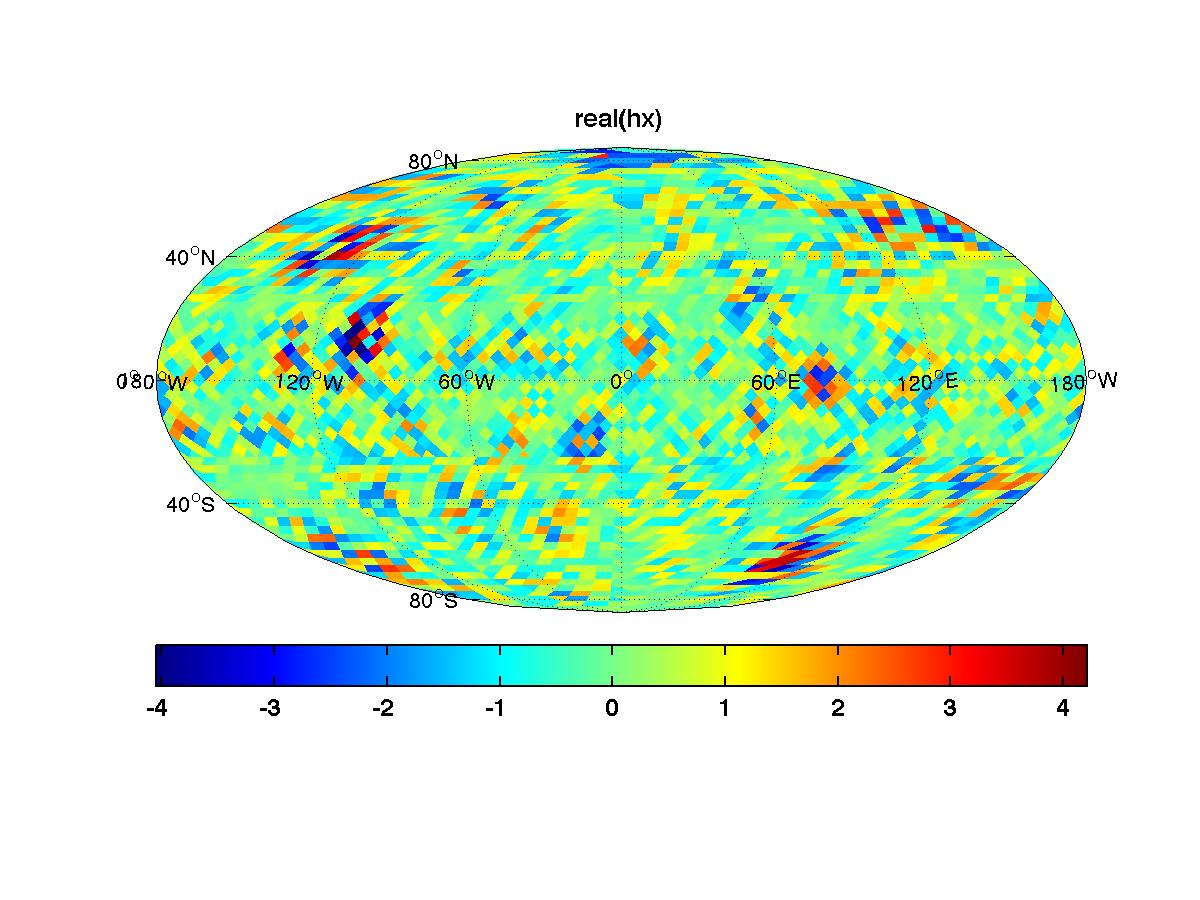}
\hspace{0.05\textwidth}
\includegraphics[trim=3cm 6.5cm 3cm 3.5cm, clip=true, angle=0, width=0.4\textwidth]{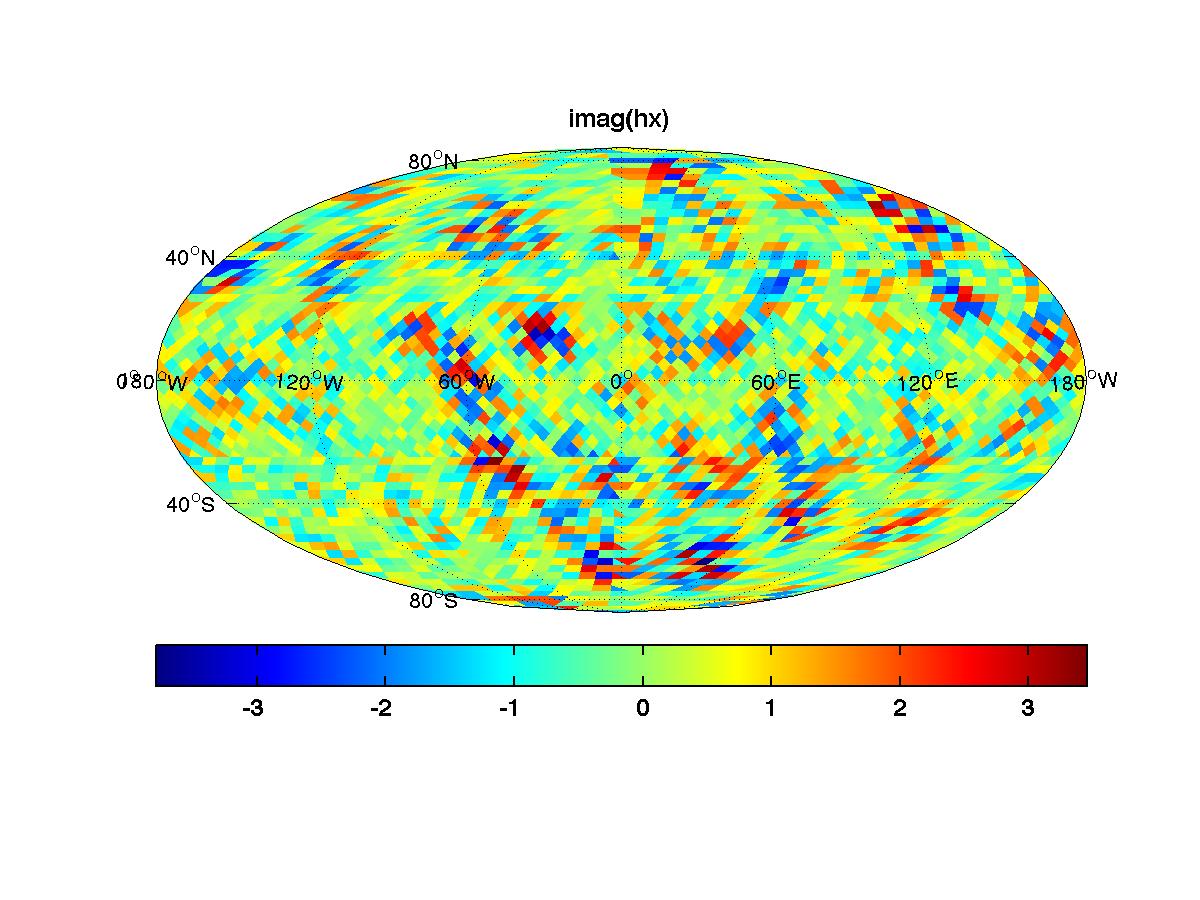}
\caption{Mollweide projections of the real and imaginary parts of 
$h_+$ and $h_\times$ for:
(a) a grad-only statistically isotropic correlated background 
with $C_l=1$ out to $l_{\rm max}=10$ (top four plots), and
(b) an anisotropic uncorrelated background that has the same power 
distribution as the statistically isotropic background but 
uncorrelated phase (bottom four plots).
Note the pixel-to-pixel variation in the maps for the uncorrelated background.}
\label{f:statiso-vs-aniso}
\end{center}
\end{figure*}

\begin{figure*}
\begin{center}
\includegraphics[trim=3cm 6.5cm 3cm 3.5cm, angle=0, width=0.5\textwidth]{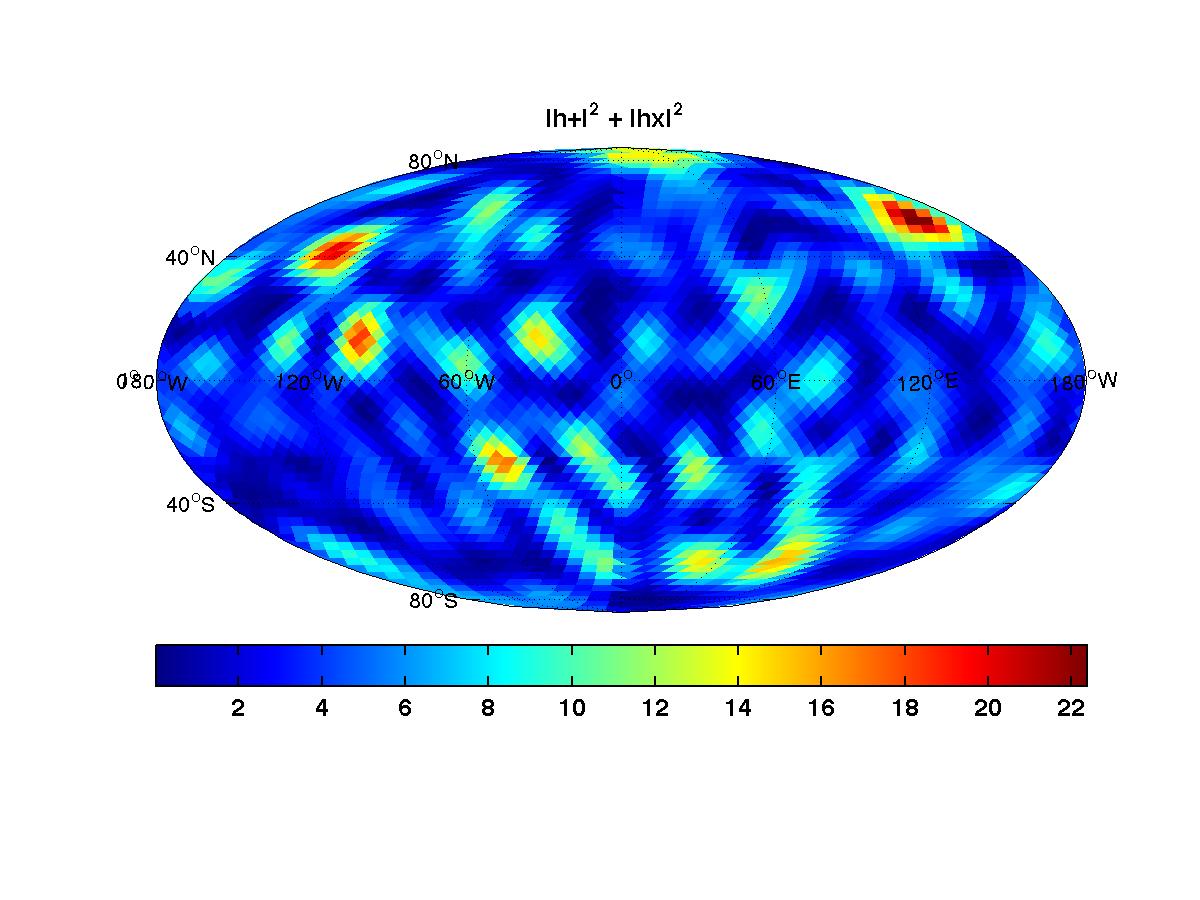}
\caption{Mollweide projection of the gravitational-wave power 
on the sky, $|h_+|^2 + |h_\times|^2$, for both the 
statistically isotropic correlated background and the anisotropic 
uncorrelated background shown in Fig.~\ref{f:statiso-vs-aniso}.}
\label{f:power}
\end{center}
\end{figure*}

To illustrate these statements, we have simulated two gravitational-wave
backgrounds that have exactly the same power distribution but different 
correlation properties:
(a) a grad-only statistically isotropic correlated 
background with angular power $C_l=1$ for $l=2,3,\cdots, l_{\rm max}$, 
where $l_{\rm max}=10$, and 
(b) an anisotropic uncorrelated background whose power distribution 
is the same as that of the statistically isotropic background but 
has uncorrelated phase. 
(Both of these simulated backgrounds are statistically unpolarised.)
The statistically isotropic background is the same as the 
gradient component of the simulated background that we analysed 
in Sec.~\ref{s:example_recovery}.
(We consider only the gradient component for this discussion, 
since PTAs are insensitive to the curl component.)  
The anisotropic uncorrelated background is constructed from the 
statistically isotropic background by randomly rotating the 
polarisation angle of the waves at each pixel on the sky.
Maps of the real and imaginary parts of $h_+$ and $h_\times$
for both types of backgrounds are shown in Fig.~\ref{f:statiso-vs-aniso}.
A sky map of the gravitational-wave power, 
$P=|h_+|^2 + |h_\times|^2$, which is 
the same for both backgrounds, is shown in Fig.~\ref{f:power}.

The size of the pixels in these sky maps is the same as that 
used for the maps in Secs.~\ref{sec:detresponse} and 
\ref{s:example_recovery}, having an angular 
resolution $\Delta\Omega\sim 50~{\rm deg}^2$.  
This angular resolution 
corresponds to $l\sim 25$, which is substantially larger than the
$l_{\rm max}\ (=10)$ used for the simulated statistically isotropic 
correlated background.  
To actually resolve individual pixels of this size using timing
residual measurements would require a PTA with $N\gtrsim 700$ pulsars, 
which is the number of modes corresponding to $l_{\rm max}\sim 25$.
This particular example is thus relevant for a PTA from the 
advanced SKA era. 
But the same qualitative conclusions that we will make here can 
also be made for a less ambitious PTA having $N\gtrsim 100$ 
pulsars and $l\sim 10$, provided $l_{\rm max}$ for the correlated 
background is sufficiently small (e.g., $l_{\rm max}\lesssim 5$) 
in order to ensure that the angular scale over which the phase is 
correlated is resolved by the measurements.

\begin{figure*}
\begin{center}
\subfigure[\ Statistically isotropic background]
{\includegraphics[angle=0, width=0.49\textwidth]{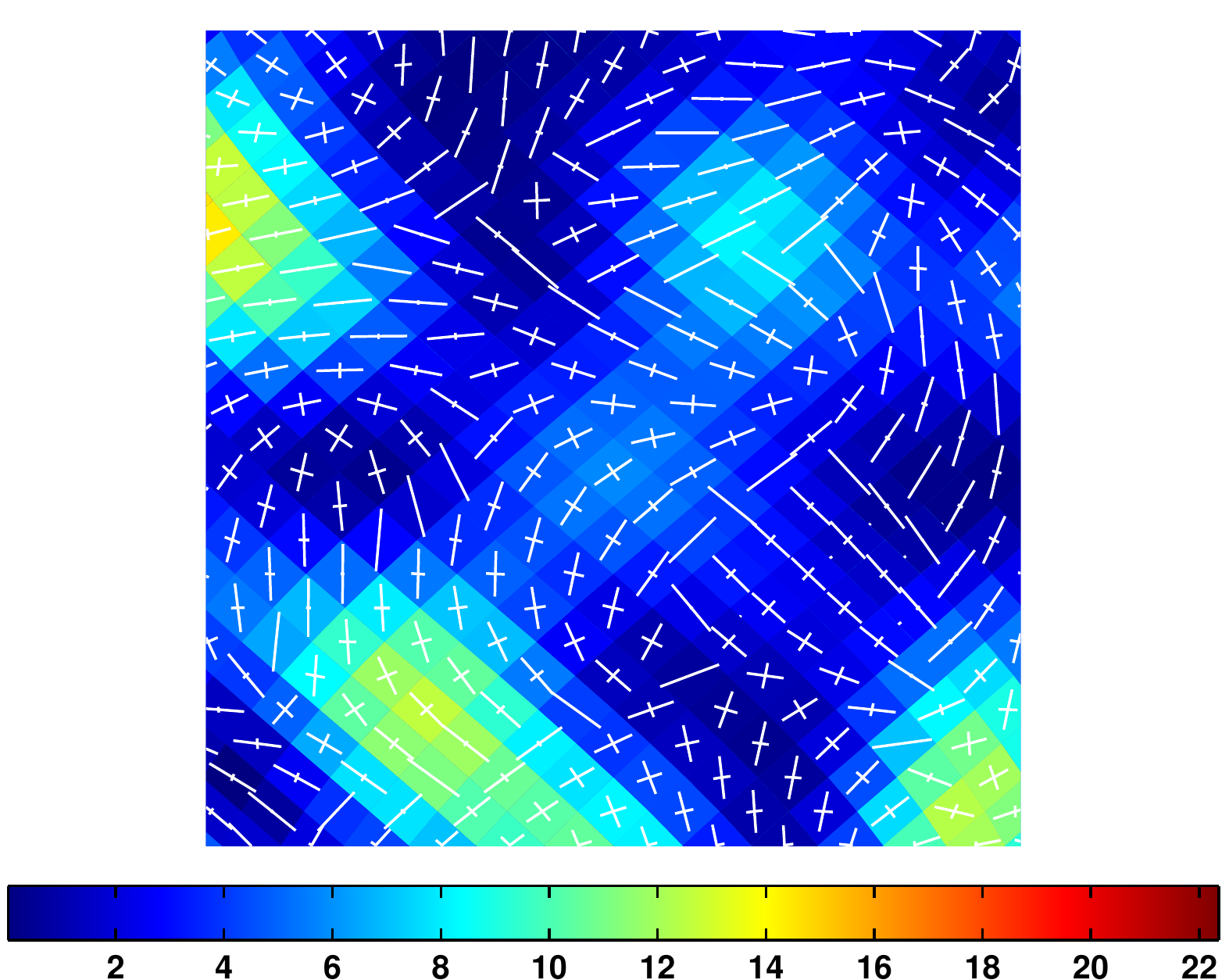}}
\subfigure[\ Anisotropic uncorrelated background]
{\includegraphics[angle=0, width=0.49\textwidth]{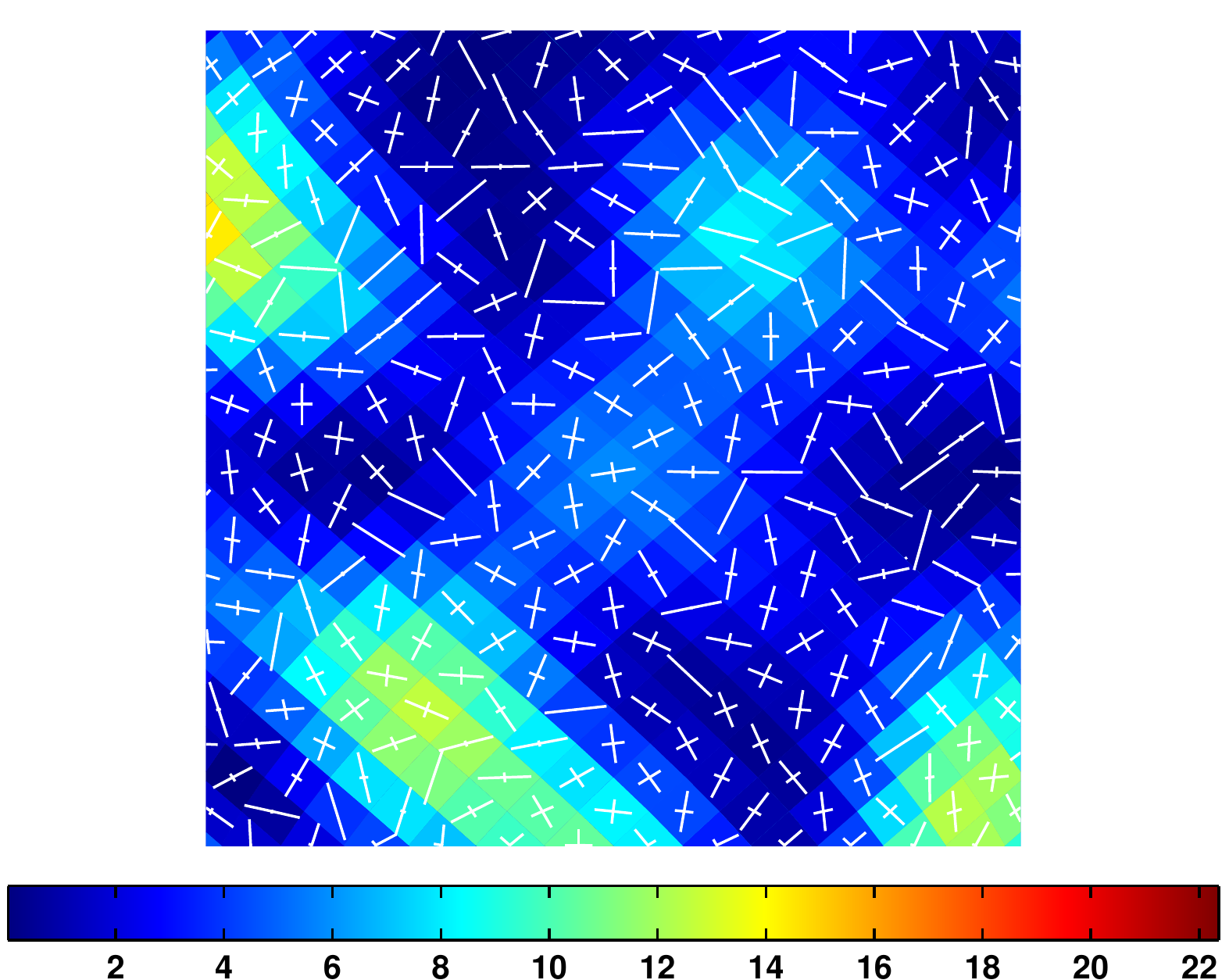}}
\caption{Polarisation ellipses for a patch of sky (centered at $0^\circ$ latitude and
$0^\circ$ longitude) for:
(a) the statistically isotropic correlated background, and 
(b) the anisotropic uncorrelated background 
shown in Fig.~\ref{f:statiso-vs-aniso}.
The polarisation ellipses are superimposed on the (common) map of 
gravitational-wave power shown in Fig.~\ref{f:power}.
The crosses indicate the direction and principal axes of the $h_+$, $h_\times$
polarisation ellipses.  
Linear polarisation in a particular direction is represented by a line; 
circular polarisation by a cross with equal-length axes.
Note that the polarisation ellipses vary smoothly over the sky for the 
statistically isotropic correlated background, as compared to that for the 
anisotropic uncorrelated background, which have randomly oriented phase angles.}
\label{f:comparison_polarization}
\end{center}
\end{figure*}

The additional information contained in the phase of $h_+$ and $h_\times$ 
at each pixel can also be represented graphically by plotting 
polarisation ellipses for the two different backgrounds, as
shown in Fig.~\ref{f:comparison_polarization}.
The polarisation ellipses are constructed as follows, using an
approach similar to that for electromagnetic waves, see, e.g., 
Sec.~7.2 of \cite{jackson}.  Given the complex values of $h_+$ and
$h_\times$ at each pixel for the two backgrounds, we first calculate
the left- and right-circularly polarised combinations $h_{L,R} \equiv
(h_+\pm i h_\times)/\sqrt{2}$.  If $|h_L/h_R|<1$, the polarisation
ellipse has semi-major and semi-minor axes $a=1+r$, $b=1-r$, where
$r=|h_L/h_R|$, and it is rotated by an angle $\psi\equiv\alpha/2$ with
respect to the horizontal (i.e., $\theta={\rm const}$), where
$\alpha={\rm phase}(h_L/h_R)$.  If $|h_L/h_R|>1$, we need to switch
$h_L$ and $h_R$ in the above expressions for $r$ and $\alpha$.  The
angle $\psi$ is the polarisation angle.  Circular and linear
polarization are special cases: $r=0$ corresponds to circular
polarisation, represented by a cross with equal-length axes; $r=1$
corresponds to linear polarisation, represented by a line in a
particular direction.  Note that the polarisation ellipses vary
smoothly over the sky for the statistically isotropic correlated 
background, as compared to that for the anisotropic uncorrelated background.
We therefore see that sky maps of $h_+$ and $h_\times$ have the potential 
to yield much more information than a map of just the total 
gravitational-wave power.

\subsection{Outlook}

The framework described here provides a completely generic approach to
mapping the gravitational-wave sky using pulsar timing arrays. Our
results suggest that it can be used to map the expected uncorrelated
gravitational-wave background with almost as much sensitivity as the
standard approach, while providing sensitivity to new physics. Further
work is required to fully assess the computational costs and
sensitivity of the approach, but we expect this will be a useful
framework for future analyses. While we have focused on pulsar timing
arrays in this work, the same approach can also be used to
characterise gravitational-wave backgrounds in other frequency bands,
relevant to ground-based or space-based detectors. We gave the overlap
reduction function for a static interferometer in
Appendix~\ref{s:LIGO_response}. The extension to more realistic moving detectors is more involved but work in this direction is underway.

\acknowledgments JG's work is supported by the Royal Society.  
JDR acknowledges support from NSF Awards PHY-1205585 and CREST
HRD-1242090.  SRT acknowledges the support of the STFC and the RAS.
CMFM acknowledges the support of the RAS, Universitas 21 and a Marie 
Curie International Outgoing Fellowship within the 7th European 
Community Framework Programme.
We thank Sanjit Mitra for initial discussions, and Bruce Allen and Chiara
Caprini for many e-mail exchanges regarding models for correlated
backgrounds. CMFM thanks Gary Hinshaw for pointing her to
Ref.~\cite{Challinor:2000}, and JDR thanks Graham Woan for useful
feedback regarding an earlier draft of the paper.  The authors also
thank the anonymous referee for useful comments on the manuscript.
This research has made use of Python and its standard libraries: numpy
and matplotlib.  We have also made use of MEALPix (a Matlab
implementation of HEALPix~\cite{HEALPix}), developed by the GWAstro
Research Group and available from {\tt http://gwastro.psu.edu}.  This
work was performed using the Darwin Supercomputer of the University of
Cambridge High Performance Computing Service
(http://www.hpc.cam.ac.uk/), provided by Dell Inc.~using Strategic
Research Infrastructure Funding from the Higher Education Funding
Council for England and funding from the Science and Technology
Facilities Council.

\begin{widetext}

\begin{appendix}
\section{Spin-weighted spherical harmonics}
\label{s:spinweightedY}

The following is a list of some useful relations involving spin-weighted
and ordinary spherical harmonics, ${}_sY_{lm}(\hat k)$ and $Y_{lm}(\hat k)$.
For more details, see e.g., 
\cite{Goldberg:1967} and \cite{delCastillo}.
Note that we use a slightly different normalization convention than in
\cite{Goldberg:1967}.
Namely, we put the Condon-Shortley factor $(-1)^m$ in the definition of the
associated Legendre functions $P_l^m(x)$, and thus do not explicitly 
include it in the definition of the spherical harmonics.
Also, for our analysis, we can restrict attention to spin-weighted 
spherical harmonics having {\em integral} spin weight $s$, even though 
spin-weighted spherical harmonics with half-integral spin weight exist.
\\

\noindent
Ordinary spherical harmonics:
\be
Y_{lm}(\hat k)=
Y_{lm}(\theta,\phi) = N_l^m
P_l^m(\cos\theta)e^{im\phi}\,,
\quad
{\rm where}\ 
N_l^m = \sqrt{\frac{2l+1}{4\pi}\frac{(l-m)!}{(l+m)!}}.
\label{e:Nlm}
\ee
Relation of spin-weighted spherical harmonics 
to ordinary spherical harmonics:
\be
\begin{aligned}
{}_sY_{lm}(\theta,\phi) 
&=\sqrt{\frac{(l-s)!}{(l+s)!}}\,
\edth^s Y_{lm}(\theta,\phi)
\quad{\rm for}\quad
0\le s\le l\,,
\\
{}_sY_{lm}(\theta,\phi) 
&=\sqrt{\frac{(l+s)!}{(l-s)!}}\,
(-1)^s
\overline{\edth}{}^{-s} Y_{lm}(\theta,\phi)
\quad{\rm for}\quad
-l\le s\le 0\,,
\end{aligned}
\ee
where
\be
\begin{aligned}
\edth \eta
&=-(\sin\theta)^s
\left[\frac{\partial}{\partial\theta}
+i\csc\theta\frac{\partial}{\partial\phi}\right]
(\sin\theta)^{-s}\eta\,,
\\
\overline{\edth} \eta
&=-(\sin\theta)^{-s}
\left[\frac{\partial}{\partial\theta}
-i\csc\theta\frac{\partial}{\partial\phi}\right]
(\sin\theta)^{s}\eta\,,
\end{aligned}
\ee
and $\eta=\eta(\theta,\phi)$ is a spin-$s$ scalar field.\\

\noindent
%
%
Series representation:
\be
{}_sY_{lm}(\theta,\phi)
=(-1)^m\left[\frac{(l+m)!(l-m)!}{(l+s)!(l-s)!}\frac{2l+1}{4\pi}
\right]^{1/2}(\sin\theta/2)^{2l}
\sum_{k=0}^{l-s}
\binom{l-s}{k}\binom{l+s}{k+s-m}(-1)^{l-k-s} 
e^{im\phi}(\cot\theta/2)^{2k+s-m}.
\ee
Complex conjugate:
\be
{}_sY_{lm}^*(\theta,\phi) = (-1)^{m+s}\,{}_{-s}Y_{l,-m}(\theta, \phi).
\label{e:conj(Y)}
\ee
Relation to Wigner rotation matrices:
\be
D^l{}_{m'm}(\phi,\theta,\psi)=
(-1)^{m'} \sqrt{\frac{4\pi}{2l+1}}\,{}_mY_{l,-m'}(\theta,\phi)e^{-im\psi}
\label{e:WignerD}
\ee
or
\be
\left[D^l{}_{m'm}(\phi,\theta,\psi)\right]^*=
(-1)^{m} \sqrt{\frac{4\pi}{2l+1}}\,{}_{-m}Y_{l,m'}(\theta,\phi)e^{im\psi}.
\label{e:WignerD_CC}
\ee
Parity transformation:
\be
{}_sY_{lm}(\pi-\theta,\phi+\pi) = (-1)^l\,{}_{-s}Y_{lm}(\theta, \phi).
\ee
\noindent
Orthonormality (for fixed $s$):
\be
\int_{S^2} {\rm d}^2\Omega_{\hat k}\>
{}_sY_{lm}(\hat k) \,{}_{s}Y_{l'm'}^*(\hat k)
\equiv\int_0^{2\pi} {\rm d}\phi\int_0^\pi \sin\theta\, {\rm d}\theta\>
{}_sY_{lm}(\theta,\phi) \,{}_sY_{l'm'}^*(\theta,\phi)
= \delta_{ll'}\delta_{mm'}.
\ee
Addition theorem for spin-weighted spherical harmonics:
\be
\sum_{m=-l}^l {}_sY_{lm}(\theta_1,\phi_1)\,{}_{s'}Y_{lm}^*(\theta_2,\phi_2)
=(-1)^{-s'}\sqrt{\frac{2l+1}{4\pi}}\,{}_{-s'}Y_{ls}(\theta_3,\phi_3)
e^{is'\chi_3}
\ee
where
\be
\cos\theta_3 = \cos\theta_1\cos\theta_2 + \sin\theta_1\sin\theta_2\cos(\phi_2-\phi_1)
\ee
and
\be
\begin{aligned}
e^{-i(\phi_3+\chi_3)/2}
&=\frac{\cos\frac{1}{2}(\phi_2-\phi_1)\cos\frac{1}{2}(\theta_2-\theta_1)
-i\sin\frac{1}{2}(\phi_2-\phi_1)\cos\frac{1}{2}(\theta_1+\theta_2)}
{\sqrt{\cos^2\frac{1}{2}(\phi_2-\phi_1)\cos^2\frac{1}{2}(\theta_2-\theta_1)
+\sin^2\frac{1}{2}(\phi_2-\phi_1)\cos^2\frac{1}{2}(\theta_1+\theta_2)}}
\\
e^{i(\phi_3-\chi_3)/2}
&=\frac{\cos\frac{1}{2}(\phi_2-\phi_1)\sin\frac{1}{2}(\theta_2-\theta_1)
+i\sin\frac{1}{2}(\phi_2-\phi_1)\sin\frac{1}{2}(\theta_1+\theta_2)}
{\sqrt{\cos^2\frac{1}{2}(\phi_2-\phi_1)\sin^2\frac{1}{2}(\theta_2-\theta_1)
+\sin^2\frac{1}{2}(\phi_2-\phi_1)\sin^2\frac{1}{2}(\theta_1+\theta_2)}}.
\end{aligned}
\ee
Addition theorem for ordinary spherical harmonics:
\be
\sum_{m=-l}^l Y_{lm}(\hat k_1) Y_{lm}^*(\hat k_2)
=\frac{2l+1}{4\pi}\,P_l(\hat k_1\cdot\hat k_2).
\label{e:addition_theorem_0Y}
\ee
Integral of a product of spin-weighted spherical harmonics:
\be
\int_{S^2} {\rm d}^2\Omega_{\hat k}\>
{}_{s_1}Y_{l_1m_1}(\hat k) \,
{}_{s_2}Y_{l_2m_3}(\hat k) \,
{}_{s_3}Y_{l_3m_3}(\hat k) \,
= \sqrt{\frac{(2l_1+1)(2l_2+1)(2l_3+1)}{4\pi}}
\left( \begin{array}{ccc}l_1&l_2&l_3\\m_1&m_2&m_3\end{array} \right)
\left( \begin{array}{ccc}l_1&l_2&l_3\\-s_1&-s_2&-s_3\end{array} \right)
\ee
where 
$\left( \begin{array}{ccc}l_1&l_2&l_3\\m_1&m_2&m_3 \end{array} \right)$
is a Wigner 3-$j$ symbol.
It can be written as
\begin{multline}
\left(\begin{array}{ccc} l&l'&L\\m&m'&M \end{array} \right) 
=\sqrt{\frac{(l+l'-L)!(l-l'+L)!(-l+l'+L)! (l+m)!(l-m)!(l'+m')!(l'-m')!(L+M)!(L-M)!}{(l+l'+L+1)!} } 
\\
\times \sum_{z \in \mathbb{Z}} 
\frac{(-1)^{z+l+l'-M}}{z!(l+l'-L-z)!(l-m-z)!(l'+m'-z)!(L-l'+m+z)!(L-l-m'+z)!}.
\end{multline}
See e.g.~\cite{Wigner:1959}, \cite{Messiah:1962} \cite{LLnonrelQM}, and references therein.
Note that although this sum is over all integers it contains only a 
finite number of non-zero terms since the factorial of a negative 
number is defined to be infinite.

\section{Legendre polynomials and associated Legendre functions}
\label{s:legendre_polynomials}

The following is a list of some useful relations involving 
Legendre polynomials $P_l(x)$ and associated Legendre functions 
$P_l^m(x)$.
For additional properties, see e.g., \cite{AbramowitzStegun}.
\\

\noindent
Differential equation:
\be
(1-x^2)\frac{{\rm d}^2}{{\rm d}x^2}P_l^m(x) 
- 2x \frac{{\rm d}}{{\rm d}x}P_l^m(x)
+\left[l(l+1)-\frac{m^2}{(1-x^2)}\right]\, P_l^m(x) = 0.
\ee
A useful recurrence relation:
\be
(x^2-1)\frac{{\rm d}}{{\rm d}x}P_l^m(x) = lxP_l^m(x) - (l+m)P_{l-1}^m(x).
\ee
Relation to ordinary Legendre polynomials, for $m=0,1,\cdots, l$:
\be
\begin{aligned}
P_l^m(x) &= (-1)^m (1-x^2)^{m/2}\frac{{\rm d}^m}{{\rm d}x^m}P_l(x)\,,
\\
P_l^{-m}(x) &= (-1)^m \frac{(l-m)!}{(l+m)!}P_l^m(x) .
\label{eq:Legendre}
\end{aligned}
\ee
Rodrigues' formula for $P_l(x)$:
\be
P_l(x) = \frac{1}{2^l l!}\frac{{\rm d}^l}{{\rm d}x^l}\left[(x^2-1)^l\right].
\ee
Series representation of Legendre polynomials:
\begin{equation}
P_l(x) 
=\sum_{k=0}^{l} (-1)^k \frac{ (l+k)! }{ (k!)^2 (l-k)! } 
\left(\frac{1-x}{2}\right)^k 
=\sum_{k=0}^l (-1)^{l+k} \frac{ (l+k)! }{ (k!)^2 (l-k)! } 
\left(\frac{1+x}{2}\right)^k.
\end{equation}
Orthogonality relation (for fixed $m$):
\be
\label{eq:PlmOrth}
\int_{-1}^1 {\rm d}x\> P_l^m(x)P_{l'}^m(x) = \frac{2(l+m)!}{(2l+1)(l-m)!}\,\delta_{ll'}.
\ee
Shifted Legendre polynomials (defined for $x\in[0,1]$):
\be
\tilde P_l(x) \equiv P_l(1-2x)
=\frac{1}{l!}\frac{{\rm d}^l}{{\rm d}x^l}\left((x-x^2)^l\right).
\ee

\section{Oscillatory behavior of the pulsar term}
\label{s:pulsar_term}

In Section \ref{sec:detresponse}, we assumed that the contribution of
the oscillatory term
$-2\cos[2\pi fL(1 + \hat{k} \cdot \hat{u})/c]$ 
to the integral for the auto-correlation was small, as it would 
be suppressed by at least a factor of $\sim\! 1/(fL/c)$.  
Here we calculate the exact expression for the contribution from this term 
for an isotropic, unpolarised and uncorrelated gravitational-wave background.  
Analogous calculations may be carried out for higher multipole moments.

If we include the pulsar term, the auto-correlation function for an
isotropic, unpolarised and uncorrelated background is given by
\be
\Gamma_{\rm auto}(f) 
= \frac{1}{(2\pi f)^2}
\int_{S^2}{\rm d}^2\Omega_{\hat k}\>
\left(2 - 2\cos[2\pi fL(1+\hat k\cdot \hat u)/c]\right)
\left[|F^+(\hat k)|^2 + |F^\times(\hat k)|^2\right]\,,
\ee
where $F^{+,\times}(\hat k)$ are given in Eq.~(\ref{e:RA_pulsars}).
(This is the same expression as in \cite{Mingarelli:2013}, 
but with an additional overall factor of $1/(2\pi f)^{2}$, 
which comes from the relation between the timing residual 
and redshift response functions.)
If we choose coordinates so that the pulsar is located along the 
$z$-axis, then
\be
F^+(\hat k) = \frac{1}{2}(1-\cos\theta)\,,
\qquad
F^\times(\hat k)=0\,,
\ee
for which
\be
\begin{aligned}
\Gamma_{\rm auto}(f)
&=\frac{1}{(2\pi f)^2}
\int_{S^2}{\rm d}^2\Omega_{\hat k}\>
\left(2 - 2\cos[2\pi fL\left(1+\cos\theta)/c\right]\right)
\frac{1}{4}(1-\cos\theta)^2
\\
&= \frac{\pi}{(2\pi f)^2}
\int_{-1}^1 {\rm d}x\>
\left(1 - \cos[2\pi fL(1+x)/c]\right)(1-x)^2\,.
\end{aligned}
\ee
This last integral is fairly simple to evaluate, the result being
\begin{align}
\Gamma_{\rm auto}(f) 
= \frac{1}{(2\pi f)^2}
\left\{\frac{8\pi}{3} - \frac{1}{\pi(fL/c)^2}\left[1-j_0(4\pi fL/c)\right]
\right\}\,,
\end{align}
where $j_0(x)\equiv \sin x /x$ is a spherical Bessel function of the
first kind.  Thus, for an isotropic, unpolarised and uncorrelated background, 
the oscillatory term is actually suppressed by a factor of $1/(fL/c)^2$.  
The pulsar-term contribution to the auto-correlation is therefore 
well-approximated by multiplying the overlap reduction function
by a factor of 2 and neglecting the oscillatory term.  For a
correlated gravitational-wave background, there is no factor of
2 contribution from the pulsar term, as discussed in
Sec.~\ref{s:discuss}.

\section{Response of a static interferometer}
\label{s:LIGO_response}

In the point-detector limit the strain response of a static 
interferometer whose vertex is located at the origin of
coordinates may be approximated by
\begin{equation}
R^A(f,k) = \frac{1}{2} e^A_{ab}(\hat{k}) 
\left(u_1^a u_1^b  - u_2^a u_2^b\right)\,,
\end{equation}
where $\hat u_1$ and $\hat u_2$ are unit vectors pointing 
along the two arms of the detector. 
We can evaluate this response in a computational frame
in which $\hat u_1$ is in the $\hat{z}$-direction and 
$\hat u_2$ is in the $\hat{x}$-direction. 
We consider the $\hat u_1$ term of the response only
to start with, which we denote by $R^P_{1(lm)}(f)$. In the
computational frame 
$e^+_{ab}(\hat{k})u_1^a u_1^b = \sin^2\theta$ 
and $e^\times_{ab}(\hat{k})u_1^a u_1^b = 0$.
This is independent of $\phi$ so as before
we need $m=0$ and since $X_{(l0)}(\hat k)=0$,
we find $R^C_{1(lm)}=0$ for all $l,m$. 
We also have $R^G_{1(lm)}=0$ for $m\neq0$ and
\be
\begin{aligned}
R^G_{1(l0)}(f) 
&= \frac{N_l}{2}\sqrt{(2l+1)\pi} 
\int_{-1}^1 {\rm d}x\>(1-x^2) G_{(lm)}^+(x) 
\\
&= 
\frac{N_l}{2}
\sqrt{(2l+1)\pi} 
\int_{-1}^1 {\rm d}x\>
\left[ -\left(\frac{1}{2} l(l-1)(1-x^2) +l\right) P_l(x) 
+ l x P_{l-1}(x) \right] 
\\
&= 2 \sqrt{\frac{\pi}{15}}\, \delta_{l,2}\,,
\end{aligned}
\ee
where the last line follows from orthogonality of the Legendre
polynomials and the fact that we have grad modes with $l \geq 2$
only. To do the $\hat u_2$ part of the integral we can carry out a
coordinate transformation to put $\hat u_2$ in the $\hat{z}$-direction 
(an anticlockwise rotation of $\pi/2$ radians about the
$\hat{y}$-axis). This transforms the values of the $a^P_{(lm)}$
coefficients. In the new frame the integral takes the same form as
before and as curl modes are transformed into curl modes, the curl
response is still zero. For the grad response we need only the
coefficient of $Y^G_{(20)}$ in the transformed coordinates. Using
Eq.~(\ref{e:harmtrans}), we see this is zero for $l \neq 2$ and
obtain the final result
\begin{equation}
R^G_{(lm)}(f) = 2\sqrt{\frac{\pi}{15}}\, \delta_{l,2} 
\left[1 - \sqrt{\frac{4\pi}{5}} Y_{2m}(\pi/2,0) \right]\,.
\end{equation}
In the same way we can
deduce that the response of an arbitrary two-arm detector in a frame
in which the arms point in directions with spherical coordinates
$(\theta_1, \phi_1)$ and $(\theta_2, \phi_2)$ is simply
\begin{equation}
R^G_{(lm)}(f) = \frac{4\pi}{5} \sqrt{\frac{1}{3}}\, \delta_{l,2} 
\left(Y_{2m}(\theta_1,\phi_1) - Y_{2m}(\theta_2,\phi_2) \right).
\end{equation}

If we drop the point-detector approximation but keep the vertex 
of the interferometer at the origin, the response function takes 
the form 
\begin{equation}
R^A(f,k) = \frac{1}{2} e^A_{ab}(\hat{k}) 
\left(
u_1^a u_1^b \,{\mathcal T}(f,\hat{k}\cdot \hat{u}_1) 
- u_2^a u_2^b \,{\mathcal T}(f,\hat{k}\cdot \hat{u}_2)\right)\,,
\end{equation}
where ${\mathcal T}$ is the transfer function. In the transformed reference
frames used above, the quantities $\hat{k}\cdot \hat u_1$ and 
$\hat{k}\cdot \hat u_2$ depend only on the transformed 
polar coordinate and are independent of the transformed azimuthal coordinate. 
The detector
therefore still has no response to curl modes, although there will now
be sensitivity to grad modes with $l > 2$. We leave a full treatment
of the response of an extended and moving interferometer to future
work.

\section{Correlation curves for anisotropic uncorrelated backgrounds}
\label{app:UncorrAniso}

Expressions for the spherical harmonic components of the 
frequency-independent part of the pulsar timing overlap reduction 
function for anisotropic, unpolarised and uncorrelated backgrounds 
are given in~\cite{Mingarelli:2013}:
\be
\bar\Gamma_{12,lm} =\int_{S^2}{\rm d}^2\Omega_{\hat k}\>
Y_{lm}(\hat k)\left[F_1^+(\hat k)F_2^+(\hat k)+
F_1^\times(\hat k)F_2^\times(\hat k)\right]\,.
\label{e:Gamma12_anisotropic}
\ee
Analytic expressions for the quadrupole and lower terms are derived 
in that paper.  Here we derive analytic expressions for 
$\bar\Gamma_{12,lm}$ for {\em all} values of $l$ and $m$, evaluating 
the integral in the computational 
frame, where pulsar 1 is located along the $z$-axis and pulsar 2 is 
located in the $xz$-plane, making an angle $\zeta$ with the $z$-axis:
\be
\begin{aligned}
\hat u_1&=(0,0,1)\,,
\\
\hat u_2&=(\sin\zeta,0,\cos\zeta)\,.
\end{aligned}
\ee
The calculation presented here differs from that presented 
in Secs.~\ref{s:alternative_derivation} and in Sec.~\ref{sec:overlap_general}, 
where the overlap reduction function is given by a sum of
{\em products} of integrals involving the detector response 
functions for each pulsar separately.
There we were able to evaluate the integrals over $\hat k$
and $\hat k'$ separately by rotating coordinates so that 
the relevant pulsar was located along the transformed $z$-axis.
Since the above expression for $\bar\Gamma_{12,lm}$ involves a
{\em single} integral of a product of detector response
functions, we can only rotate coordinates so that one pulsar
is located along the transformed $z$-azis.

In this frame, we find that:
\be
\begin{aligned}
F_1^+(\hat k) &= \frac{1}{2}(1-\cos\theta)\,,
\\
F_1^\times(\hat k) &= 0\,,
\\
F_2^+(\hat k) &= \frac{1}{2}
\left[(1-\cos\zeta\cos\theta-\sin\zeta\sin\theta\cos\phi)
-\frac{2\sin^2\zeta \sin^2\phi}
{1+\cos\zeta\cos\theta+\sin\zeta\sin\theta\cos\phi}
\right]\,,
\\
F_2^\times(\hat k) &= -\frac{1}{2}
\left(
\frac{\sin^2\zeta \cos\theta\sin(2\phi) - \sin(2\zeta)\sin\theta\sin\phi}
{1+\cos\zeta\cos\theta+\sin\zeta\sin\theta\cos\phi}
\right)\,.
\end{aligned}
\ee
Thus, the expression for $\bar\Gamma_{12,lm}$ simplifies to
\begin{align}
\bar\Gamma_{12,lm} 
=\int_{S^2}{\rm d}^2\Omega_{\hat k}\>
Y_{lm}(\hat k)F_1^+(\hat k)F_2^+(\hat k)
=\frac{1}{2}N_l^m\int_{-1}^1 {\rm d}x\>
(1-x)P_l^m(x)I_m(x)\,,
\end{align}
where
\be
I_m(x)\equiv 
\int_0^{2\pi}{\rm d}\phi\> F_2^+(\hat k) e^{im\phi}\,.
\ee
The integral for $I_m(x)$ can be evaluated using contour 
integration, making the substitutions $z=e^{i\phi}$,
$\cos\phi = (z+z^{-1})/2$, etc.
In general, the integrand has poles at 0, $z_+$, $z_-$, where
\begin{align}
z_+ \equiv -\frac{(1\mp\cos\zeta)(1\mp\cos\theta)}{\sin\zeta\sin\theta}
=-\sqrt{
\left(\frac{1\mp\cos\zeta}{1\pm\cos\zeta}\right)
\left(\frac{1\mp\cos\theta}{1\pm\cos\theta}\right)}\,,
\quad
z_- \equiv \frac{1}{z_+}\,.
\end{align}
In the above expressions, the top (bottom) sign corresponds 
to the region 
$-\cos\zeta\le \cos\theta\le 1$ 
($-1\le \cos\theta\le -\cos\zeta$).
One can show that for both of these regions, $z_+$ is 
inside the unit circle and hence contributes to the 
contour integral, while $z_-$ is outside the unit circle.
The results of the $\phi$-integrations 
for $m=0,1,2,\cdots$ are:
\be
\begin{aligned}
I_0(x) &= \pi\left[(1-x\cos\zeta) - 2\frac{(1\mp\cos\zeta)}{(1\pm x)}\right]\,,
\\
I_1(x) &=I_{-1}(x) = \pi\left[-\frac{1}{2}\sin\zeta(1-x^2)^{\frac{1}{2}} 
+ \frac{(1\mp\cos\zeta)^{\frac{3}{2}}}{(1\pm\cos\zeta)^{\frac{1}{2}}}
\frac{(1\mp x)^{\frac{1}{2}}}{(1\pm x)^{\frac{3}{2}}}\right]\,,
\\
I_m(x) &=I_{-m}(x)= (-1)^{m+1}\pi\left[
\frac{(1\mp\cos\zeta)^{\frac{m}{2}+1}}{(1\pm\cos\zeta)^{\frac{m}{2}}}
\frac{(1\mp x)^{\frac{m}{2}}}{(1\pm x)^{\frac{m}{2}+1}}
-
\frac{(1\mp\cos\zeta)^{\frac{m}{2}}}{(1\pm\cos\zeta)^{\frac{m}{2}-1}}
\frac{(1\mp x)^{\frac{m}{2}-1}}{(1\pm x)^{\frac{m}{2}}}
\right]\,,
\quad m=2,3,\cdots .
\end{aligned}
\ee
Since 
\be
P_l^m(x) = (-1)^m (1-x)^{m/2}(1+x)^{m/2}\frac{{\rm d}^m}{{\rm d}x^m}P_l(x)\,,
\ee
it follows that $\bar\Gamma_{12,lm}$ can be written in terms of sums of
integrals of the form
\be
\begin{aligned}
{\mathcal F}^-_{q,r,l,m}(\cos\zeta)
&\equiv\int_{-1}^{-\cos\zeta} {\rm d}x\> \frac{(1+x)^q}{(1-x)^r}
\frac{{\rm d}^m}{{\rm d}x^m}P_l(x)\,,
\\
{\mathcal F}^+_{q,r,l,m}(\cos\zeta)
&\equiv\int_{-\cos\zeta}^1 {\rm d}x\> \frac{(1-x)^q}{(1+x)^r}
\frac{{\rm d}^m}{{\rm d}x^m}P_l(x)\,,
\end{aligned}
\ee
where $q=m, m-1$ and $r=0, -1$ for ${\mathcal F}^-_{q,r,l,m}(\cos\zeta)$, and 
$q=m, m+1$ and $r=0, 1$ for ${\mathcal F}^+_{q,r,l,m}(\cos\zeta)$.
(The integrals for ${\mathcal F}^\pm_{q,r,l,m}(\cos\zeta)$ are explicitly
evaluated in the following subsection.)\\
 
\noindent
Explicit expressions for $\bar\Gamma_{12,lm}$ are given below:\\
\noindent
$m=0$:
\be
\begin{aligned}
\bar\Gamma_{12,l0}
&= \frac{\pi}{2}\sqrt{\frac{2l+1}{4\pi}}
\Bigg\{\int_{-1}^1 {\rm d}x\>(1-x)(1-x\cos\zeta)P_l(x)
\\
&\hspace{1in}-2(1+\cos\zeta)\int_{-1}^{-\cos\zeta}{\rm d}x\> P_l(x)
-2(1-\cos\zeta)\int_{-\cos\zeta}^{1}{\rm d}x\> \frac{(1-x)}{(1+x)}P_l(x)
\Bigg\}
\\
&= \frac{1}{2}\sqrt{(2l+1)\pi}
\Bigg\{\left(1+\frac{1}{3}\cos\zeta\right)\,\delta_{l0}
-\frac{1}{3}\left(1+\cos\zeta\right)\,\delta_{l1}
+\frac{2}{15}\cos\zeta\,\delta_{l2}
\\
&\hspace{1.5in}-(1+\cos\zeta){\mathcal F}^-_{0,0,l,0}(\cos\zeta)
-(1-\cos\zeta){\mathcal F}^+_{1,1,l,0}(\cos\zeta)
\Bigg\}\,,
\end{aligned}
\ee
where we substituted combinations of Legendre polynomials for $1,x,x^2$ 
and then used the orthogonality relation, Eq.~\eqref{eq:PlmOrth}, 
to evaluate the integral from $-1$ to $1$.\\

\noindent
Similarly for $m=1$:
\be
\begin{aligned}
\bar\Gamma_{12,l1}
&= \frac{\pi}{2}\sqrt{\frac{2l+1}{4\pi}\frac{(l-1)!}{(l+1)!}}
\Bigg\{\frac{1}{2}\sin\zeta\int_{-1}^1 {\rm d}x\>(1-x)(1-x^2)
\frac{{\rm d}}{{\rm d}x}P_l(x)
\\
&\hspace{.75in}-\frac{(1+\cos\zeta)^{3/2}}{(1-\cos\zeta)^{1/2}}
\int_{-1}^{-\cos\zeta}{\rm d}x\> (1+x)\frac{{\rm d}}{{\rm d}x}P_l(x)
-\frac{(1-\cos\zeta)^{3/2}}{(1+\cos\zeta)^{1/2}}
\int_{-\cos\zeta}^{1}{\rm d}x\> \frac{(1-x)^2}{(1+x)}
\frac{{\rm d}}{{\rm d}x}P_l(x)
\Bigg\}
\\
&= \frac{1}{4}\sqrt{(2l+1)\pi}\sqrt{\frac{(l-1)!}{(l+1)!}}
\Bigg\{
2\sin\zeta\left(\frac{1}{3}\,\delta_{l1}-\frac{1}{5}\,\delta_{l2}\right)
\\
&\hspace{1.75in}-\frac{(1+\cos\zeta)^{3/2}}{(1-\cos\zeta)^{1/2}}{\mathcal F}^-_{1,0,l,1}(\cos\zeta)
-\frac{(1-\cos\zeta)^{3/2}}{(1+\cos\zeta)^{1/2}}{\mathcal F}^+_{2,1,l,1}(\cos\zeta)
\Bigg\}\,,
\end{aligned}
\ee
where we used integration by parts and orthogonality of Legendre
polynomials, Eq.~\eqref{eq:PlmOrth}, to evaluate the integral from $-1$ to $1$.\\

\noindent
$m=2,3,\cdots$:
\be
\begin{aligned}
\bar\Gamma_{12,lm}
&= -\frac{\pi}{2}\sqrt{\frac{2l+1}{4\pi}\frac{(l-m)!}{(l+m)!}}
\Bigg\{
\frac{(1+\cos\zeta)^{\frac{m}{2}+1}}{(1-\cos\zeta)^{\frac{m}{2}}}
\int_{-1}^{-\cos\zeta}{\rm dx}\>(1+x)^m\frac{{\rm d}^m}{{\rm d}x^m}P_l(x)
\\
&\hspace{1.35in}
-\frac{(1+\cos\zeta)^{\frac{m}{2}}}{(1-\cos\zeta)^{\frac{m}{2}-1}}
\int_{-1}^{-\cos\zeta}{\rm dx}\>(1+x)^{m-1}(1-x)\frac{{\rm d}^m}{{\rm d}x^m}P_l(x)
\\
&\hspace{1.35in}
+\frac{(1-\cos\zeta)^{\frac{m}{2}+1}}{(1+\cos\zeta)^{\frac{m}{2}}}
\int_{-\cos\zeta}^{1}{\rm dx}\>\frac{(1-x)^{m+1}}{(1+x)}\frac{{\rm d}^m}{{\rm d}x^m}P_l(x)
\\
&\hspace{1.35in}
-\frac{(1-\cos\zeta)^{\frac{m}{2}}}{(1+\cos\zeta)^{\frac{m}{2}-1}}
\int_{-\cos\zeta}^{1}{\rm dx}\>(1-x)^{m}\frac{{\rm d}^m}{{\rm d}x^m}P_l(x)
\Bigg\}
\\
&= -\frac{1}{4}\sqrt{(2l+1)\pi}\sqrt{\frac{(l-m)!}{(l+m)!}}
\Bigg\{
\frac{(1+\cos\zeta)^{\frac{m}{2}+1}}{(1-\cos\zeta)^{\frac{m}{2}}}{\mathcal F}^-_{m,0,l,m}(\cos\zeta)
-\frac{(1+\cos\zeta)^{\frac{m}{2}}}{(1-\cos\zeta)^{\frac{m}{2}-1}}{\mathcal F}^-_{m-1,-1,l,m}(\cos\zeta)
\\
&\hspace{1.6in}+\frac{(1-\cos\zeta)^{\frac{m}{2}+1}}{(1+\cos\zeta)^{\frac{m}{2}}}{\mathcal F}^+_{m+1,1,l,m}(\cos\zeta)
-\frac{(1-\cos\zeta)^{\frac{m}{2}}}{(1+\cos\zeta)^{\frac{m}{2}-1}}{\mathcal F}^+_{m,0,l,m}(\cos\zeta)
\Bigg\}.
\end{aligned}
\ee
For $m<0$,
\be
\bar\Gamma_{12,lm} = (-1)^m\bar\Gamma_{12,l, -m}\,,
\ee
which follows from $Y_{lm}(\hat k)= (-1)^m Y_{l,-m}(\hat k)$.\\

Figure~\ref{f:gammaLM} in the main text shows plots of 
$\bar\Gamma_{12,lm}$ for $l=0,1,\cdots,5$, calculated using the above formulas.  
The plots for $l=0,1,2,3$ are identical to those in 
Figure~2 of \cite{Mingarelli:2013}, as expected.
From the plots, we see that when the two pulsars are coaligned
(i.e., when $\zeta=0$ or, equivalently, when $\hat u_1=\hat u_2$), 
the only non-zero contributions to the overlap reduction function
come from $m=0$ and $l=0$, 1, or 2.
To verify that this is the case for all $l\ge 3$,
we note that in the computational frame with the co-aligned pulsars 
located  along the $z$-axis:
\be
\begin{aligned}
\bar\Gamma_{11,lm} 
&=\int_{S^2} {\rm d}^2\Omega_{\hat k}\>
Y_{lm}(\hat k)\left[|F_1^+(\hat k)|^2 + |F_1^\times(\hat k)|^2\right]
\\
&=\frac{1}{4}\int_0^{2\pi} {\rm d}\phi\int_{-1}^1{\rm d}\cos\theta\>
Y_{lm}(\theta,\phi)(1-\cos\theta)^2
\\
&=\delta_{m0}\,\frac{\pi}{2}\sqrt{\frac{2l+1}{4\pi}}\int_{-1}^1{\rm d}x\>
P_l(x)(1-x)^2\,.
\end{aligned}
\ee
Since
\be
(1-x)^2 = 1-2x+x^2 = \frac{4}{3}P_0(x) - 2P_1(x) + \frac{2}{3}P_2(x)\,,
\ee
orthogonality of the Legendre polynomials, Eq.~\eqref{eq:PlmOrth}, 
implies
\be
\bar\Gamma_{11,lm} 
=\delta_{m0}\,\frac{\pi}{2}\sqrt{\frac{2l+1}{4\pi}}
\left[\frac{8}{3}\delta_{l0} - \frac{4}{3}\delta_{l1} + \frac{4}{15}\delta_{l2}\right]
=\delta_{m0}\left[\frac{\sqrt{4\pi}}{3}\,\delta_{l0}
-\sqrt{\frac{\pi}{3}}\,\delta_{l1}
+\frac{1}{3}\sqrt{\frac{\pi}{5}}\,\delta_{l2}\right]\,,
\ee
which is zero for $l\ge 3$ or $m\ne 0$.

\subsection{Evaluating the ${\mathcal F}^{\pm}_{q,r,l,m}(\cos\zeta)$ integrals}
\label{s:calF}

The integrals for ${\mathcal F}^\pm_{q,r,L,m}(\cos\zeta)$ can be 
evaluated using the series representation of the Legendre polynomials
\begin{equation}
P_l(x) = \sum_{k=0}^{l} (-1)^k \frac{ (l+k)! }{ (k!)^2 (l-k)! } 
\left(\frac{1-x}{2}\right)^k 
=\sum_{k=0}^l (-1)^{l+k} \frac{ (l+k)! }{ (k!)^2 (l-k)! } 
\left(\frac{1+x}{2}\right)^k .
\end{equation}
Explicitly, we find
\be
\begin{aligned}
{\mathcal F}^-_{q,r,L,m}(\cos\zeta)
&\equiv\int_{-1}^{-\cos\zeta} {\rm d}x\> 
\frac{(1+x)^q}{(1-x)^r}\frac{{\rm d}^m}{{\rm d}x^m}P_L(x) 
\\ 
&= \sum_{i=0}^q \sum_{j=m}^L 2^{i-j} (-1)^{q-i+j+m} 
\frac{ q! (L+j)!} {i! (q-i)! j! (L-j)! (j-m)! } 
\int_{-1}^{-\cos\zeta} {\rm d}x\>(1-x)^{q-i-r+j-m}\,,
\end{aligned}
\ee
for which
\be
\begin{aligned}
{\mathcal F}^-_{q,0,L,m}(\cos\zeta) 
&=\sum_{i=0}^q \sum_{j=m}^L 2^{i-j} (-1)^{q-i+j+m} 
\frac{ q! (L+j)! \left(2^{q-i+j-m+1} - (1+\cos\zeta)^{q-i+j-m+1}\right)} {i! (q-i)! j! (L-j)! (j-m)! (q-i+j-m+1)}\,,
\\
{\mathcal F}^-_{q,1,L,m}(\cos\zeta) 
&=\sum_{i=0}^{q-1} \sum_{j=m}^L 2^{i-j} (-1)^{q-i+j+m} 
\frac{ q! (L+j)! \left(2^{q-i+j-m} - (1+\cos\zeta)^{q-i+j-m}\right)} {i! (q-i)! j! (L-j)! (j-m)! (q-i+j-m)} 
\\
& \hspace{0.4cm} + \sum_{j=m+1}^L 2^{q-j} (-1)^{j+m} 
\frac{ (L+j)! \left(2^{j-m} - (1+\cos\zeta)^{j-m}\right)} {j! (L-j)! (j-m)! (j-m)} 
\\
& \hspace{0.8cm} + \frac{2^{q-m} (L+m)!}{m! (L-m)!} \ln\left(\frac{2}{1+\cos\zeta}\right)\,.
\end{aligned}
\ee
Similarly,
\be
\begin{aligned}
{\mathcal F}^+_{q,r,L,m}(\cos\zeta)
&\equiv\int_{-\cos\zeta}^1 {\rm d}x\> 
\frac{(1-x)^q}{(1+x)^r}\frac{{\rm d}^m}{{\rm d}x^m}P_L(x)
\\ 
&= \sum_{i=0}^q \sum_{j=m}^L 2^{i-j} (-1)^{L+q-i+j} \frac{ q! (L+j)!} {i! (q-i)! j! (L-j)! (j-m)! } 
\int_{-\cos\zeta}^1 {\rm d}x\>(1+x)^{q-i-r+j-m}\,,
\end{aligned}
\ee
for which
\be
\begin{aligned}
{\mathcal F}^+_{q,0,L,m}(\cos\zeta) 
&=\sum_{i=0}^q \sum_{j=m}^L 2^{i-j} (-1)^{L+q-i+j} 
\frac{ q! (L+j)! \left(2^{q-i+j-m+1} - (1-\cos\zeta)^{q-i+j-m+1}\right)} {i! (q-i)! j! (L-j)! (j-m)! (q-i+j-m+1)}\,,
\\
{\mathcal F}^+_{q,1,L,m}(\cos\zeta) 
&=\sum_{i=0}^{q-1} \sum_{j=m}^L 2^{i-j} (-1)^{L+q-i+j} 
\frac{ q! (L+j)! \left(2^{q-i+j-m} - (1-\cos\zeta)^{q-i+j-m}\right)} {i! (q-i)! j! (L-j)! (j-m)! (q-i+j-m)} 
\\
& \hspace{0.4cm} + \sum_{j=m+1}^L 2^{q-j} (-1)^{L+j} 
\frac{ (L+j)! \left(2^{j-m} - (1-\cos\zeta)^{j-m}\right)} {j! (L-j)! (j-m)! (j-m)} 
\\
& \hspace{0.8cm} + \frac{(-1)^{L+m} 2^{q-m} (L+m)!}{m! (L-m)!} \ln\left(\frac{2}{1-\cos\zeta}\right)\,.
\end{aligned}
\ee
We also need to evaluate ${\mathcal F}^-_{q,r,l,m}(\cos\zeta)$ for $r=-1$.
This can be reduced to combinations
of ${\mathcal F}^-_{q,0,l,m}(\cos\zeta)$ and 
${\mathcal F}^-_{q+1,0,l,m}(\cos\zeta)$ by writing $(1-x)=2-(1+x)$:
\be
{\mathcal F}^-_{q,-1,l,m}(\cos\zeta) =
2{\mathcal F}^-_{q,0,l,m}(\cos\zeta) -{\mathcal F}^-_{q+1,0,l,m}(\cos\zeta)\,.
\ee
Alternatively, we can just evaluate this integral directly, finding
\be
{\mathcal F}^-_{q,-1,l,m}(\cos\zeta) 
=\sum_{i=0}^q \sum_{j=m}^l 2^{i-j} (-1)^{q-i+j+m} 
\frac{ q! (l+j)! \left(2^{q-i+j-m+2} - (1+\cos\zeta)^{q-i+j-m+2}\right)} 
{i! (q-i)! j! (l-j)! (j-m)! (q-i+j-m+2)}\,.
\ee
\end{appendix}

\end{widetext}

\bibliography{manuscript}

\end{document}